\documentclass[usenatbib,usegraphicx,usedcolumn]{mn2e}
\usepackage{times,amssymb,amsmath}
\title[Galerkin methods for astrophysics]{A discontinuous Galerkin method for solving the fluid and MHD equations in astrophysical simulations}
\author[P. Mocz et. al.]{Philip Mocz$^{1}$\thanks{E-mail: pmocz@cfa.harvard.edu (PM)}, Mark Vogelsberger$^{1}$\thanks{Hubble Fellow}, Debora Sijacki$^{1,2}$, R\"udiger Pakmor$^{3}$, and Lars Hernquist$^{1}$ \\
$^{1}$Harvard University, Cambridge, MA 02138, USA\\
$^{2}$Kavli Institute for Cosmology, Cambridge and Institute of Astronomy, Madingley Road, Cambridge, CB3 0HA\\
$^{3}$Heidelberger Institut f\"ur Theoretische Studien, Schloss-Wolfsbrunnenweg 35, 69118 Heidelberg, Germany\\}
\begin{document}

\date{subm. to MNRAS, xx xxx 2013}

\pagerange{\pageref{firstpage}--\pageref{lastpage}} \pubyear{2013}

\maketitle

\label{firstpage}

\begin{abstract}

A discontinuous Galerkin (DG) method suitable for large-scale astrophysical simulations
on Cartesian meshes as well as arbitrary static and moving Voronoi meshes is presented. Most major astrophysical fluid
dynamics codes use a finite volume (FV) approach.  We demonstrate that
the DG technique offers distinct advantages over FV formulations on
both static and moving meshes.  The DG method is also easily
generalized to higher than second-order accuracy without requiring the
use of extended stencils to estimate derivatives (thereby making the
scheme highly parallelizable).  We implement the technique in the {\sc
Arepo} code for solving the fluid and the magnetohydrodynamic (MHD)
equations.  By examining various test problems, we show that our new
formulation provides improved accuracy over FV approaches of the same
order, and reduces post-shock oscillations and artificial diffusion of
angular momentum.  In addition, the DG method makes it possible to
represent magnetic fields in a locally divergence-free way, improving
the stability of MHD simulations and moderating global divergence
errors, and is a viable alternative for solving the MHD equations on
meshes where Constrained-Transport (CT) cannot be applied.  We find
that the DG procedure on a moving mesh is more sensitive to the choice
of slope limiter than is its FV method counterpart.  Therefore, future
work to improve the performance of the DG scheme even further will
likely involve the design of optimal slope limiters.  As presently
constructed, our technique offers the potential of improved accuracy
in astrophysical simulations using the moving mesh {\sc Arepo} code as
well as those employing adaptive mesh refinement (AMR).

\end{abstract}


\section{Introduction}\label{sec:intro}

Discontinuous Galerkin (DG) methods have recently been implemented for
solving systems of conservation laws to arbitrary orders of accuracy,
and have been shown to be competitive with more established and
traditional finite volume (FV) approaches
\citep{Bassi:1997:HAD:254433.254435,Bassi:1997:HAD:274112.274113,Cockburn:2004:LDD:1008428.1008437,Luo:2008:DGM:1410469.1410594,Li:2005:LDD:1057321.1057352}.
In this paper, we develop a second-order DG formulation for arbitrary
moving and static meshes that is appropriate for even the largest
astrophysical simulations.  DG techniques offer numerous advantages
over FV methods, as summarized by \cite{Luo:2008:DGM:1410469.1410594}.
In particular, DG procedures can be applied to arbitrary meshes
(moving meshes, adaptive mesh refinement (AMR)) and the method is
``compact'' in the sense that each cell is treated independently and
elements communicate only with adjacent elements having a common face
irrespective of the order of accuracy.  The DG method is conservative
and requires solving the Riemann problem across cell interfaces,
similar to FV schemes.  The main challenge with DG implementations
lies in minimizing unphysical post-shock oscillations (e.g. with slope
limiting, flux limiting, shock capturing, or weighted essentially
non-oscillatory (WENO) approaches)
\citep{Luo:2008:DGM:1410469.1410594}, which is also an issue for FV
methods.  Some DG formulations are found to be more sensitive to
certain shock limiters than their FV counterparts, but techniques
exist to prevent unphysical oscillatory solutions in high-order DG
methods \citep{NME:NME1172,FLD:FLD1823,Luo:2008:DGM:1410469.1410594}.

Our DG implementation falls into the class of centroidal Taylor basis
procedures developed by \cite{Luo:2008:DGM:1410469.1410594}.  This
formulation of DG is relatively new and is quite different from the
more widespread approach that employs nodal basis value functions
(which would not be generalizable to a moving Voronoi mesh where the
number of faces per cell can change with time).  The primary
difference between our centroidal DG method and FV schemes is in the
manner in which gradients (as well as higher order derivatives) are
computed for the solution of fluid variables in a cell.  FV methods
require the use of an extended stencil (which become spatially broad
for estimating higher order terms).  DG techniques, on the other hand,
evolve the coefficients of a set of basis functions that describe the
solution local to a cell in the same way that cell-averages are
evolved in the FV approach.  This localizes the solution within a
given cell, which can lead to reduced numerical errors and makes
codes highly parallelizable.  The centroidal DG approach may thus be
viewed as an extension of the FV method.

Moreover, the DG procedure allows for a locally divergence-free
representation of the solution in a cell
\citep{Luo:2008:DGM:1410469.1410594,Li:2005:LDD:1057321.1057352}.
This not only reduces the amount of memory required to store the
result but is also a useful for improving the accuracy of
magnetohydrodynamic (MHD) simulations.  The continuum equations of
ideal MHD impose the condition $\nabla\cdot \mathbf{B}=0$, but
discretized versions of the equations do not necessarily preserve the
zero divergence constraint. The locally divergence-free DG method
keeps divergences to zero within cell domains unlike FV schemes, but
does not guarantee a globally divergence-free solution (equivalent to
continuous transverse magnetic field components across cell
interfaces) due to the local discontinuous representation of the
result (see \S~\ref{sec:divB}).

The strictest approach for preserving $\nabla\cdot \mathbf{B}=0$ at the
discretized level to machine precision is the Constrained-Transport
(CT) framework, developed for the MHD equations by
\cite{1988ApJ...332..659E}.  The CT method uses Stoke's theorem to
represent the magnetic fields by face-averaged rather than
cell-averaged quantities.  However, while the CT method can be easily
implemented on fixed rectangular grids when a single, fixed timestep
is used, it is presently not known whether CT can be adopted for
meshes of arbitrary structure, moving meshes, or general time-stepping
schemes.  The CT approach has been implemented in AMR codes by using
synchronized time-stepping and restriction and prolongation operators
\citep{2001JCoPh.174..614B,2006AA...457..371F,2011ApJS..195....5M},
although this makes the original AMR formulation significantly more
complicated. In addition, sometimes CT schemes coupled with Godunov
methods need to be modified to prevent pressures from becoming negative
at the cost of maintaining conservation of energy to machine precision
\citep{1999JCoPh.149..270B}. A number of divergence cleaning schemes,
such as the Dedner hyperbolic cleaning method and the Powell 8-wave
technique, have been developed for controlling global divergence
errors in situations where the CT algorithm cannot be employed
\citep{1999JCoPh.154..284P,Toth:2000:DBC:349920.349997,2002JCoPh.175..645D}. The
locally divergence-free DG implementation, either on its own, or
coupled to a cleaning scheme, may improve the divergence-free
constraint.

Widely-used grid-based codes for solving fluid flows in astrophysical
systems, such as {\sc Flash} \citep{2000ApJS..131..273F}, {\sc Enzo}
\citep{2004astro.ph..3044O}, {\sc Ramses} \citep{2006AA...457..371F},
{\sc Athena} \citep{2008ApJS..178..137S}, and {\sc Arepo}
\citep{2010MNRAS.401..791S} all employ the FV approach.  These codes
have had numerous successes in simulating cosmological structure
formation, galaxy interactions, the interstellar medium, and
protoplanetary and accretion discs.  Here we investigate whether the
DG procedure offers a viable alternative for designing future
generations of simulation codes by directly comparing second-order DG
and FV methods with the same time integration scheme.

Our goal is to develop a DG algorithm for arbitrary meshes that is
efficient and simple in its implementation and, to the extent
possible, minimizes numerical errors, artificial diffusion of angular
momentum, and global inaccuracies in the magnetic field.  We are
particularly interested in the application of the method to moving
mesh algorithms, such as the {\sc Arepo} code written by
\cite{2010MNRAS.401..791S}.  The moving mesh technique is a novel type
of fluid solver that is essentially a hybrid of traditional static
Eulerian codes and the pseudo-Lagrangian, mesh-free smoothed particle
hydrodynamics (SPH) method.  In the moving mesh approach, fluid
elements move with the local velocity flow, rendering the method
quasi-Lagrangian.  This greatly reduces advection errors arising from
large bulk velocity motions of the flow, making the code well-suited
for simulating galaxy collisions.  {\sc Arepo} has been generalized to
solve the Navier-Stokes equations \citep{2013MNRAS.428..254M}, as well
the MHD equations using the Dedner and Powell divergence cleaning
schemes \citep{2011MNRAS.418.1392P,2012arXiv1212.1452P}. We show in
what follows that the DG scheme can improve the accuracy of the
current version of {\sc Arepo} as well as other FV codes.

Our paper is organized as follows. In \S~\ref{sec:methods} we describe
the DG method with centroidal Taylor basis functions and demonstrate
how it is a natural generalization of the FV method.  In
\S~\ref{sec:results} we present the results of numerical tests in
which we compare the DG and FV methods.  In \S~\ref{sec:disc1} we
summarize the main findings of these tests. In \S~\ref{sec:disc2} we
discuss the advantages of DG methods for astrophysical applications.
In \S~\ref{sec:disc3} we briefly offer possible ways of refining the
slope limiting technique, which could improve the accuracy of the DG
method even further. In \S~\ref{sec:conclusion} we provide
conclusions.

\section{Discontinuous Galerkin formulation}\label{sec:methods}

\subsection{Governing equations}\label{sec:eqns}

The ideal MHD equations can be written in conservative form as:
\begin{equation}
\frac{\partial\mathbf{U}}{\partial t} + \nabla \cdot \mathbf{F} = 0,
\label{eqn:cons}
\end{equation}
where $\mathbf{U}$ is the conservative state vector and $\mathbf{F}(\mathbf{U})$ is the flux function:
\begin{equation}
\mathbf{U} = \begin{pmatrix}\rho \\ \rho\mathbf{v} \\ \rho e \\ \mathbf{B}\end{pmatrix},\,\,\,\,\,\,\,\,\,
\mathbf{F} = \begin{pmatrix} \rho\mathbf{v} \\ \rho\mathbf{v}\mathbf{v}^T + p - \mathbf{B}\mathbf{B}^T \\ \rho e \mathbf{v} + p\mathbf{v} - \mathbf{B}(\mathbf{v}\cdot \mathbf{B}) \\ \mathbf{B}\mathbf{v}^T - \mathbf{v}\mathbf{B}^T\end{pmatrix}.
\end{equation}
Here, $\rho$ is the gas density, $p=p_{\rm gas} + \frac{1}{2}\mathbf{B}^2$ is the total gas pressure, $e=\rho u+\frac{1}{2}\mathbf{v}^2+\frac{1}{2\rho}\mathbf{B}^2$ is the total energy per unit mass, and $u$ is the thermal energy per unit mass.  In the numerical examples described in this paper, we consider an equation of state of the form $p_{\rm gas} = (\gamma - 1)\rho u$, where $\gamma$ is the adiabatic index.

The above equations reduce to the Euler equations (which describe compressible, inviscid flows) in the case that $\mathbf{B}=0$.

\subsection{Discontinuous Galerkin method}\label{sec:dg}

The DG method is defined by first considering the weak formulation of the conservation equations (\ref{eqn:cons}) obtained by multiplying by a test function $\mathbf{W}$, integrating over the domain ($\Omega$), and performing an integration by parts:
\begin{equation}
  \int_\Omega \frac{\partial \mathbf{U}}{\partial t} \mathbf{W}\,d\Omega 
+ \int_\Gamma \mathbf{F}\cdot\hat{\mathbf{n}} \mathbf{W}\,d\Gamma
+ \int_\Omega \mathbf{F}\cdot\nabla\mathbf{W}\,d\Omega = 0 ,
\label{eqn:dg}
\end{equation}
where $\Gamma = \partial \Omega$ is the boundary of $\Omega$ and $\hat{\mathbf{n}}$ is the outward unit normal vector of the boundary.

We seek to discretize Equation~(\ref{eqn:dg}). We begin by writing the solution in each cell $e$ as a second-order accurate Taylor series expansion about the centroid $(x_c, y_c)$, and employing coordinates $(x,y)$ with origin at $(x_c,y_c)$.  For example, in 2D:
\begin{equation}
\mathbf{U}_e = \tilde{\mathbf{U}}_e + \frac{\partial \mathbf{U}_e}{\partial x}|_cx + \frac{\partial \mathbf{U}_e}{\partial y}|_cy ,
\end{equation}
where $\tilde{\mathbf{U}}_e$ is the cell average of the fluid variables. 
Our local basis functions for each fluid variable are:
\begin{equation}
V_1 = 1,\, V_2 = x,\, V_3 = y,
\end{equation}
and the unknowns in this problem are the cell averages and the cell derivatives.

If we use the basis functions $V_i$ each as possible test functions $\mathbf{W}$, we obtain a set of evolution equations for the cell averages and derivatives (see also \cite{Luo:2008:DGM:1410469.1410594} for a formal mathematical presentation of centroidal DG methods):
\begin{equation}
\frac{d}{dt} \int_{\Omega_e} \tilde{\mathbf{U}}_e \,d\Omega + \int_{\Gamma_e} \mathbf{F}(\mathbf{U}_e)\cdot \hat{\mathbf{n}}\,d\Gamma = 0 ,
\label{eqn:fv}
\end{equation}

\begin{equation}
\begin{split}
\frac{d}{dt} \int_{\Omega_e} \begin{pmatrix} x^2 & xy \\ xy & y^2 \end{pmatrix}
\begin{pmatrix} \frac{\partial U_{e,i}}{\partial x}|_c \\[1em]  \frac{\partial U_{e,i}}{\partial y}|_c \end{pmatrix} \,d\Omega \\
+ \int_{\Gamma_e} \mathbf{F}(U_{e,i})\cdot \hat{\mathbf{n}} \begin{pmatrix} x \\ y \end{pmatrix}\,d\Gamma  \\
- \int_{\Omega_e}  \mathbf{F}(U_{e,i}) \cdot \nabla \begin{pmatrix}  x \\  y \end{pmatrix}  \,d\Omega \\
= 0 ,
\label{eqn:dgderiv}
\end{split}
\end{equation}
where $U_{e,i}$ is a single component of $\mathbf{U}_e$.  Thus, the $x$ and $y$ derivatives are coupled and 
it is necessary to invert a $2$ by $2$ matrix to obtain the derivatives.

We see from Equations~(\ref{eqn:fv}) and (\ref{eqn:dgderiv}) that the cell-averages and cell derivatives decouple for the choice of the Taylor-basis function. In fact, Equation~(\ref{eqn:fv}) is the same equation used to update a FV scheme. Thus the centroidal DG method is a natural higher order generalization of the FV approach if one asserts that cells are only allowed to communicate with their nearest neighbors.

The matrix 
\begin{equation}
M = \int_{\Gamma_e} \begin{pmatrix} x^2 & xy \\ xy & y^2 \end{pmatrix}\,d\Omega
\end{equation}
in Equation~(\ref{eqn:dgderiv}) is called the mass matrix. It stores second-order moments of the cell (which have to be computed for every active cell at every timestep in a moving mesh approach, and every time a cell is refined in an AMR approach) and is symmetric. The moments for each cell are calculated exactly using Gaussian quadrature. 

In the 3D case, using a Taylor basis we have the following weak formulation 
of the Euler equations 
for a cell $e$ (again, coordinates $(x,y,z)$ in the notation below have origin $(x_c,y_c,z_c)$): 
\begin{equation}
\frac{d}{dt} \int_{\Omega_e} \tilde{\mathbf{U}}_e \,d\Omega + \int_{\Gamma_e} \mathbf{F}(\mathbf{U}_e)\cdot \hat{\mathbf{n}}\,d\Gamma = 0 ,
\label{eqn:fv3d}
\end{equation}

\begin{equation}
\begin{split}
\frac{d}{dt} \int_{\Omega_e} \begin{pmatrix} x^2 & xy & xz \\ xy & y^2 & yz \\ xz & yz & z^2 \end{pmatrix}
\begin{pmatrix} \frac{\partial U_{e,i}}{\partial x}|_c \\[1em] \frac{\partial U_{e,i}}{\partial y}|_c \\[1em] \frac{\partial U_{e,i}}{\partial z}|_c \end{pmatrix} \,d\Omega \\
+ \int_{\Gamma_e} \begin{pmatrix} \mathbf{F}(U_{e,i})\cdot \hat{\mathbf{n}}\,x \\  \mathbf{F}(U_{e,i})\cdot \hat{\mathbf{n}}\,y \\ \mathbf{F}(U_{e,i})\cdot \hat{\mathbf{n}}\,z \end{pmatrix} \,d\Gamma  \\
- \int_{\Omega_e}  \begin{pmatrix}  \mathbf{F}_x(U_{e,i}) \\ \mathbf{F}_y(U_{e,i}) \\  \mathbf{F}_z(U_{e,i}) \end{pmatrix}  \,d\Omega \\
= 0 .
\label{eqn:dgderiv3d}
\end{split}
\end{equation}

Now if we define the volume and moment-averaged quantities $\mathbf{Q}_e$ and $\mathbf{R}_e$ as:
\begin{equation}
\mathbf{Q}_e = \int_{\Omega_e}\mathbf{U}_e\,d\Omega ,
\end{equation}
\begin{equation}
\mathbf{R}_{e,i} = \int_{\Omega_e} \begin{pmatrix} x^2 & xy & xz \\ xy & y^2 & yz \\ xz & yz & z^2 \end{pmatrix}
\begin{pmatrix} \frac{\partial U_{e,i}}{\partial x}|_c \\[1em] \frac{\partial U_{e,i}}{\partial y}|_c \\[1em] \frac{\partial U_{e,i}}{\partial z}|_c \end{pmatrix} \,d\Omega
\end{equation}
then we can write a second-order conservative discretization in time of Equations~(\ref{eqn:fv3d}) and (\ref{eqn:dgderiv3d}):
\begin{equation}
\mathbf{Q}_e^{(n+1)} = \mathbf{Q}_e^{(n)} - \Delta t \sum_f A_{ef} \hat{\mathbf{F}}_{ef}^{(n+1/2)},
\end{equation}

\begin{equation}
\begin{split}
\mathbf{R}_{e,i}^{(n+1)} & = \mathbf{R}_{e,i}^{(n)} \\ 
&- \Delta t \sum_f A_{ef} \hat{\mathbf{F}}_{ef}(U_{e,i})^{(n+1/2)}  \begin{pmatrix} c_{ef,x} \\
c_{ef,y} \\ c_{ef,z} \\  \end{pmatrix} \\
&+\Delta t \int_{\Omega_e}  \begin{pmatrix}  \mathbf{F}_x(U_{e,i})^{(n+1/2)} \\ \mathbf{F}_y(U_{e,i})^{(n+1/2)} \\  \mathbf{F}_z(U_{e,i})^{(n+1/2)} \end{pmatrix}  \,d\Omega ,
\end{split}
\label{eqn:GalerkUpdate}
\end{equation}
where $\hat{\mathbf{F}}_{ef}^{(n+1/2)}$ is an appropriately time-averaged approximation to the true flux $\mathbf{F}_{ef}$ across a cell face between cells $e$ and $f$, $A_{ef}$ is the area of the cell face, and $(c_{ef,x},c_{ef,y},c_{ef,z})$ is the location of the centroid of the cell face in the coordinate system local to cell $e$. The volume integral in the interior of the cell is carried out with Gaussian quadrature.

The basic idea of the DG method is to update $\mathbf{Q}_e$ and $\mathbf{R}_e$ for each active cell during each timestep. One can then obtain the cell averages of conserved fluid variables at the end of the step by dividing the $\mathbf{Q}_e$ by the cell volume (and consequently translated to primitive variables). Derivative information is obtained by matrix inversion of the mass matrix applied to $\mathbf{R}_e$. Derivatives of primitive variables may then be calculated by expanding the derivatives and solving for the primitive gradients, as, for example:
\begin{equation}
\begin{split}
\frac{\partial(\rho v_x)}{\partial y} = \rho \frac{\partial(v_x)}{\partial y}  + v_x\frac{\partial(\rho )}{\partial y}
\\
\Rightarrow
 \frac{\partial(v_x)}{\partial y} = \frac{1}{\rho}\left(\frac{\partial(\rho v_x)}{\partial y} - v_x\frac{\partial(\rho )}{\partial y}\right)
 \end{split}.
\end{equation}

Finally, in order to represent magnetic fields, we use a locally divergence-free basis (e.g. in 2D) for $\begin{pmatrix} B_x \\ B_y \end{pmatrix}$ instead of a Taylor basis, in particular:
\begin{equation}
\mathbf{V}_1 = \begin{pmatrix} 1 \\ 0 \end{pmatrix},\, 
\mathbf{V}_2 = \begin{pmatrix} 0 \\ 1 \end{pmatrix},\,
\mathbf{V}_3 = \begin{pmatrix} y \\ 0 \end{pmatrix},\, 
\mathbf{V}_4 = \begin{pmatrix} 0 \\ x \end{pmatrix},\,
\mathbf{V}_5 = \begin{pmatrix} x \\ -y \end{pmatrix}.
\end{equation}
The number of required basis functions is reduced from $6$ to $5$ owing to 
the divergence-free constraint. We derive an equation for the evolution of the magnetic field gradients analogously to Equation~(\ref{eqn:dgderiv});
namely:
\begin{equation}
\begin{split}
\frac{d}{dt} \int_{\Omega_e} \begin{pmatrix} y^2 & 0 & xy \\ 0 & x^2 & -xy \\ xy & -xy & x^2+y^2 \end{pmatrix}
\begin{pmatrix} \alpha_1 \\ \alpha_2 \\ \alpha_3 \end{pmatrix} \\
+ \int_{\Gamma_e} \begin{pmatrix} \mathbf{F}(B_x)\cdot \hat{\mathbf{n}}\,y \\  \mathbf{F}(B_y)\cdot \hat{\mathbf{n}}\,x  \\ \mathbf{F}(B_x)\cdot \hat{\mathbf{n}}\,x - \mathbf{F}(B_y)\cdot \hat{\mathbf{n}}\,y  \end{pmatrix} \,d\Gamma  \\
- \int_{\Omega_e}  \begin{pmatrix}  \mathbf{F}_y(B_x) \\ \mathbf{F}_x(B_y) \\  \mathbf{F}_x(B_x)-\mathbf{F}_y(B_y) \end{pmatrix}  \,d\Omega \\
=0 ,
\end{split}
\end{equation}
where
\begin{equation}
\alpha_1 = \frac{\partial B_x}{\partial y},
\end{equation}
\begin{equation}
\alpha_2 = \frac{\partial B_y}{\partial x},
\end{equation}
\begin{equation}
\alpha_3 = \frac{\partial B_x}{\partial x}=- \frac{\partial B_x}{\partial x}, 
\end{equation}
and the coordinates $(x,y)$ have origin $(x_c,y_c)$.
In this case, a $3$ by $3$ matrix has to be inverted to directly obtain all the magnetic field gradients.

In 3D, locally divergence-free basis functions may be chosen as:
\begin{equation}
\begin{split}
\mathbf{V}_1 = \begin{pmatrix} 1 \\ 0 \\ 0 \end{pmatrix},\, 
\mathbf{V}_2 = \begin{pmatrix} 0 \\ 1 \\ 0 \end{pmatrix},\, 
\mathbf{V}_3 = \begin{pmatrix} 0 \\ 0 \\ 1 \end{pmatrix},\,      \\
\mathbf{V}_4 = \begin{pmatrix} y \\ 0 \\ 0\end{pmatrix},\, 
\mathbf{V}_5 = \begin{pmatrix} z \\ 0 \\ 0\end{pmatrix},\,       \\
\mathbf{V}_6 = \begin{pmatrix} 0 \\ x \\ 0\end{pmatrix},\, 
\mathbf{V}_7 = \begin{pmatrix} 0 \\ z \\ 0\end{pmatrix},\,       \\
\mathbf{V}_8 = \begin{pmatrix} 0 \\ 0 \\ x\end{pmatrix},\, 
\mathbf{V}_9 = \begin{pmatrix} 0 \\ 0 \\ y\end{pmatrix},\,       \\
\mathbf{V}_{10} = \begin{pmatrix} x \\ -y \\ 0 \end{pmatrix},\,
\mathbf{V}_{11} = \begin{pmatrix} x \\ 0 \\ -z \end{pmatrix} .
\end{split}
\end{equation}
In which case one obtains the weak formulation 
of the evolution equation for the coefficients of the bases that determine the derivatives ($\alpha_3,\ldots,\alpha_{11}$):
\begin{equation}
\begin{split}
\frac{d}{dt} \int_{\Omega_e} 
\mathbf{M}_B
\begin{pmatrix} \alpha_3 \\ \alpha_4 \\ \alpha_5 \\ \alpha_6 \\ \alpha_7 \\ \alpha_8 \\ \alpha_9 \\ \alpha_{10} \\ \alpha_{11} \end{pmatrix} \\
+ \int_{\Gamma_e} 
\begin{pmatrix} 
\mathbf{F}(B_x)\cdot \hat{\mathbf{n}}\,y \\ 
\mathbf{F}(B_x)\cdot \hat{\mathbf{n}}\,z \\ 
\mathbf{F}(B_y)\cdot \hat{\mathbf{n}}\,x \\ 
\mathbf{F}(B_y)\cdot \hat{\mathbf{n}}\,z \\ 
\mathbf{F}(B_z)\cdot \hat{\mathbf{n}}\,x \\ 
\mathbf{F}(B_z)\cdot \hat{\mathbf{n}}\,y \\ 
\mathbf{F}(B_x)\cdot \hat{\mathbf{n}}\,x - \mathbf{F}(B_y)\cdot \hat{\mathbf{n}}\,y \\ 
\mathbf{F}(B_x)\cdot \hat{\mathbf{n}}\,x - \mathbf{F}(B_z)\cdot \hat{\mathbf{n}}\,z
\end{pmatrix} \,d\Gamma  \\
- \int_{\Omega_e}  
\begin{pmatrix}  
\mathbf{F}_y(B_x) \\ 
\mathbf{F}_z(B_x) \\
\mathbf{F}_x(B_y) \\ 
\mathbf{F}_z(B_y) \\
\mathbf{F}_x(B_z) \\ 
\mathbf{F}_y(B_z) \\
\mathbf{F}_x(B_x) - \mathbf{F}_y(B_y) \\ 
\mathbf{F}_x(B_x) - \mathbf{F}_z(B_z)
\end{pmatrix}  \,d\Omega \\
=0 ,
\end{split}
\end{equation}
where
\begin{equation}
\mathbf{M}_B=
\begin{pmatrix} 
y^2 & yz & 0 & 0 & 0 & 0 & xy & xy \\
yz & z^2 & 0 & 0 & 0 & 0 & xz & xz \\
0 & 0 & x^2 & xz & 0 & 0 & -xy & 0 \\
0 & 0 & xz & x^2 & 0 & 0 & -zy & 0 \\
0 & 0 & 0 & 0 & x^2 & xy & 0 & -xz \\
0 & 0 & 0 & 0 & xy & y^2 & 0 & -yz \\
xy & xz & -xy & -xz & 0 & 0 & x^2-y^2 & x^2 \\
xy & xz & 0 & 0 & -xz & -yz & x^2 & x^2-z^2 
\end{pmatrix} .
\end{equation}

\subsection{Fluid dynamics on a moving mesh}\label{sec:movingmesh}

In the case of non-static meshes, the Euler and MHD equations
need to be modified to account for the 
motion of the grid. 
The flux over an interface moving at velocity $\mathbf{w}$ or inside a cell moving at velocity $\mathbf{w}$ is a combination of the static flux and an advection step due to the movement:
\begin{equation}
\mathbf{F}_{\rm m}(\mathbf{U}) = \mathbf{F}_{\rm s}(\mathbf{U}) - \mathbf{U}\mathbf{w}^T .
\end{equation}

All Riemann problems across cell interfaces are solved in the rest-frame of the face, followed by adding appropriate terms to return to the lab frame, as described in detail in \cite{2011MNRAS.418.1392P} (see their equation 17). This approach retains a stable, upwind character. The Riemann problem is solved using an exact solver for the Euler equations and an HLLD solver \citep{2005JCoPh.208..315M} for the MHD equations. 

\subsection{Time stepping}\label{sec:time}
We use a second-order accurate in time MUSCL-Hancock scheme to update the fluid variables at the next timestep, the same method 
as is used for {\sc Arepo}'s FV solver, described in \cite{2010MNRAS.401..791S}. In the MUSCL-Hancock 
procedure \citep{1974JCoPh..14..361V,toro1999riemann} the essential idea is to use cell averages to predict the values of the primitive quantities at cell edges half a timestep in advance (equation (18) of \cite{2010MNRAS.401..791S}), and use these predicted values to solve the Riemann problem and obtain $\hat{\mathbf{F}}_{ef}^{(n+1/2)}$ in order to finish updating the solution to the next timestep. The same prediction equations 
are also used to calculate the flux in the interior of the cell in the volume integral term of Equation~(\ref{eqn:GalerkUpdate}). We find that this explicit time-updating scheme works very well for our DG method. Our approach is different from traditional DG formulations that typically use explicit or implicit Runge-Kutta techniques \citep{Luo:2008:DGM:1410469.1410594}. A benefit of using the second-order MUSCL-Hancock 
integrator is that it can be coupled 
with a symplectic second-order gravity solver to treat fluids with self-gravity \citep{2010MNRAS.401..791S}. 

The fact that we use the same time integration scheme for the DG and FV methods allow us to compare the 
advantages of one over the other in a direct manner. The primary difference in the second-order DG technique compared to the FV 
approach is only in the way in which cell gradients are handled. In a timestep, it is the quantity 
\begin{equation}
\int \mathbf{R}_{e,i} = M\begin{pmatrix} \frac{\partial U_{e,i}}{\partial x}|_c \\[1em] \frac{\partial U_{e,i}}{\partial y}|_c \end{pmatrix}\,d\Omega
\label{eqn:GalerkinVar}
\end{equation}
that is evolved by Equation~(\ref{eqn:GalerkUpdate}) in a quite similar manner as volume averaged conserved quantities are evolved in a FV scheme. After each timestep update, the matrix system of equations is then inverted to obtain the gradients of the conserved variables, which are then transformed to gradients of the primitive variables (just as cell volume integrated conserved variables are converted to cell-averaged primitive variables) for the next half-timestep prediction step in the MUSCL-Hancock scheme.

\subsection{Slope limiter}\label{sec:sl}

For the static and moving FV method and the static DG method, we use the original 
slope limiter in {\sc Arepo} \citep{2010MNRAS.401..791S}. This limiter requires that the linearly reconstructed quantities on face centroids do not exceed the maxima or minima among all neighbouring cells. Each gradient is replaced with a slope-limited gradient:
\begin{equation}
\langle \nabla \phi \rangle_i^\prime = \alpha_i \langle \nabla \phi \rangle_i ,
\end{equation}
where the slope limiter coefficient $0\leq \alpha_i \leq 1$ is computed as:
\begin{equation}
\alpha_i = {\rm min}(1,\psi_{ij})
\end{equation}
\begin{equation}
\psi_{ij} = \begin{cases}
(\phi_i^{\rm max} - \phi_i)/\Delta \phi_{ij} & {\rm for\,}\Delta \phi_{ij}> 0 \\
(\phi_i^{\rm min} - \phi_i)/\Delta \phi_{ij} & {\rm for\,}\Delta \phi_{ij}< 0 \\
1 & {\rm for\,}\Delta \phi_{ij}= 0
\end{cases} .
\end{equation}
Here, $\Delta \phi_{ij} = \langle\nabla\phi\rangle_i\cdot(\mathbf{f}_{ij}-\mathbf{s}_i)$ is the estimated change between the centroid $\mathbf{f}_{ij}$ of the face and the centre of cell $i$, and $\phi_i^{\rm max} = {\rm max}(\phi_j)$ and $\phi_i^{\rm min} = {\rm min}(\phi_j)$ are the maximum and minimum values occurring for $\phi$ among all neighbouring cells of cell $i$, including $i$ itself.

We find that the DG method on a moving mesh is quite sensitive to the choice of slope limiter and the above form results in excessive post-shock oscillations in certain test problems, likely due to the fact that it is not total variation diminishing (TVD) and that it also limits local smooth extrema.  Identifying a robust limiter which is not too dissipative is still an open problem for the moving DG method. In what follows, we adopt a limiter which is similar in spirit to a WENO method \citep{Luo:2007:HWL:1276530.1276745}. We find this limiter to be robust for all of our tests and use 
it as the best, \textit{default choice} for all test problems.

For each cell $i$ we consider two candidate values for the slope of a primitive fluid variable $\phi$. One candidate is the slope-limited gradient obtained with a stencil, as in the FV method (in the case of the magnetic field gradient we project it to a divergence-free space). The other candidate is the unlimited DG local gradient. For each candidate $k$ we compute an oscillation factor:
\begin{equation}
o_{ik} = \sum_j |\Delta \phi_{ij} | + |\phi_j - (\phi_i + \Delta \phi_{ij})| .
\end{equation}
The slope is then a weighted sum of the candidates:
\begin{equation}
\langle \nabla \phi \rangle_i^\prime = \frac{ \sum_k \langle \nabla \phi \rangle_{ik}\cdot \frac{1}{(\epsilon+o_{ik})^\gamma}}{ \sum_k \frac{1}{(\epsilon+o_{ik})^\gamma}} ,
\end{equation}
where $\epsilon$ is machine epsilon and $\gamma = 0.5$. The slope-limiting occurs at the beginning of each timestep, before any prediction steps or updating take place. Following the limiting step, the quantities in Equation~(\ref{eqn:GalerkinVar}) are also re-calculated.

In all of our numerical tests, we use the slope 
limiter of \cite{2010MNRAS.401..791S} for the static and moving FV results and the static DG results. However, we use the modified WENO-type limiter as the default choice for the moving DG results because we find it is much more robust and 
reduces non-physical oscillations to a minimum. In a couple of our numerical tests with the moving DG method, we present the results of the original limiting scheme (labelled as `limiter $2$') for comparison, although we favour the alternate limiter for the moving DG method.

\subsection{Magnetic field divergence errors}\label{sec:divB}

When solving the Riemann problem across flux interfaces, a constant magnetic field perpendicular to the interface must be assumed. We use the average value of the perpendicular magnetic fields extrapolated from the left and right sides of the interface: $B_x = \frac{1}{2}(B_{x,L}+B_{x,R})$. This means that despite having a locally divergence-free representation of the magnetic field inside each cell in the DG formulation, there is still a divergence error estimated by Stokes theorem:
\begin{equation}
\nabla\cdot\mathbf{B}_i = \frac{1}{\mathcal{V}_i}\sum_{\rm faces} \mathbf{B}\cdot \hat{\mathbf{n}} A_i ,
\end{equation}
where $\mathcal{V}_i$ is the volume of cell $i$, and we sum (over the faces) the outward normal values of the magnetic field multiplied by the area of the face. However, a locally divergence free representation of the magnetic field is expected to reduce global divergence errors because the contribution to the divergence error from within each cell is exactly zero. 

\begin{figure}
\centering
\includegraphics[width=0.47\textwidth]{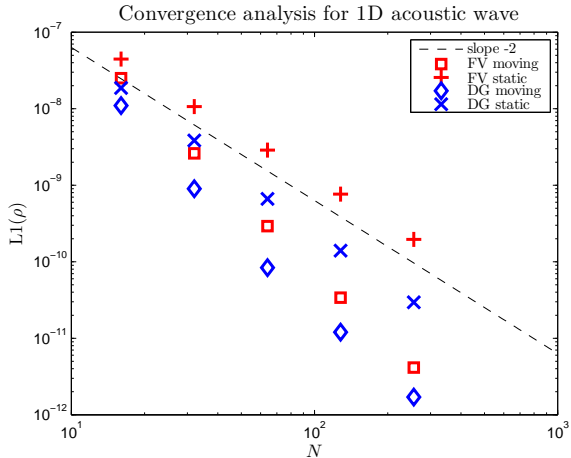}
\caption{Convergence of the 1D acoustic wave in the L1 norm. Second-order convergence is achieved, as expected (in fact the moving
mesh schemes show superconvergence). The DG and FV methods are compared on static and moving meshes. The moving DG
technique produces the smallest errors. Errors are $\sim60$ per cent smaller than the moving FV approach.
The static DG method has errors $\sim 80$ smaller than the static FV scheme.}
\label{fig:acoustic}
\end{figure}

\section{Results of numerical tests}\label{sec:results}

We perform a series of 
numerical tests documented in the literature to compare 
the static and moving DG and FV methods. The results of these tests are presented in the following subsections.

\subsection{1D acoustic wave}\label{sec:acoustic}
The first test we present is a simple 1D acoustic wave, discussed in \cite{2008ApJS..178..137S}, and also 
described in \cite{2010MNRAS.401..791S}. This setup serves as a sensitive test of the convergence rate of a code. A simple acoustic wave of
unit wavelength is
initialized with very small amplitude $\Delta\rho/\rho=10^{-6}$ and $\rho=1$ in a periodic domain of unit length. The gas has pressure $p=3/5$ and adiabatic index $\gamma=5/3$. The L1 error norm is computed when the wave returns to its original position (the analytic solution here is 
identical to the initial state). Fig.~\ref{fig:acoustic} shows the L1 error norms for the moving and static DG and FV methods as a function of mesh resolution. Second-order convergence is achieved, as expected. The moving and static DG algorithms show a $\sim 60$ per cent and $\sim 80$ per cent reduction of errors over their FV counterparts, respectively.

\subsection{1D Sod shock tube}\label{sec:sod}
\begin{figure*}
\centering
\includegraphics[width=0.47\textwidth]{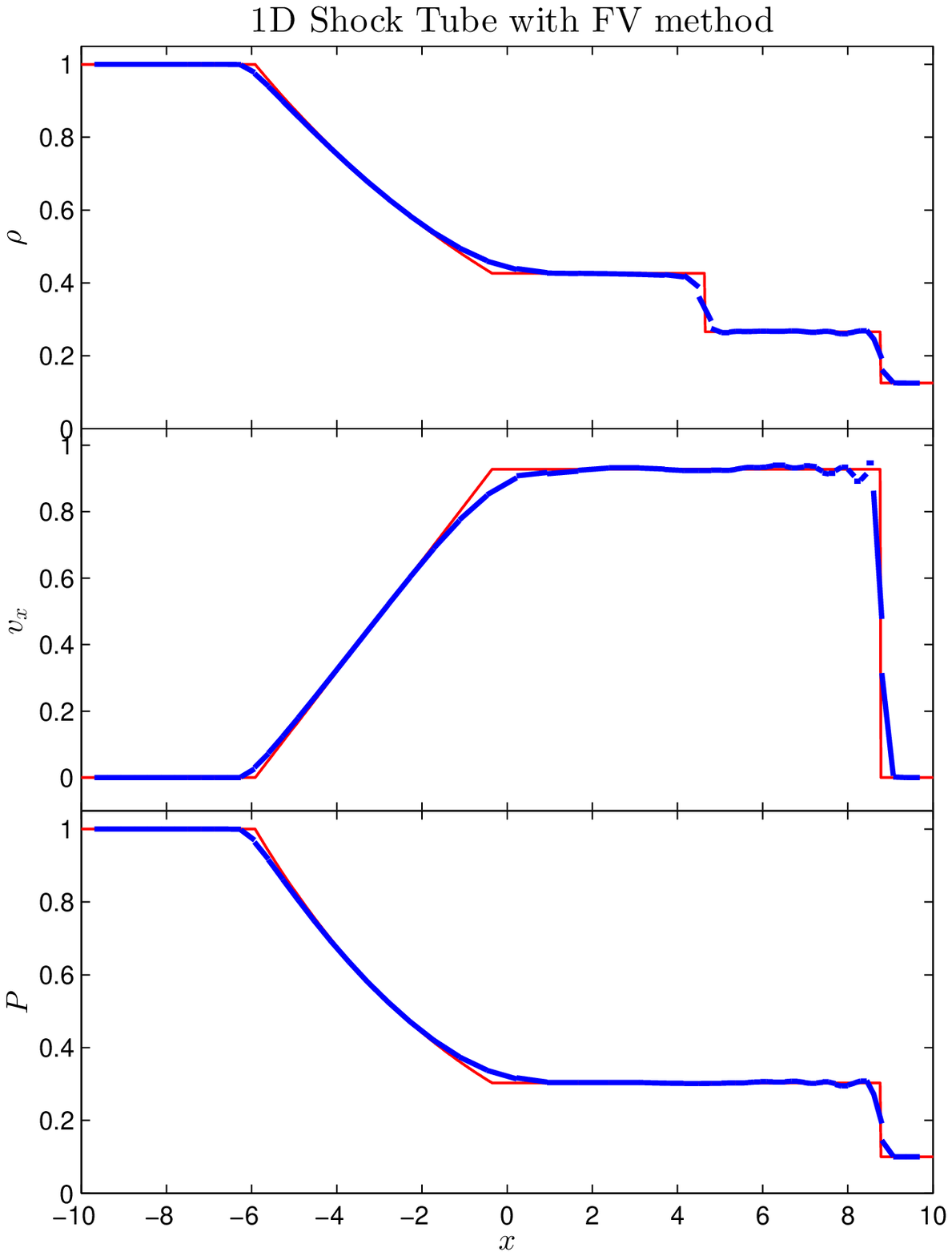}
\includegraphics[width=0.47\textwidth]{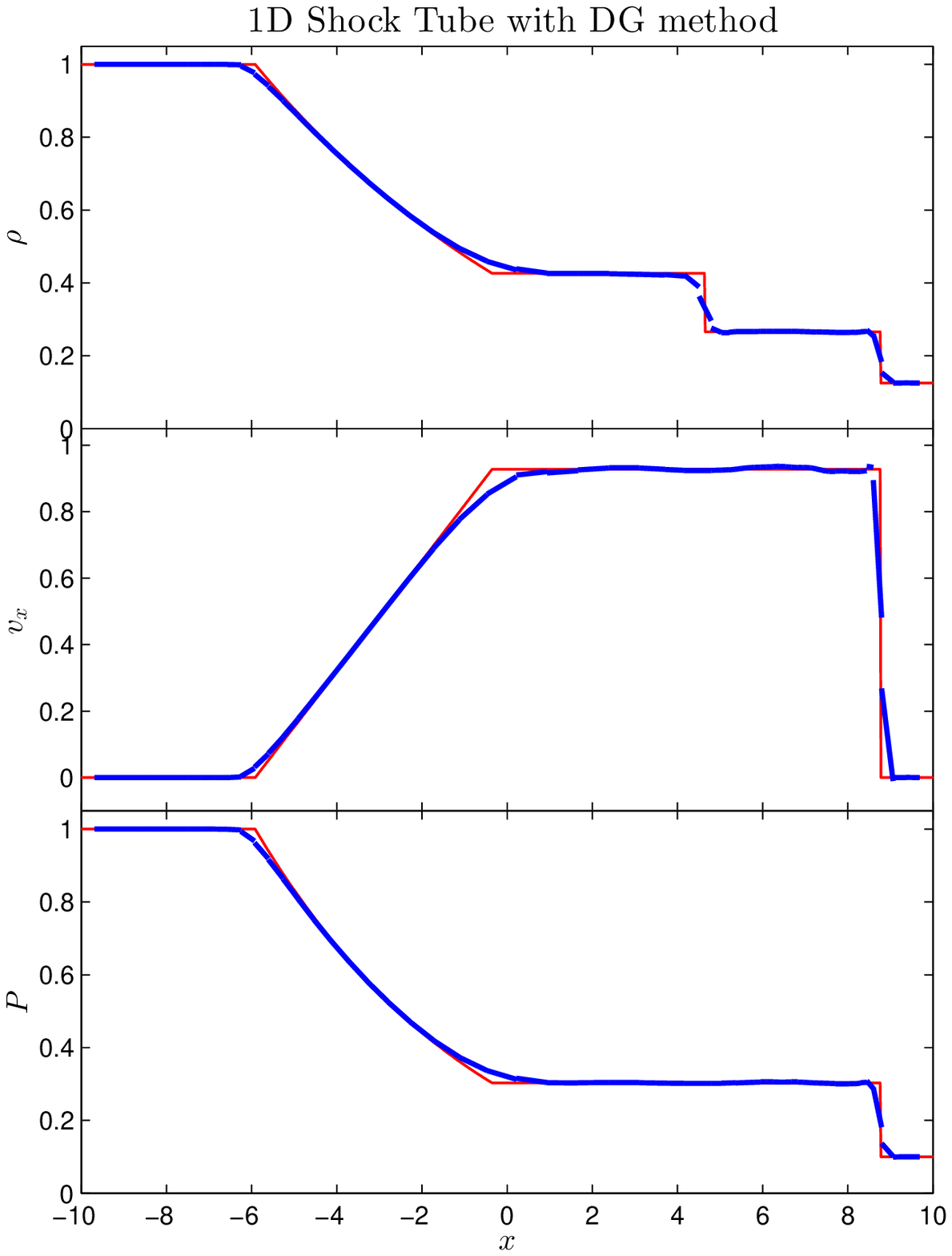}
\caption{Solution of the 1D shock tube test at $t=5.0$ (resolution $64$) with the moving FV method (left) and the moving DG method (right). The moving DG method reduces post-shock oscillations noticeably (especially in the velocity) while also being less diffusive.}
\label{fig:sodA}
\end{figure*}

We continue our investigation by simulating a 1D Sod shock tube. We adopt initial conditions 
employed in a large number of other code tests \citep{1989ApJS...70..419H,1991ApJ...377..559R,2004NewA....9..137W,2005MNRAS.364.1105S,2010MNRAS.401..791S}. The left side ($x<0$) is described by $p_L=1$, $\rho_L=1$, $v_L=0$, the right side ($x\geq 0$) is described by $p_R=0.1795$, $\rho_R=0.25$, $v_R=0$, and the gas as adiabatic index $\gamma=1.4$. We evolve the system until $t=5.0$. The solutions in the moving FV and DG formulations are shown in Fig.~\ref{fig:sodA}. 

\begin{figure}
\centering
\includegraphics[width=0.47\textwidth]{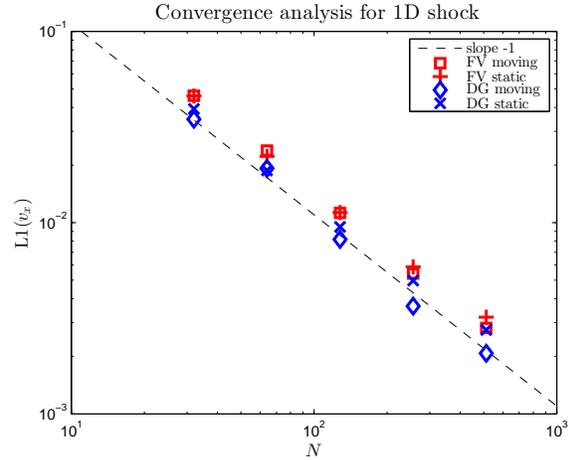}
\caption{Convergence of the 1D shock tube test in the L1 norm. First-order convergence is achieved, as expected due to the discontinuity in the solution. The DG method on a moving mesh produces the smallest errors, a $\sim 30$ per cent reduction over the moving FV method.}
\label{fig:sodB}
\end{figure}

The DG method does a superior job of maintaining sharp shock interfaces 
while at the same time reducing post-shock oscillations (especially in velocity). Typically one might expect a trade-off between sharpness and reduction of non-physical oscillations. We note that the DG method on a moving mesh uses the modified WENO-type slope limiter (the limiter of \cite{2010MNRAS.401..791S} produces overly severe non-physical oscillations). In both the moving FV and moving DG schemes, the shock discontinuities are typically broadened over two cells, but the slopes in the DG version are larger. On static grids (not shown) the shock discontinuities typically span three cells at this resolution and time in the simulation. Error convergence plots for moving and static approaches are shown in Fig.~\ref{fig:sodB}, where it is seen that DG methods again have advantages over their FV counterparts. The method is first-order accurate in this test problem because the solution has to be slope-limited owing to the presence of the shock discontinuity.

\begin{figure}
\centering
\includegraphics[width=0.47\textwidth]{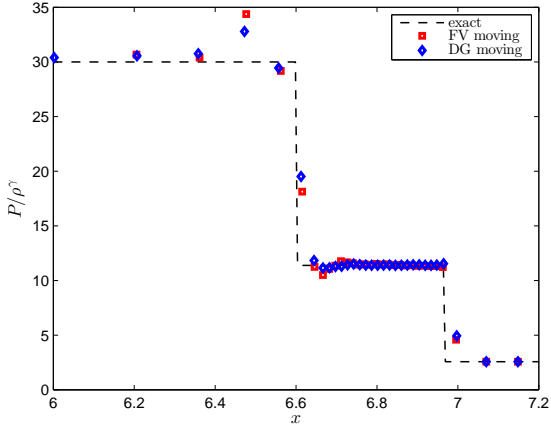}
\caption{Entropy profile of a strong shock ($M=6.3$) at $t=5.0$.}
\label{fig:ss}
\end{figure}

We also consider a strong shock version of the shock tube (Mach number
$M=6.3$), as studied in \cite{2012MNRAS.424.2999S}. The initial
conditions are $P_L=30.0$, $\rho_L=1.0$, $v_L=0$, $P_R=0.14$,
$\rho_R=0.125$, $v_R=0$, $\gamma=1.4$. Moving mesh codes (as opposed
to static mesh codes) may exhibit a `spike' in the entropy at the
contact discontinuity (which can be eliminated with smoothed initial
conditions). The feature is due to the fact that moving mesh codes
preserve contact discontinuities, present in the initial conditions,
to much higher precision than do static mesh codes. In
Fig..~\ref{fig:ss} we present the entropy profile for the strong shock
at t = 5.0, which shows that the DG method also exhibits a
`spike' feature, albeit of somewhat reduced magnitude and has
smaller post-shock oscillations after the contact discontinuity. We
note again that the DG method on a moving mesh uses the modified
WENO-type slope limiter.

\subsection{Gresho vortex}\label{sec:gresho}
We move on to a 2D test for the conservation of vorticity and angular momentum. The problem proposed by \cite{1990IJNMF..11..621G} considers a static `triangle vortex'. We adopt the initial conditions described in \cite{Liska:2003:CSD:954386.954478}. The vortex has azimuthal velocity profile:
\begin{equation}
v_\phi(r) = \begin{cases}  
5r & {\rm for\,} 0\leq r < 0.2 \\
2-5r & {\rm for\,} 0.2\leq r < 0.4 \\
0 & {\rm for\,} r\geq 0.4
\end{cases} .
\end{equation}
The gas has constant density $\rho=1$ and adiabatic index $\gamma=5/3$. The pressure profile:
\begin{equation}
p(r) = \begin{cases}  
5+\frac{25}{2}r^2 & {\rm for\,} 0\leq r < 0.2 \\
9 + \frac{25}{2}r^2 - 20r + 4\ln(r/0.2)  & {\rm for\,} 0.2\leq r < 0.4 \\
3+4\ln2 & {\rm for\,} r\geq 0.4
\end{cases}
\end{equation}
balances the centrifugal force with the pressure gradient so that the vortex is a steady-state solution. 

\begin{figure}
\centering
\includegraphics[width=0.47\textwidth]{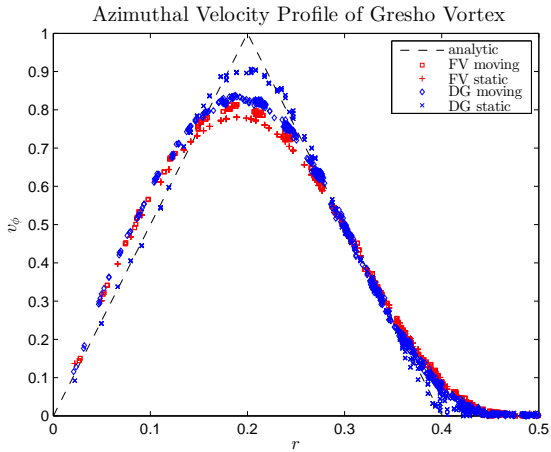}
\caption{Solution to the 2D Gresho vortex problem at $t=3.0$, at resolution $32^2$. DG methods 
exhibit reduced levels of angular momentum 
diffusion owing to 
treating the gradient of a cell in a completely local manner.
Here the static DG method shows an advantage over the moving DG method due to the differences in their slope limiters. Static DG, static FV, and moving FV all have the same slope limiter, while for the moving DG method we find it is generally better to take a weighted average of the local slope of a cell and the one obtained with a stencil to prevent spurious oscillations.}
\label{fig:greshoA}
\end{figure}

\begin{figure}
\centering
\includegraphics[width=0.47\textwidth]{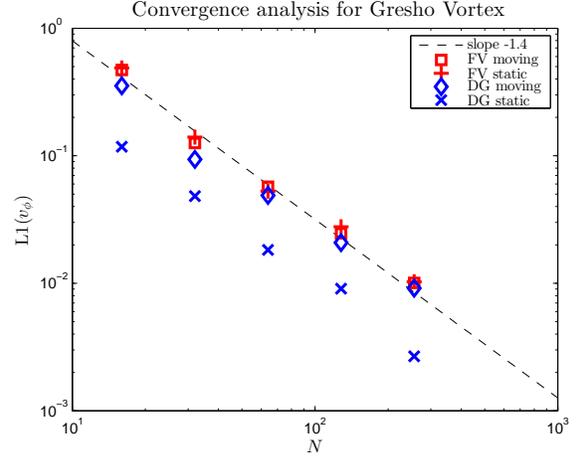}
\caption{Convergence of the 2D Gresho vortex test in the L1 norm. We 
display the error in $v_\phi$, which is a measure of angular momentum diffusion. Here the static DG method shows a $\sim 70$ per cent 
improvement over the other methods.}
\label{fig:greshoB}
\end{figure}

Developing a scheme that minimizes angular momentum diffusion for grid-based methods is important because lack of angular momentum conservation is one of the main disadvantages of grid-based methods relative to SPH, 
which conserves total angular momentum due to its pseudo-Lagrangian character \citep{2012JCoPh.231..759P}.  (We note, however, that this
``advantage'' of SPH
comes at the expense of an inaccurate handling of the mass continuity equation,
as discussed by \cite{2012MNRAS.425.3024V}.)
The solutions of the Gresho vortex problem 
are shown in Fig.~\ref{fig:greshoA} and an error convergence plot is presented in Fig.~\ref{fig:greshoB}. 
The static DG approach performs significantly better than the static FV method, likely owing
to the purely local manner in which it
handles gradients, and the moving DG approach offers 
a small improvement over the moving FV method here. 
In this test, the static DG method shows an advantage over the moving DG scheme, attributable to the differences in their slope limiters. We used the same slope limiter for the static DG, static FV, and moving FV methods, while for the moving DG approach we find it is generally better to take a weighted average of the local slope of a cell and the one obtained with a stencil to prevent spurious oscillations.  This weighting step prevents the gradient from being treated purely locally, since we use a stencil-estimated slope in the stencil.  Further refinement of the slope limiter could lead to improvements in the moving DG scheme to the level of the static DG method in this test.

We also verified with additional tests that the static DG approach maintains its strong advantage over the static FV approach on arbitrary meshes. The regularity of a Cartesian grid is not a necessary requirement for the DG method to work well.

\subsection{2D implosion}\label{sec:implosion}
Next we perform a 2D implosion test \citep{1999JCoPh.153..596H} with periodic boundary conditions, as in \cite{2012MNRAS.424.2999S}. The domain is a box of side length $0.3$. The initial pressure and density are $p=1.0$, $\rho=1.0$ for $x+y > 0.15$ and $p=0.14$, $\rho=0.125$ otherwise. The gas is initially at rest and has adiabatic index $\gamma = 1.4$. This test is well suited for 
studying interacting shocks, Richtmyer-Meshkov instabilities, diffusivity, and ability of codes to maintain a symmetric solution. 

\begin{figure*}
\centering
\includegraphics[width=0.33\textwidth]{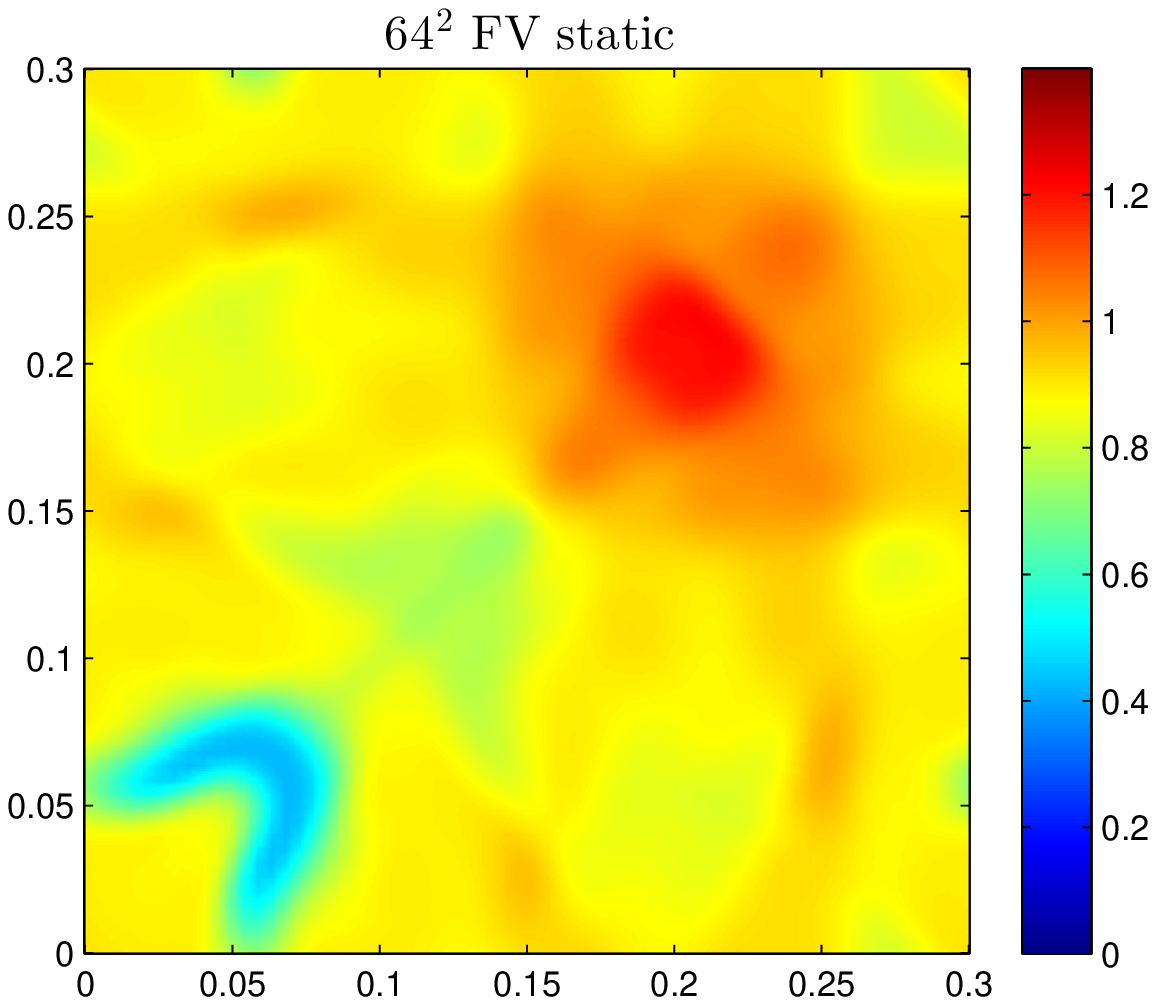}
\includegraphics[width=0.33\textwidth]{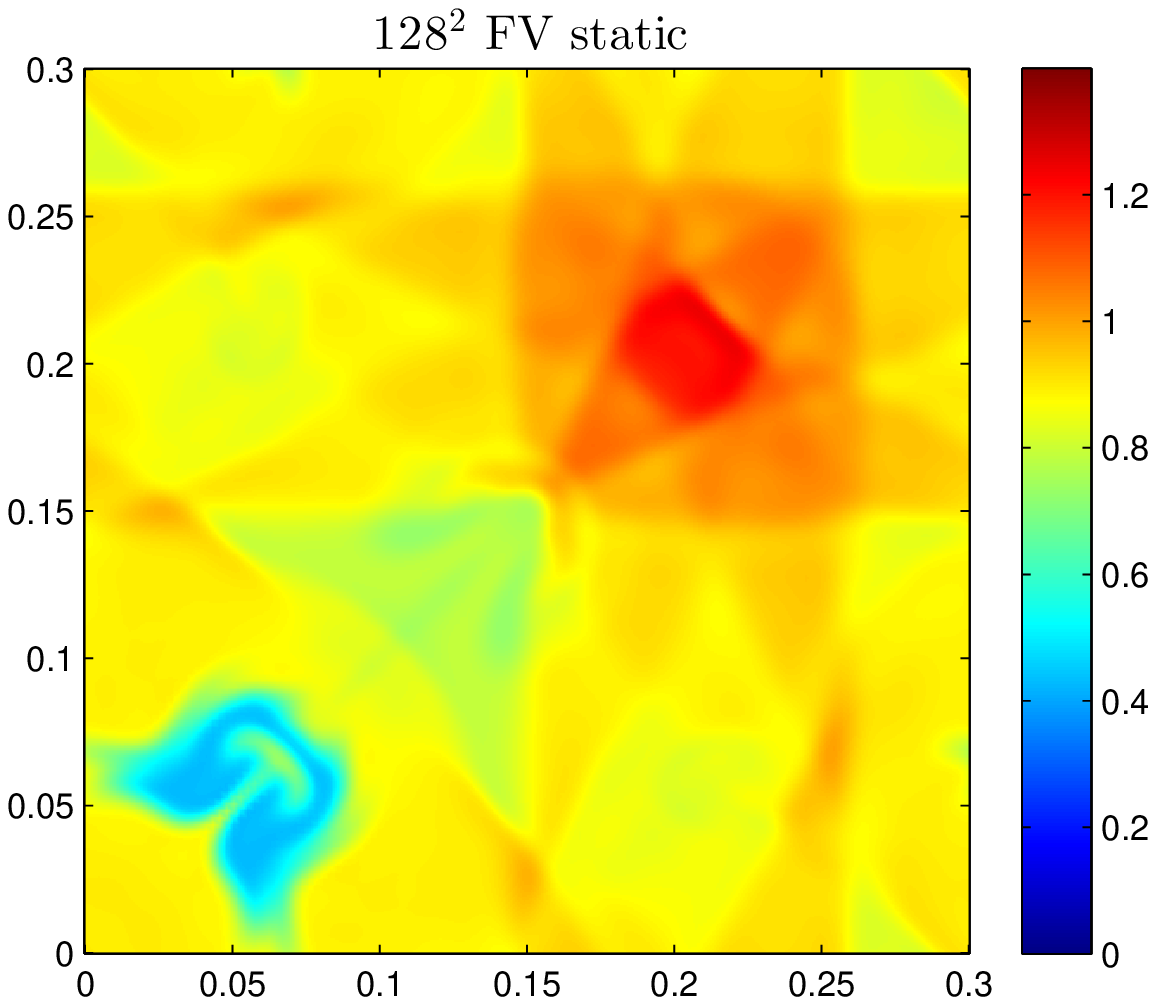}
\includegraphics[width=0.33\textwidth]{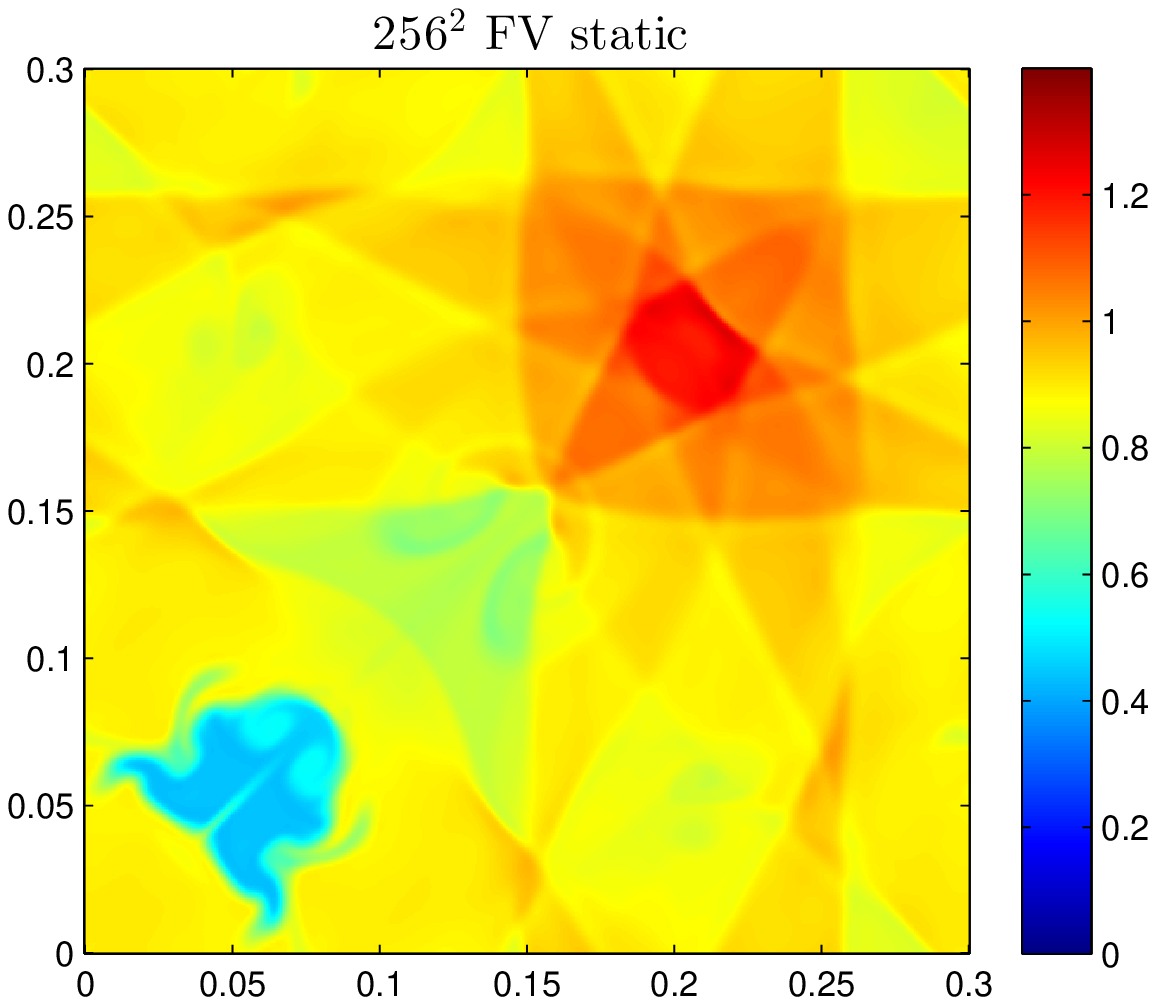}\\
\includegraphics[width=0.33\textwidth]{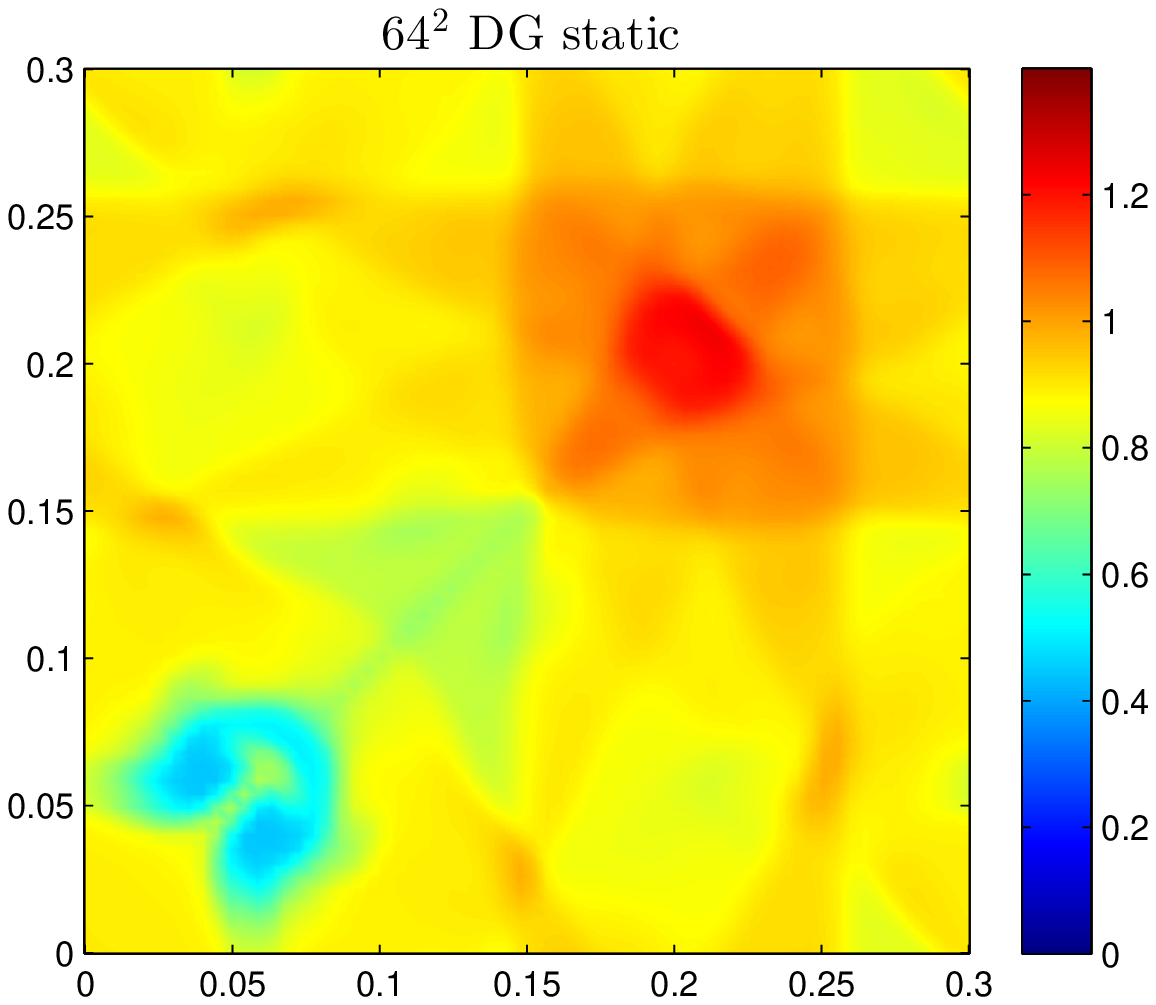}
\includegraphics[width=0.33\textwidth]{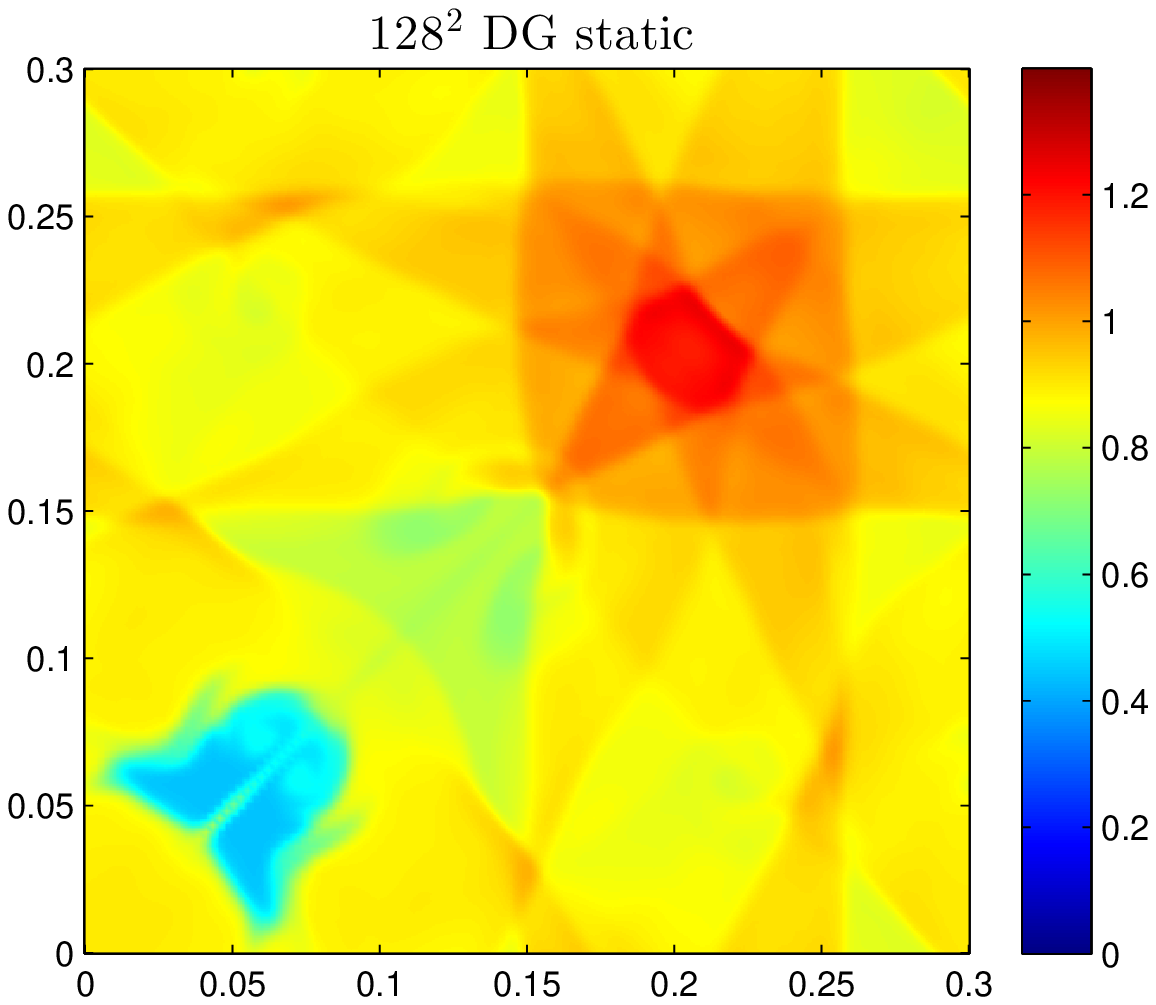}
\includegraphics[width=0.33\textwidth]{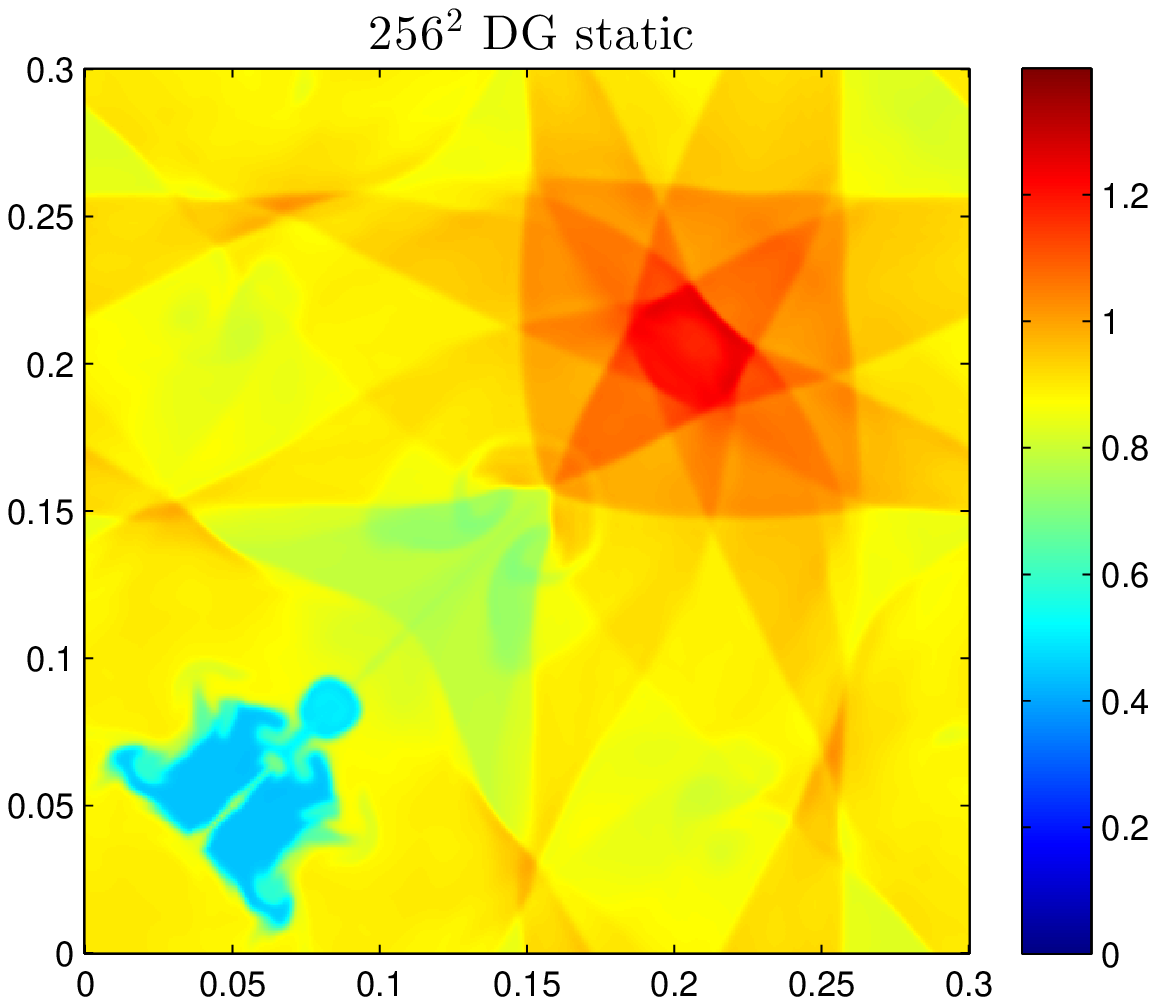}\\
\includegraphics[width=0.33\textwidth]{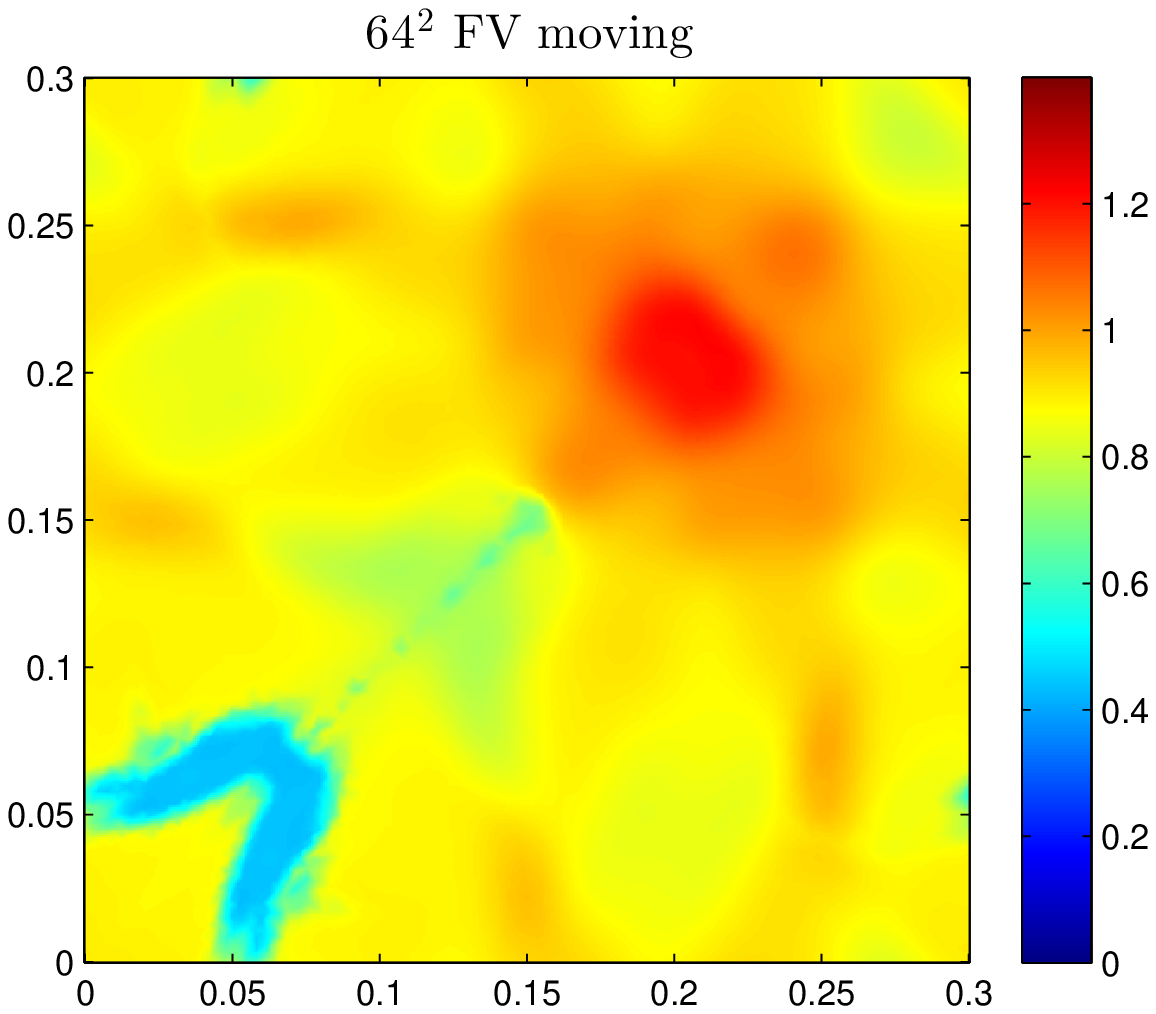}
\includegraphics[width=0.33\textwidth]{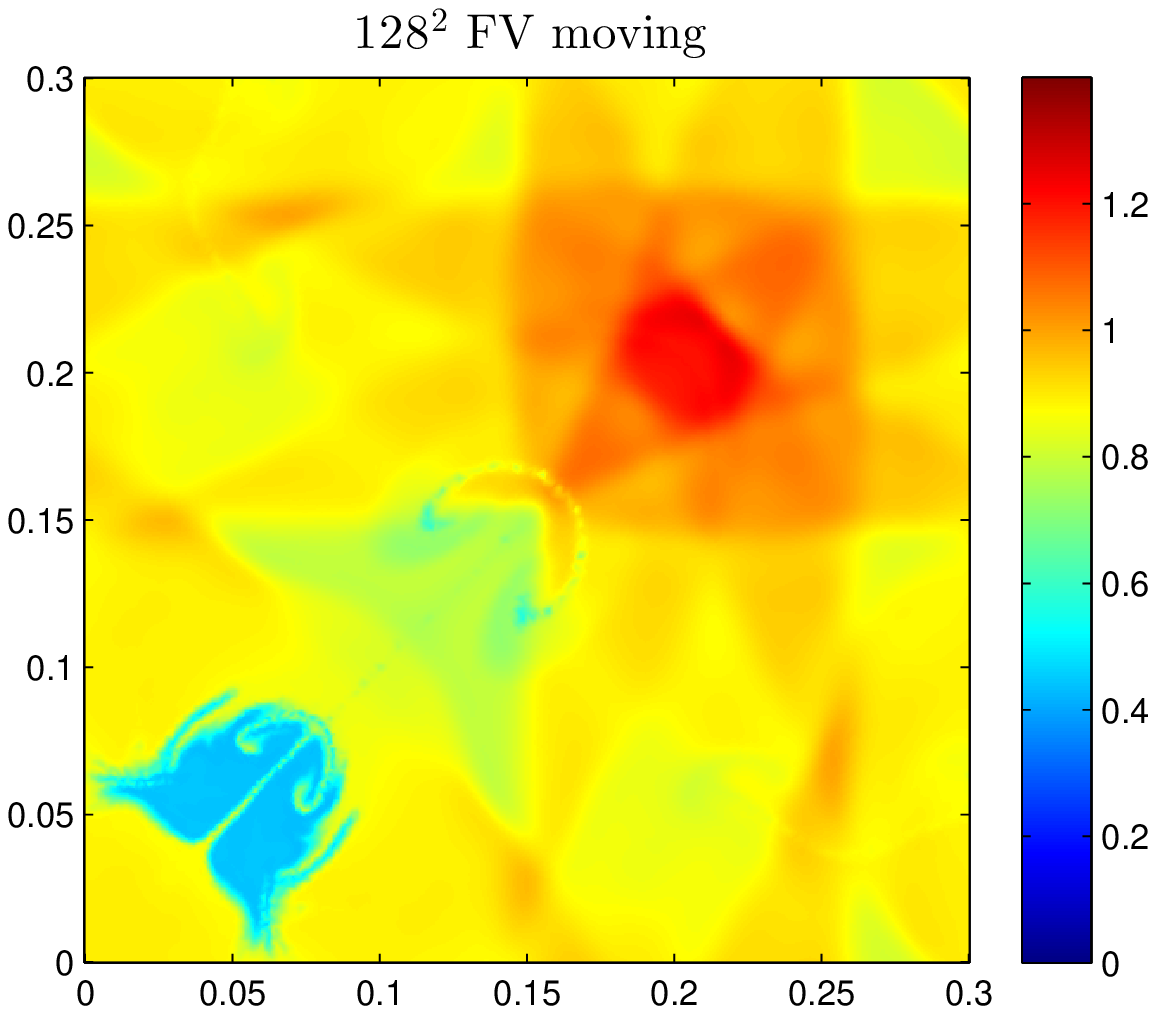}
\includegraphics[width=0.33\textwidth]{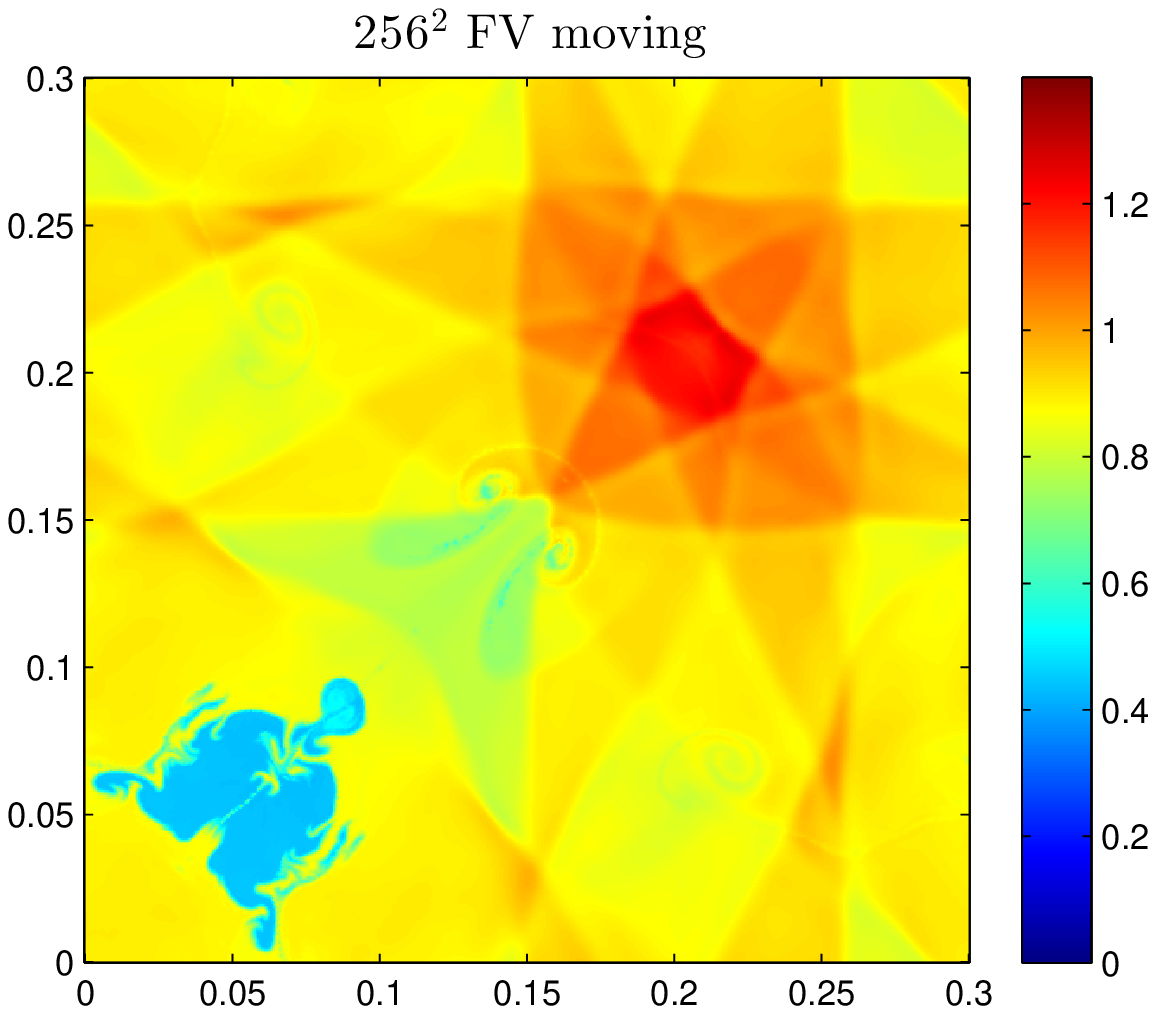}\\
\includegraphics[width=0.33\textwidth]{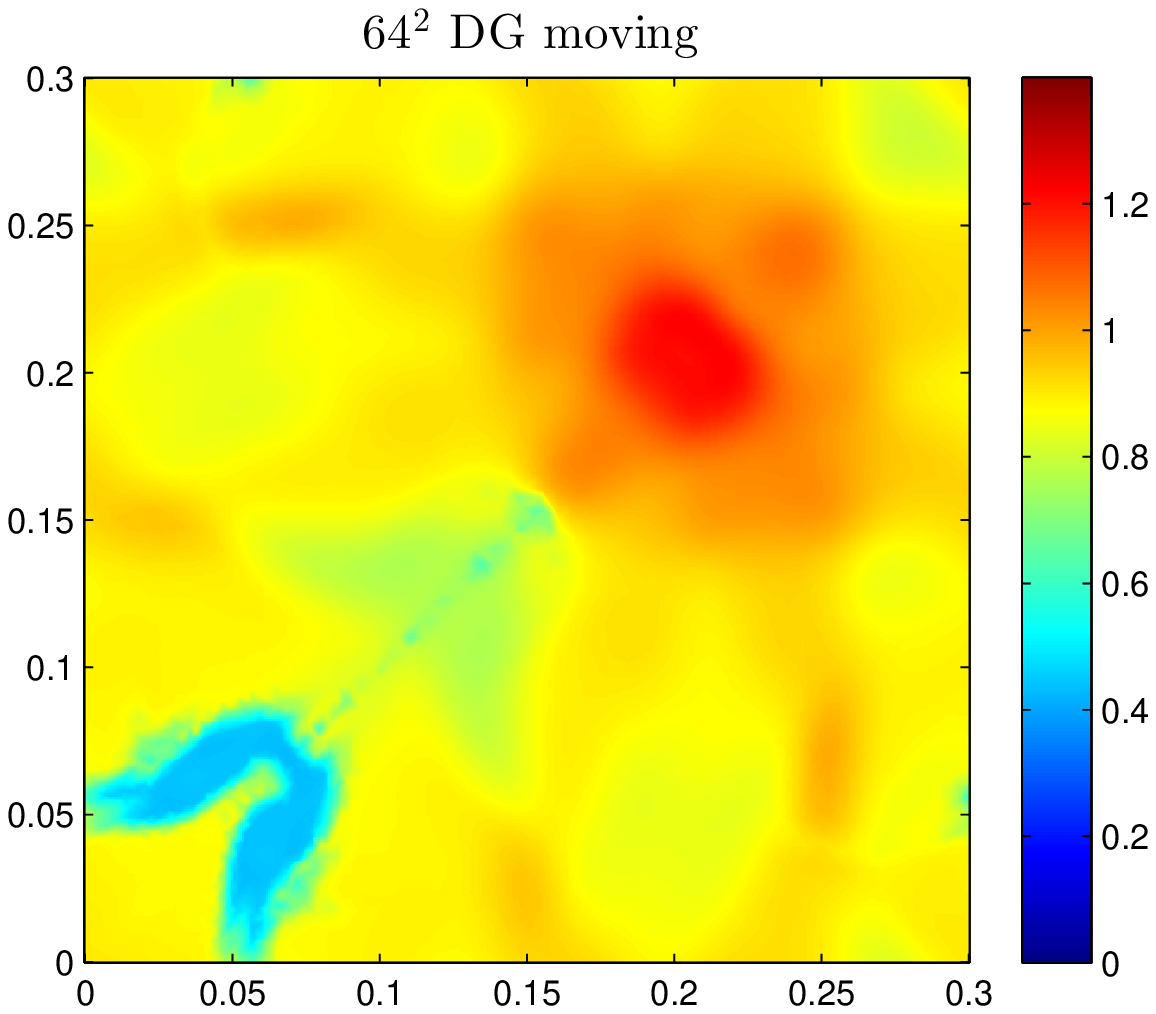}
\includegraphics[width=0.33\textwidth]{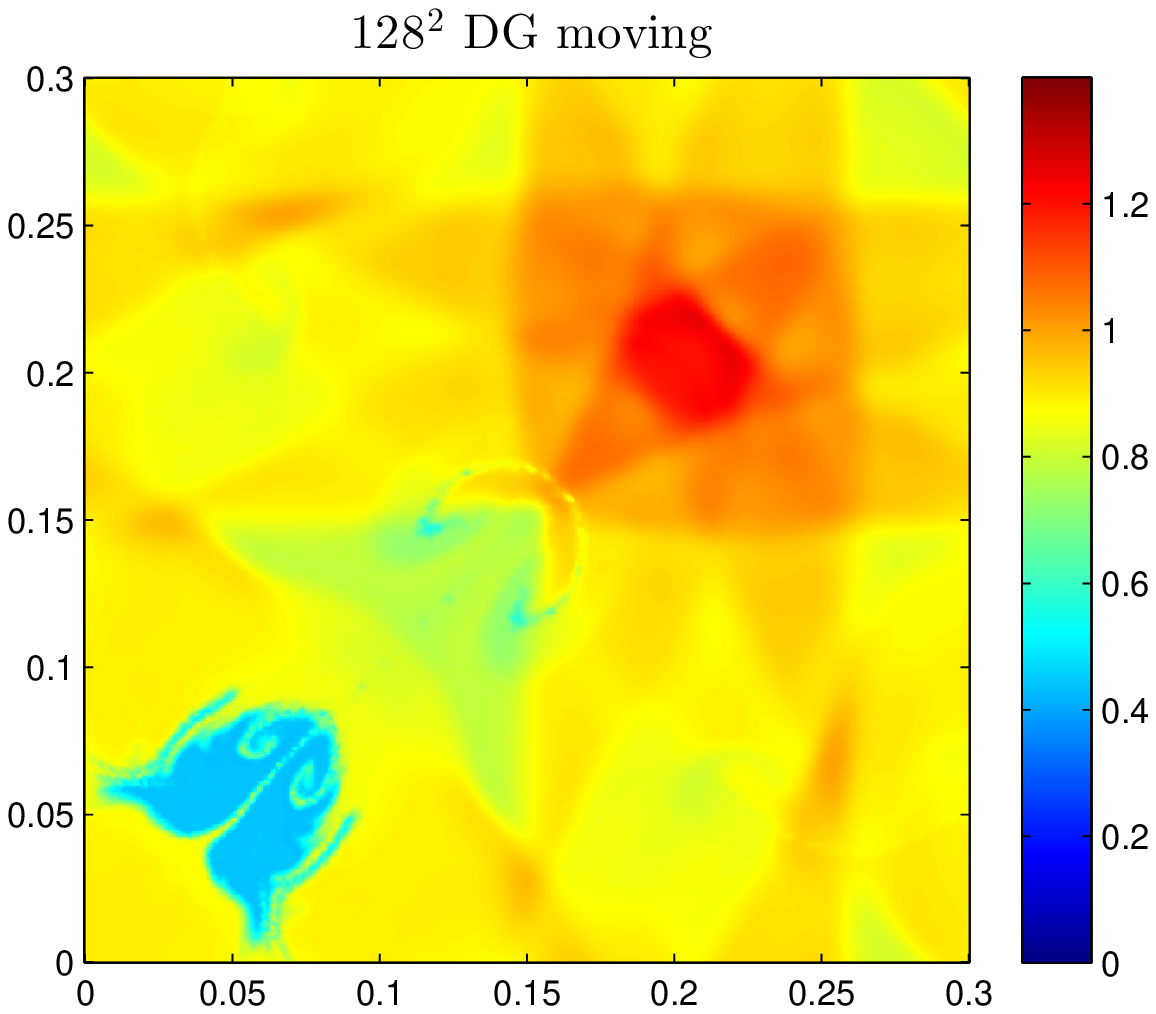}
\includegraphics[width=0.33\textwidth]{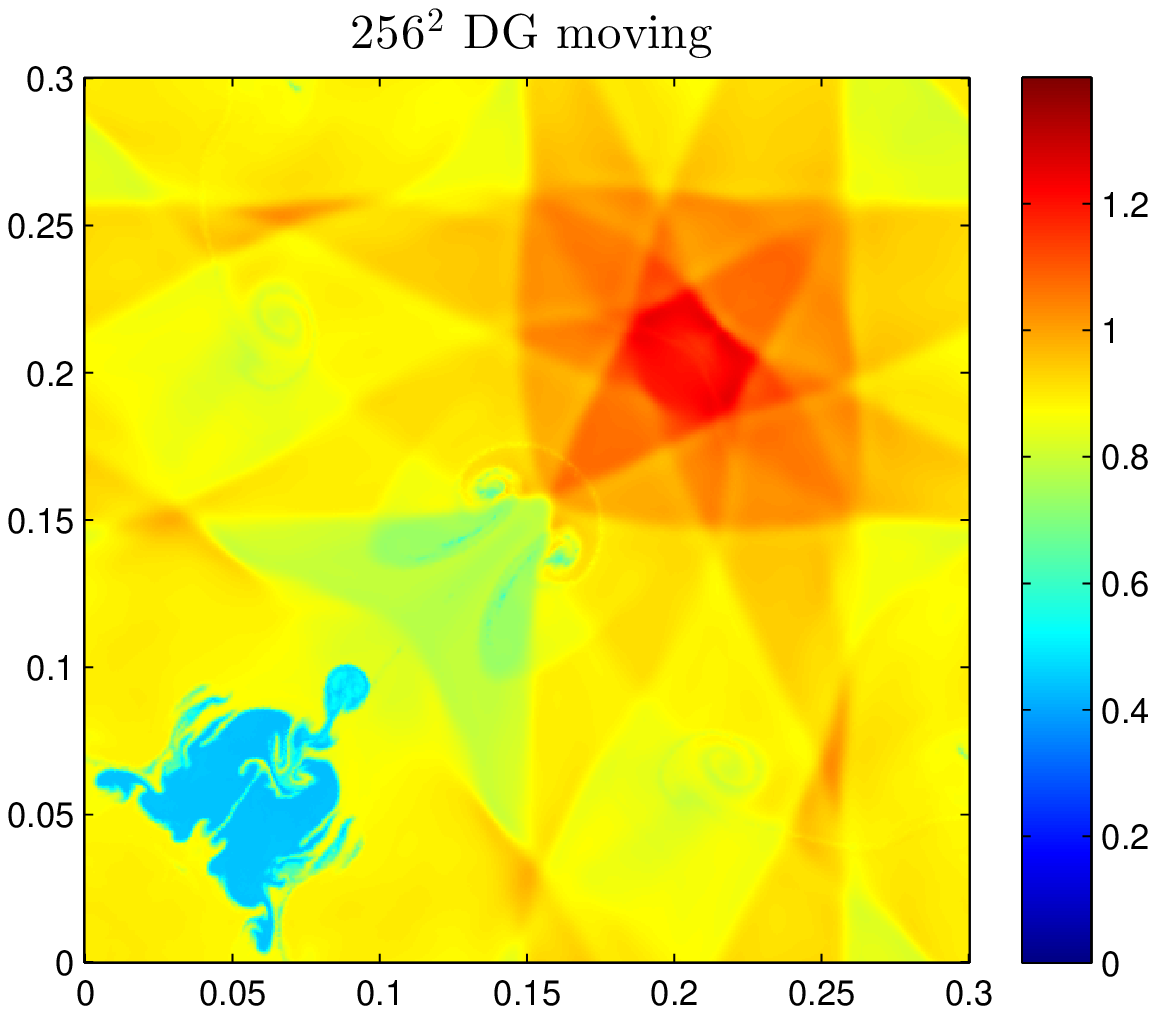}\\
\includegraphics[width=0.33\textwidth]{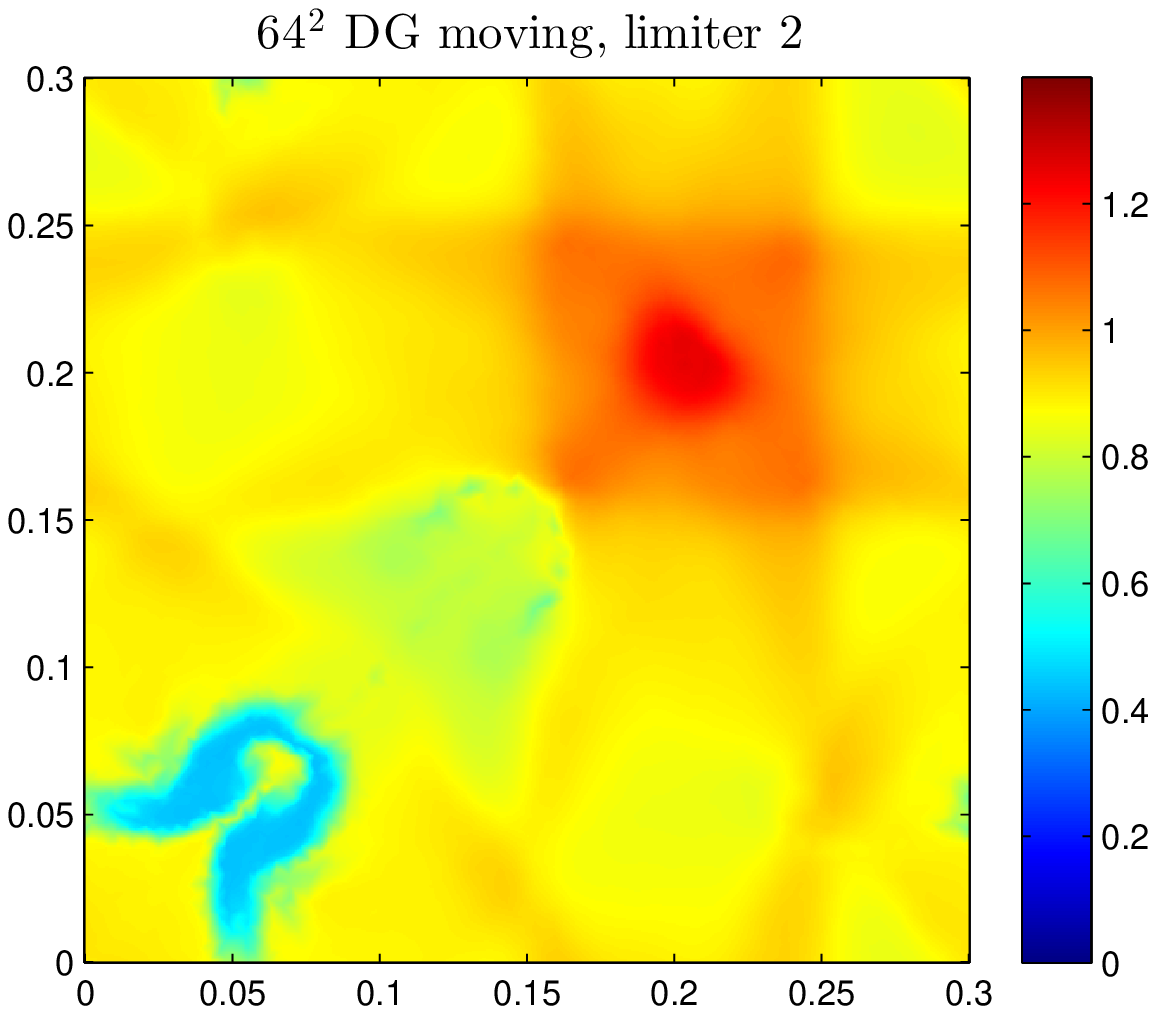}
\includegraphics[width=0.33\textwidth]{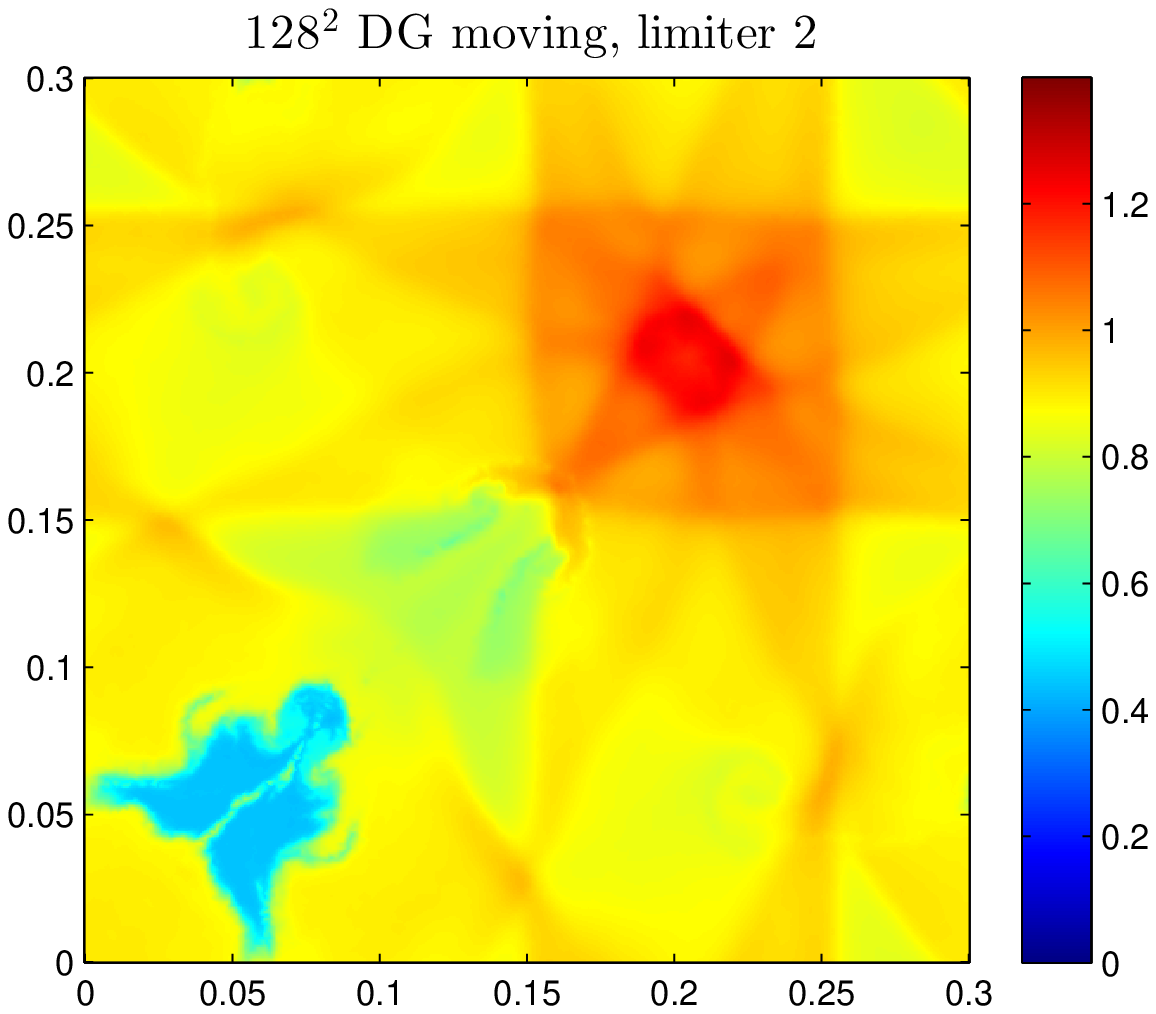}
\includegraphics[width=0.33\textwidth]{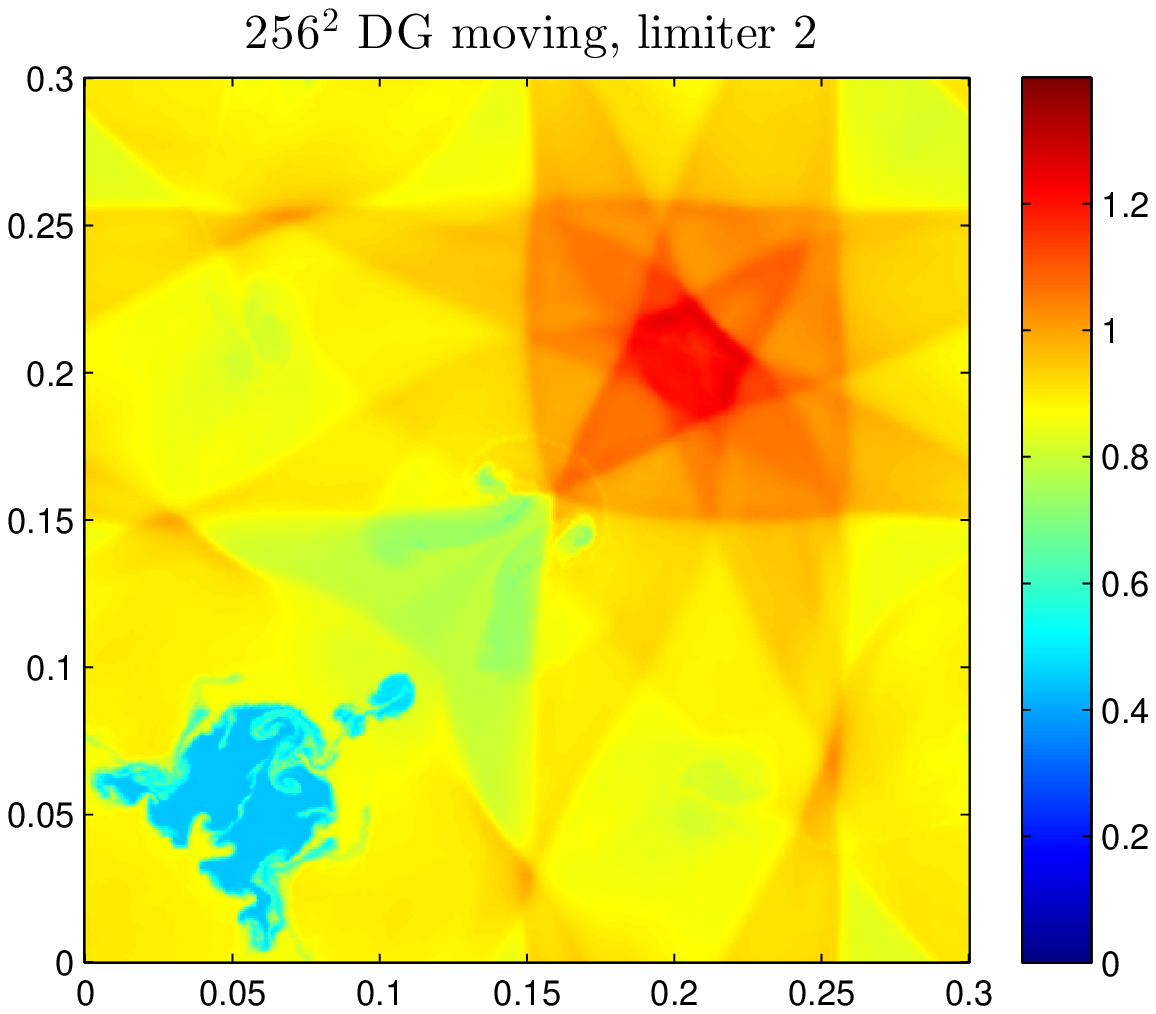}
\caption{Implosion test ($t=1.0$, periodic boundary conditions) plots of density for moving and static FV and DG methods at resolutions of $64^{\,\,2}$, $128^{2}$, and $256^{2}$. We have also included results of the moving DG method with the same limiter as static FV, moving FV and static DG (limiter 2), which shows the best results in this test, although in general we prefer the modified limiter. The presence of asymmetry in the moving mesh simulations is due to mesh noise. The static DG method and the moving DG method (limiter 2) resolve features of the instability that occur in the low density region in the lower left-hand corner which their FV counterparts resolve only at twice the resolution.}
\label{fig:implosion}
\end{figure*}

The development of the implosion is presented in Fig.~\ref{fig:implosion} at several resolutions. We see that the DG method produces less diffusive results than the FV method. In this particular example, the original limiter adopted by 
\cite{2010MNRAS.401..791S} works well for the moving DG method (labelled as `limiter 2' in the figure), producing sharp 
shock interfaces, and so we show the results of both limiters for the moving DG method. A point of interest to 
examine is the low-density region that develops in the bottom left corner of the simulations. The further along diagonally (towards the center) the structure has developed, the less numerical diffusion is present.
In the cases of static DG and moving DG with limiter 2, the region obtained at a resolution of $128^2$ resembles more closely the solution obtained by the FV approach at twice the resolution $256^2$ rather than at the same resolution $128^2$. This suggests that a purely local treatment of derivatives (no stencils) as in the DG method allows for a better treatment of fluid instabilities and increases the effective resolution of the simulation. The moving FV and moving DG results with the WENO-type limiter are fairly similar, but the DG solution shows that the low-density feature has advanced further diagonally, indicating less numerical diffusion. Note that there are some asymmetries that develop in the moving mesh approach 
owing to the fact that in our implementation the fluxes across interfaces are added in an arbitrary order and slight differences can arise from finite-precision arithmetic which can be amplified by the additional degree of freedom of the motion of the mesh (we call this effect `mesh noise'). The appearance and the magnitude of these asymmetries are sensitive to the mesh regularization options, and one could obtain more symmetric results with careful fine-tuning of the regularization parameters.

\subsection{Kelvin-Helmholtz instability}\label{sec:kh}
\begin{figure*}
\centering
\includegraphics[width=0.47\textwidth]{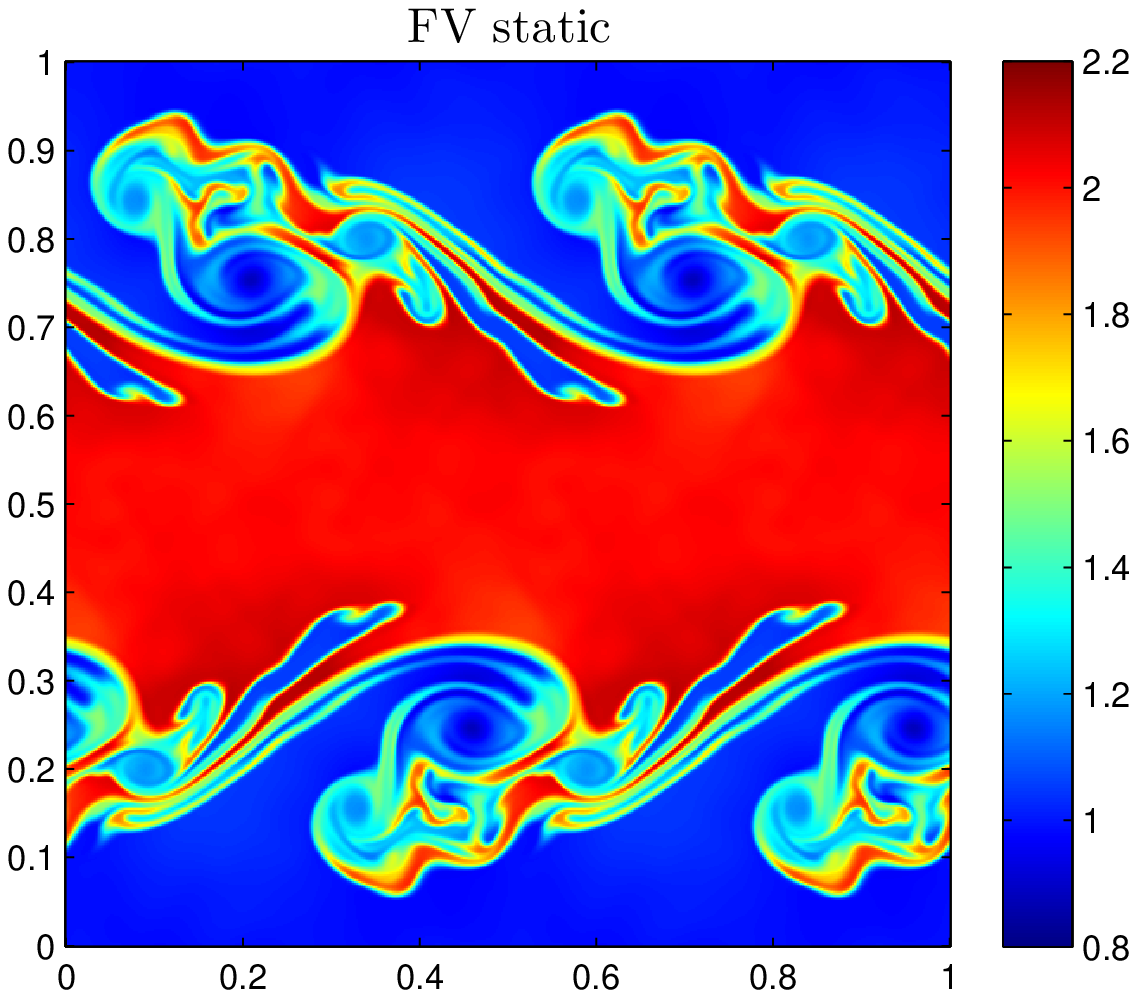}
\includegraphics[width=0.47\textwidth]{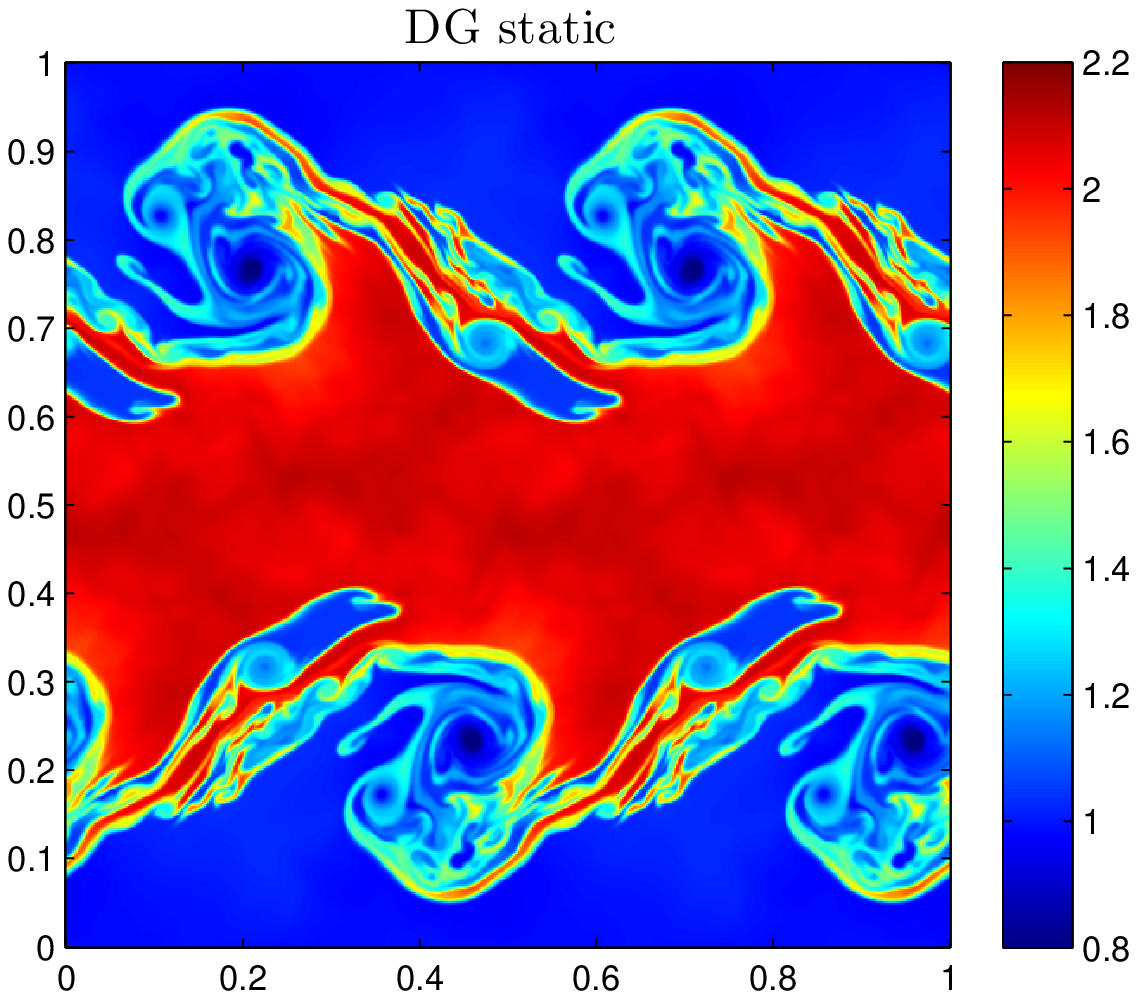}\\
\includegraphics[width=0.47\textwidth]{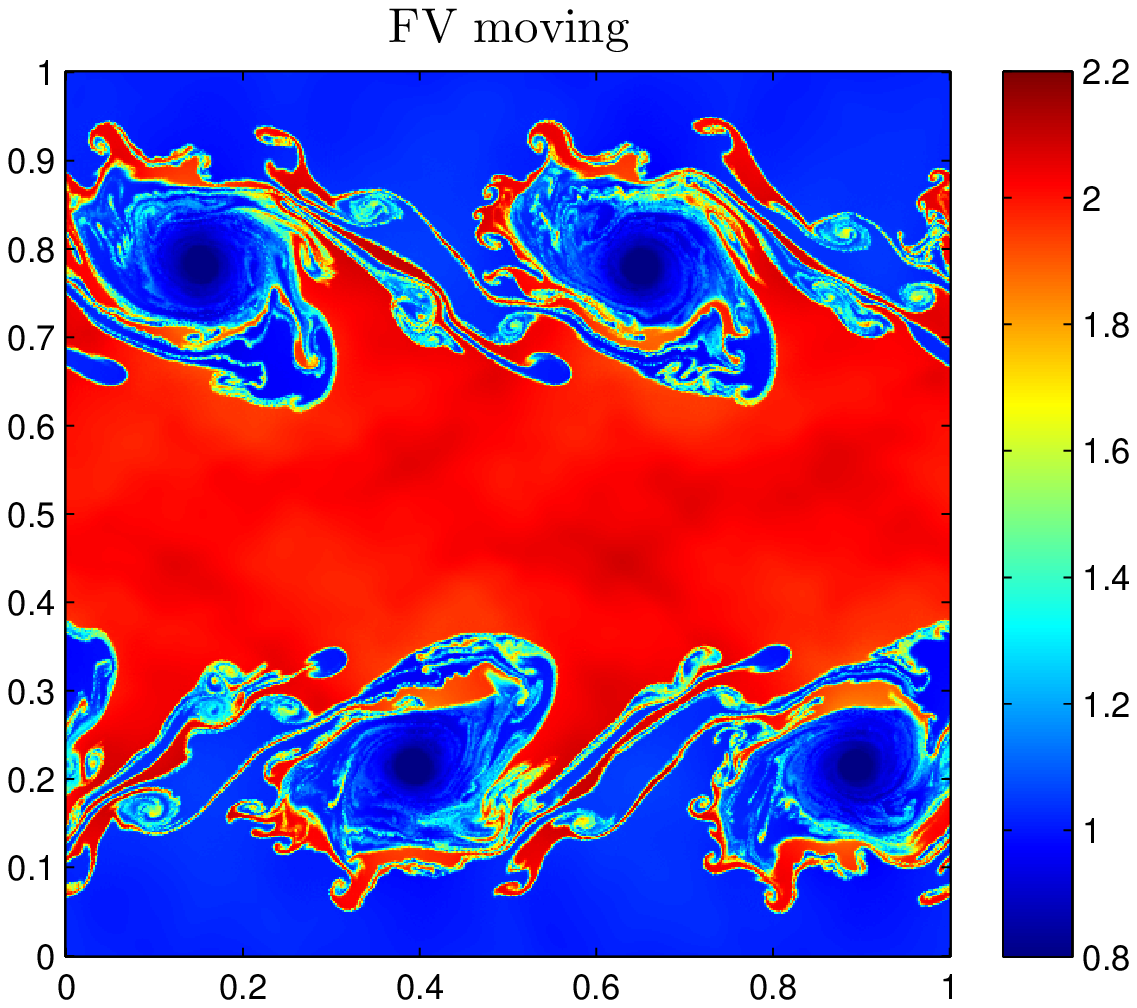}
\includegraphics[width=0.47\textwidth]{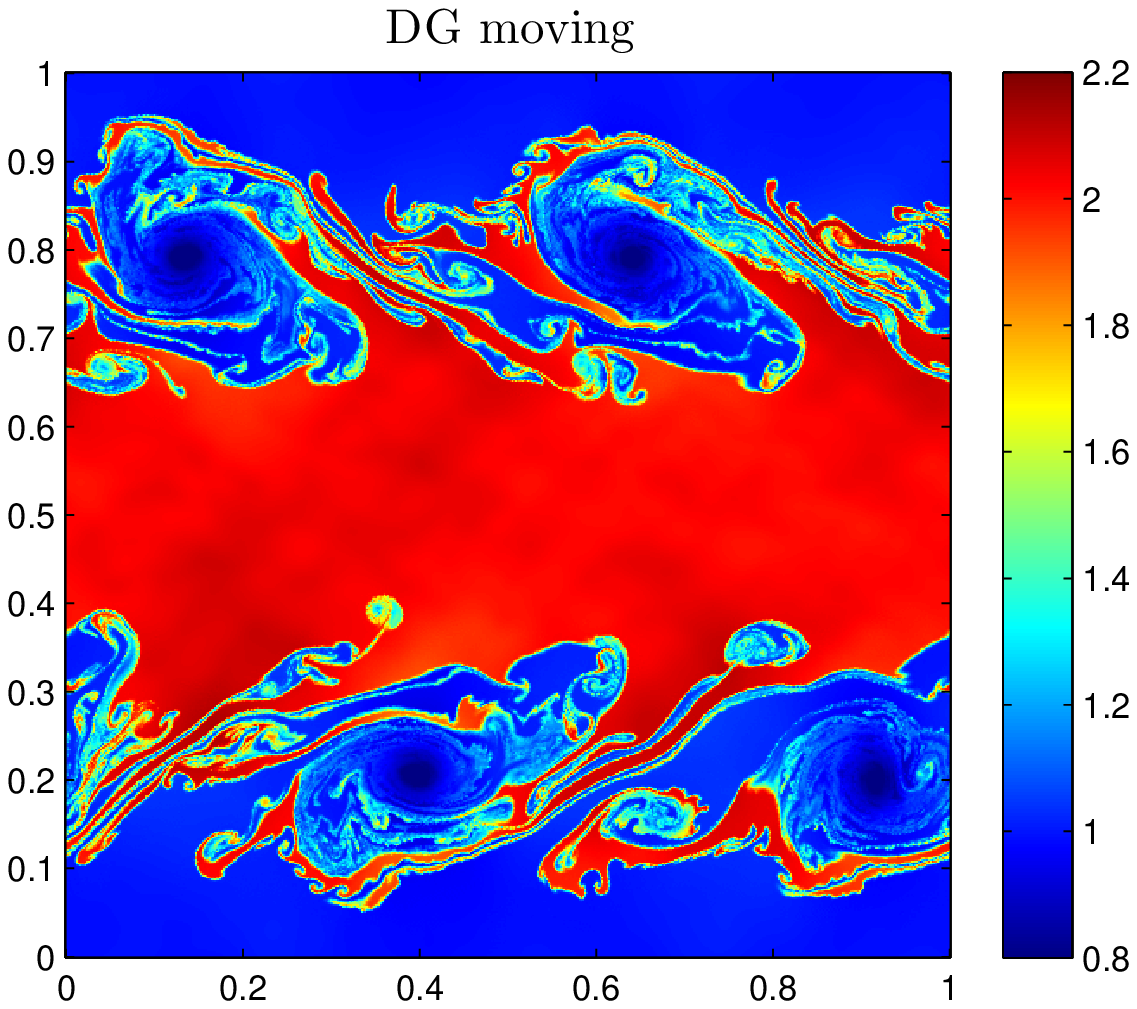}\\
\includegraphics[width=0.47\textwidth]{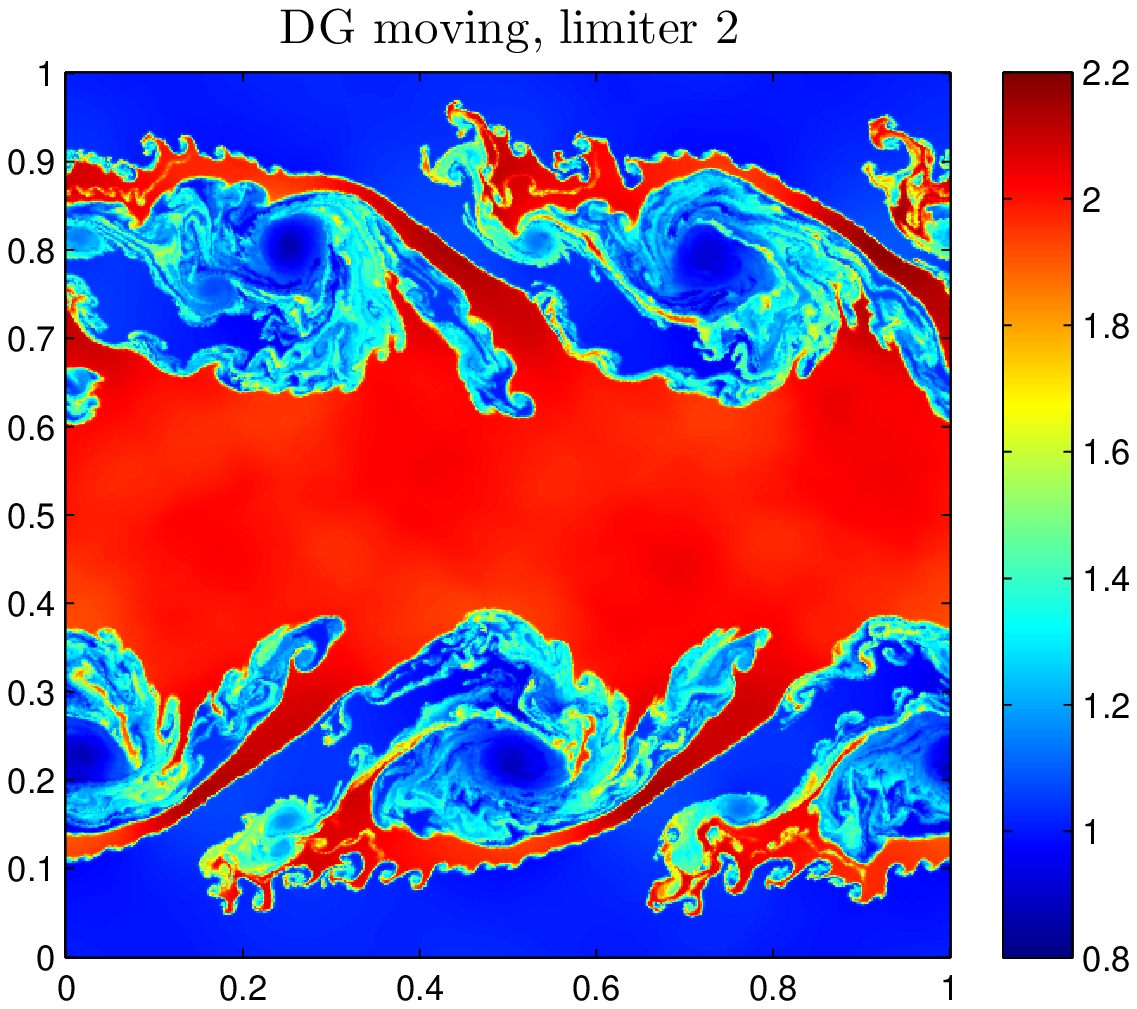}
\caption{Kelvin Helmholtz test ($t=2.0$, plots of density) for moving and static FV and DG methods at resolution $512^{\,\,2}$. As in Fig.~\ref{fig:implosion} we included the results of the moving DG method with the two different limiters.
Variations in colour reflect the local fluid density, as indicated by the colour bars to the
right of each frame.}
\label{fig:kh}
\end{figure*}

In the next test we consider shear flow in 2D which produces Kelvin
Helmholtz (KH) instabilities.  The initial conditions are those of
\cite{2010MNRAS.401..791S}.  In a periodic box of side length $1.0$ gas
is set up to have uniform 
pressure $p=2.5$ and adiabatic
index $\gamma = 5.3$.  The density is stratified vertically and has
value $\rho=2$ in the central (red) region, and $\rho=1$ in the 
regions at the top and bottom of the box (blue), as indicated 
in Fig.~\ref{fig:kh}.  As mixing
occurs, the other colours denote the corresponding 
intermediate fluid densities.
The velocity in the central horizontal strip
$|y-0.5|<0.25$ has value $v_x=0.5$ to the right, while the rest of
the box has the fluid moving to the left at $v_x=-0.5$. In addition we
add a perturbation:
\begin{equation}
\begin{split}
v_y(x,y) &= w_0 \sin(4\pi x) \\ &\times\left( {\rm exp}\left[-\frac{-(y-0.25)^2}{2\sigma^2}\right] +  {\rm exp}\left[-\frac{-(y-0.75)^2}{2\sigma^2}\right]  \right)
\end{split}
\end{equation}
with $w_0=0.1$ and $\sigma = 0.5/\sqrt{2}$ in order to excite a single mode of the instability with wavelength equal to half the box size.  The results of the various solvers are shown in Fig.~\ref{fig:kh}. The static DG method is better than the static FV method at resolving secondary KH instabilities that develop over the primary KH-billows, which the moving mesh approaches resolve. The moving FV and moving DG methods develop qualitatively similar structures, with small scale structures being well-preserved rather than mixed. The DG method that uses the original limiter (`limiter 2') resolves small features but also exhibits more diffusive mixing.

\subsection{3D subsonic driven turbulence}\label{sec:turbulence}

\begin{figure*}
\centering
\includegraphics[width=0.47\textwidth]{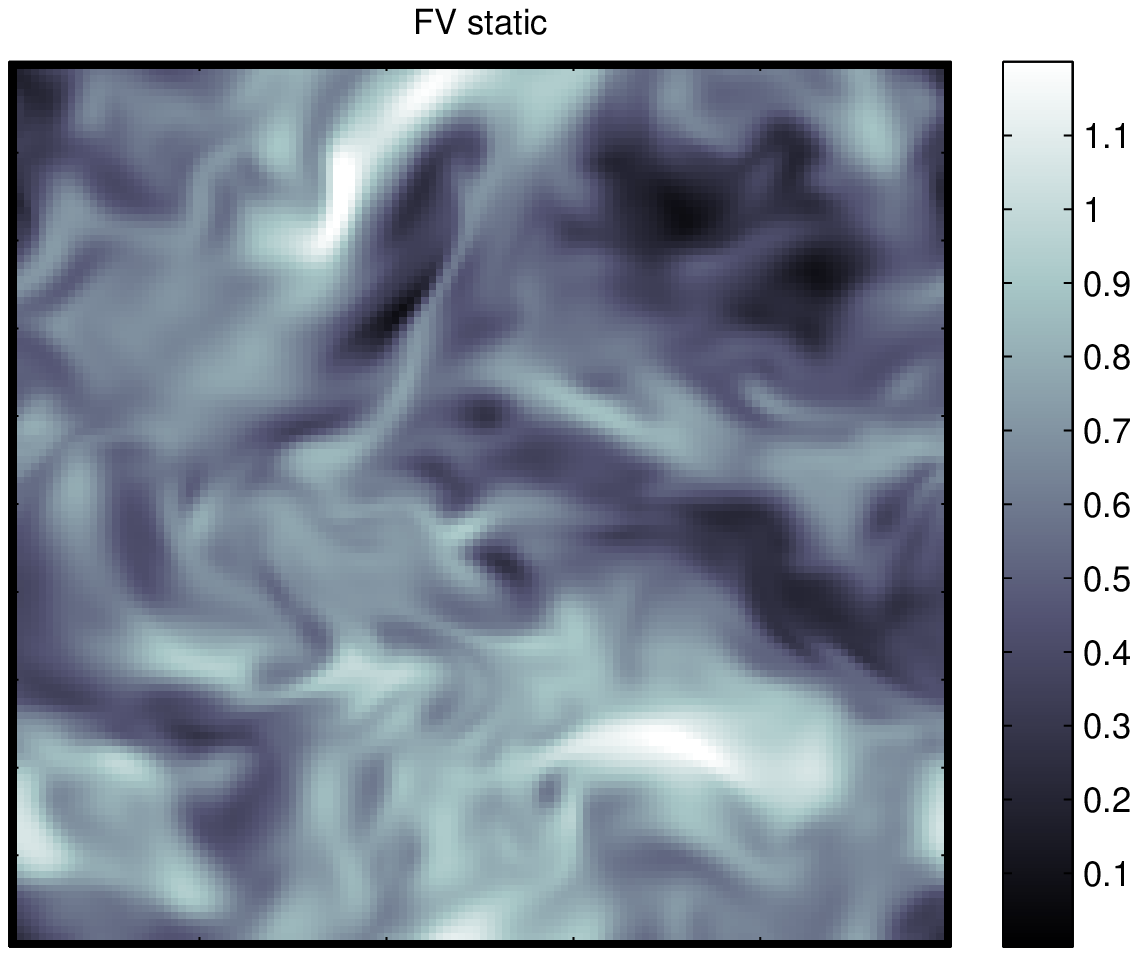}
\includegraphics[width=0.47\textwidth]{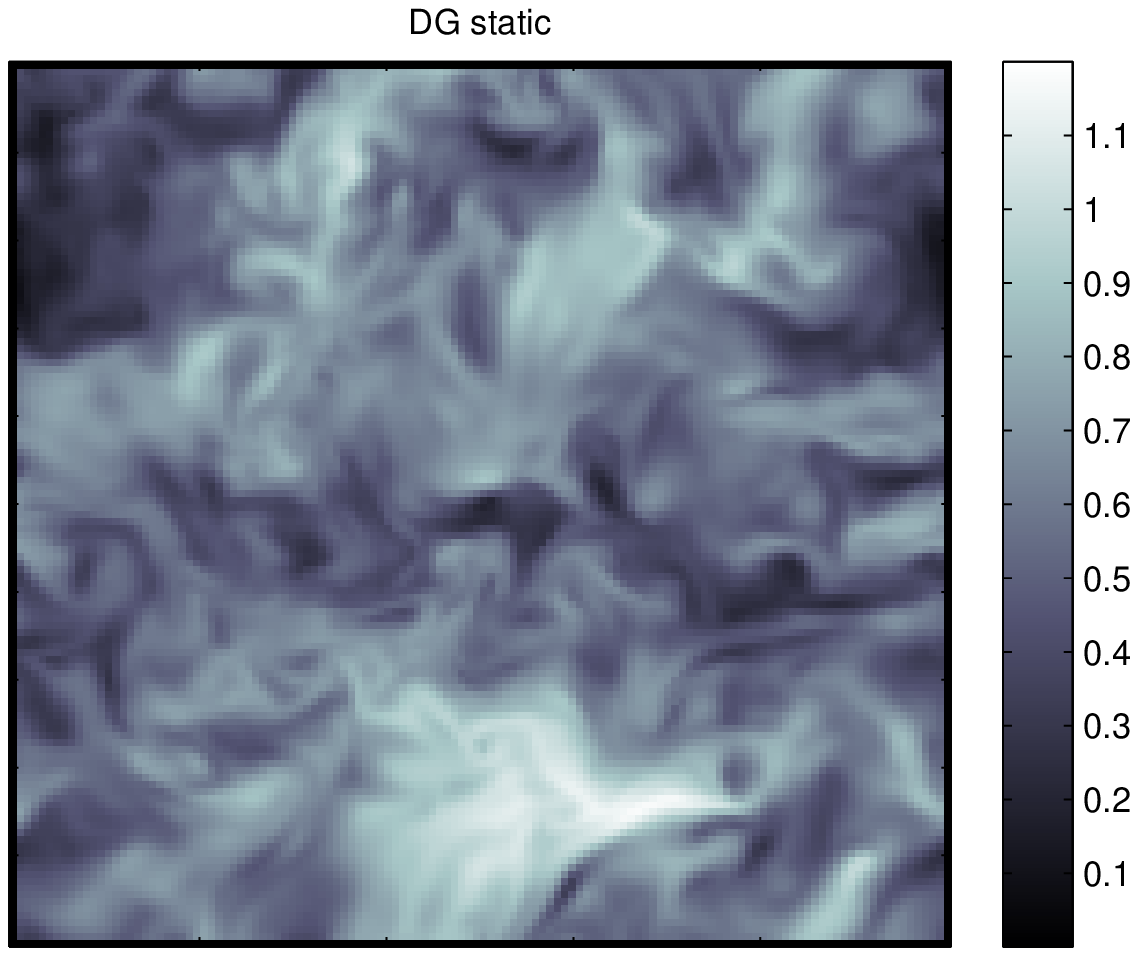}\\
\includegraphics[width=0.47\textwidth]{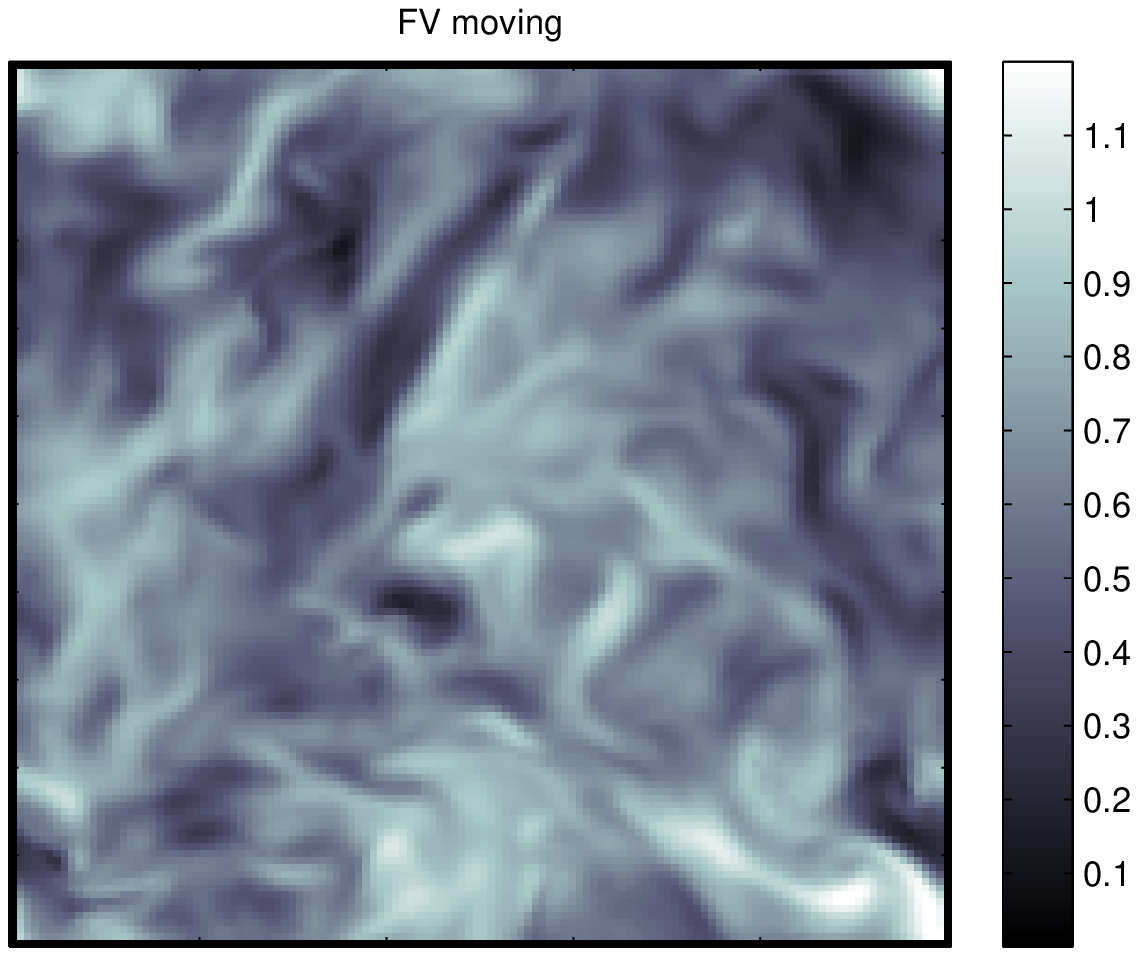}
\includegraphics[width=0.47\textwidth]{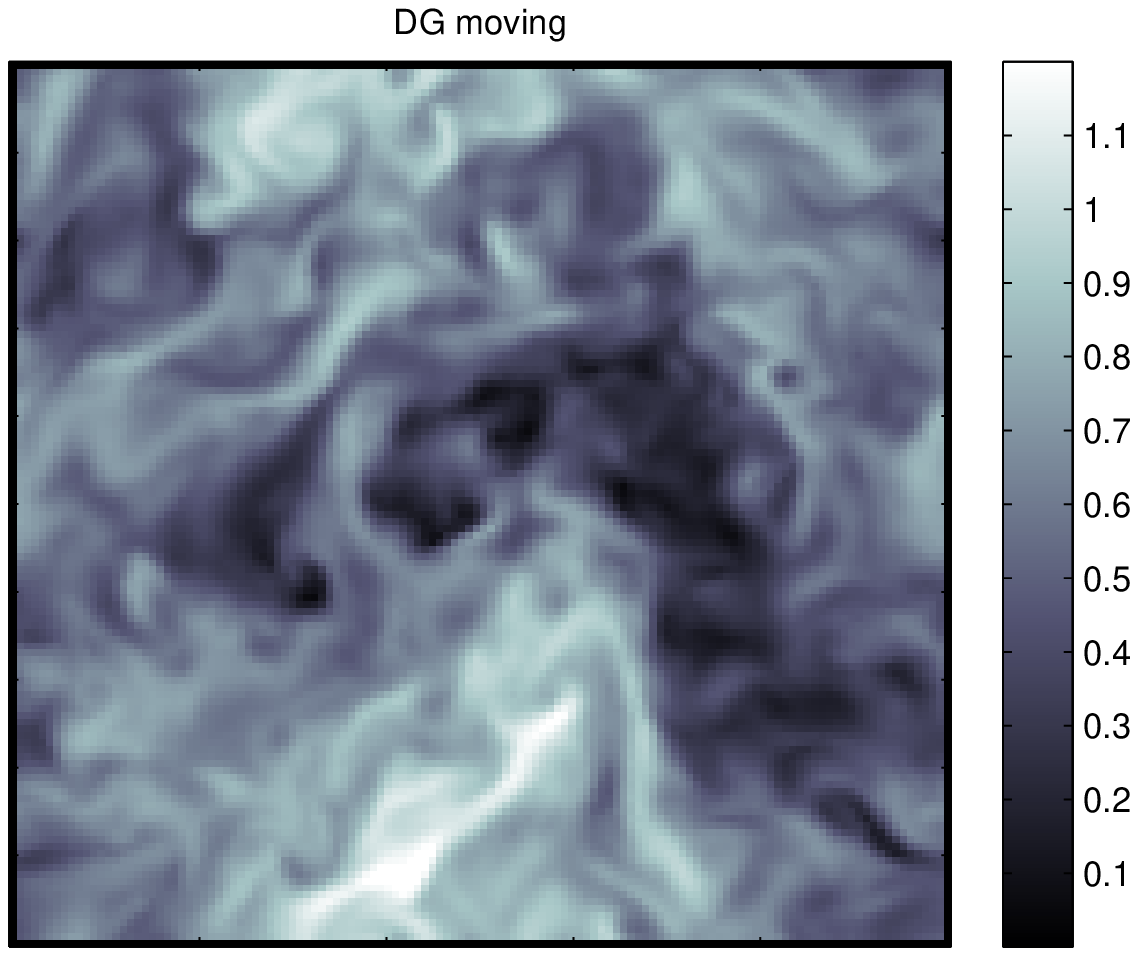}
\caption{Plots of a slice of the velocity magnitudes $|\mathbf{v}|$ in subsonic turbulently driven isothermal gas at $t=25.6$ for static and moving FV and DG methods in a 3D periodic box (resolution $128^3$).}
\label{fig:turbA}
\end{figure*}

\begin{figure}
\centering
\includegraphics[width=0.47\textwidth]{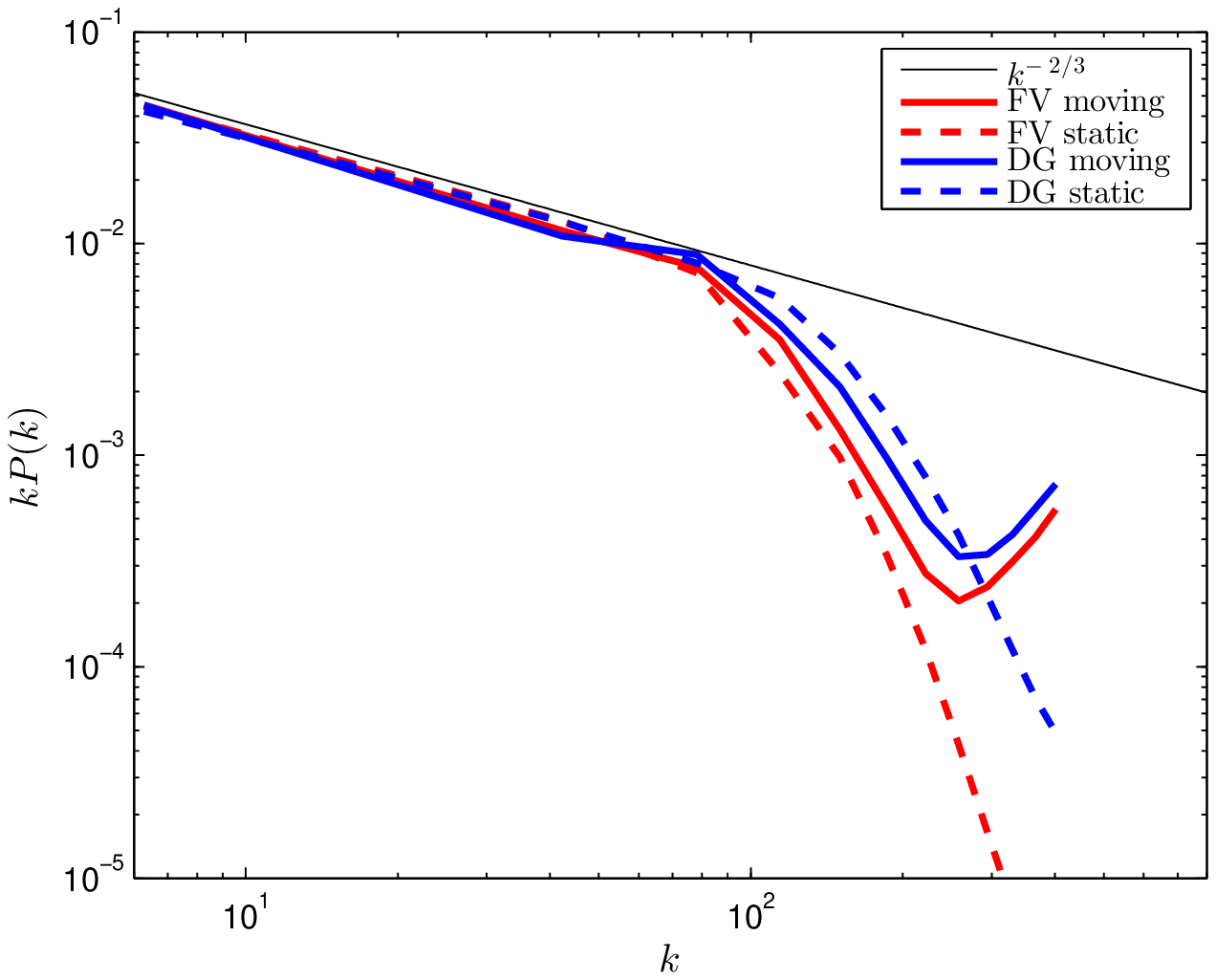}
\caption{Velocity power spectra for the subsonic turbulently driven isothermal gas computed with the static and moving FV and DG methods. On the largest spatial scales the results display a Kolmogorov cascade ($kP(k)\propto k^{-2/3}$). The DG method does an improved 
job of resolving the power to smaller spatial scales for the same number of cells.}
\label{fig:turbB}
\end{figure}

For a 3D test, we consider isothermal gas in a periodic box being turbulently driven by external stochastic forcing on large scales, as 
examined by \cite{2012MNRAS.423.2558B}. We use the driving routine and parameters for Mach number $M=0.3$ turbulence listed in Table 4 of \cite{2012MNRAS.423.2558B}. We are interested in how well the static and moving DG methods can reproduce 
a Kolmogorov-like velocity power spectrum ($P(k)\propto k^{-5/3}$). In Fig.~\ref{fig:turbA} we present plots of the velocity magnitudes with our various methods at $t=25.6$, computed at a resolution of $128^{3}$. In Fig.~\ref{fig:turbB} the accompanying velocity power spectra are presented. The results agree with the expectations for a Kolmogorov cascade on the largest spatial scales. The DG method does an improved job of resolving the power to smaller spatial scales with the same number of cells. The improvement of using the DG method over a FV method is greater than the improvement of using a moving mesh over a static one.

\subsection{Magnetic rotor}\label{sec:rotor}
We now move on to testing the MHD part of the code. First we consider the magnetic rotor test \citep{1999JCoPh.149..270B,Toth:2000:DBC:349920.349997}. The setup of this problem is as follows. A dense rotating disc of fluid is tapered off into the ambient fluid, which is at rest. The computational domain is a periodic box of side length $1$. The adiabatic index of the gas is $\gamma = 5/3$. The initial conditions are given by $p=0.5$, $B_x=2.5/\sqrt{4\pi}$, $B_y=0$, 
\begin{equation}
\rho = \begin{cases}
10 &  {\rm if\,} r \leq r_0 \\
1+f &  {\rm if\,} r_0 < r\leq r_1 \\
1 &  {\rm if\,} r > r_1
\end{cases}
\end{equation}
\begin{equation}
v_x = \begin{cases}
-(y-0.5)/r_0 &  {\rm if\,} r \leq r_0 \\
-f(y-0.5)/r &  {\rm if\,} r_0 < r\leq r_1 \\
0 &  {\rm if\,} r > r_1
\end{cases},
\end{equation}
\begin{equation}
v_y = \begin{cases}
(x-0.5)/r_0 &  {\rm if\,} r \leq r_0 \\
f(x-0.5)/r &  {\rm if\,} r_0 < r\leq r_1 \\
0 &  {\rm if\,} r > r_1
\end{cases},
\end{equation}
where $r_0 = 0.1$, $r_1 = 0.115$, $f=(r_1-r)/(r_1-r_0)$, $r^2=(x-0.5)^2+(y-0.5)^2$. In the rotor problem, centrifugal forces are not balanced so in the evolution the magnetic field confines the rotating dense fluid into an oblate rotor shape. 

The rotor problem can be a sensitive test for unphysical features that occur if the global divergence of the magnetic field is not 
sufficiently well-constrained.
\cite{1999JCoPh.149..270B,Toth:2000:DBC:349920.349997,Li:2005:LDD:1057321.1057352} find that the Mach number $M=|\mathbf{v}|/c$, where $c=\sqrt{\gamma p /\rho}$ is the sound speed, can suffer serious unphysical distortions around the central rotating area. 

\begin{figure*}
\centering
\includegraphics[width=0.47\textwidth]{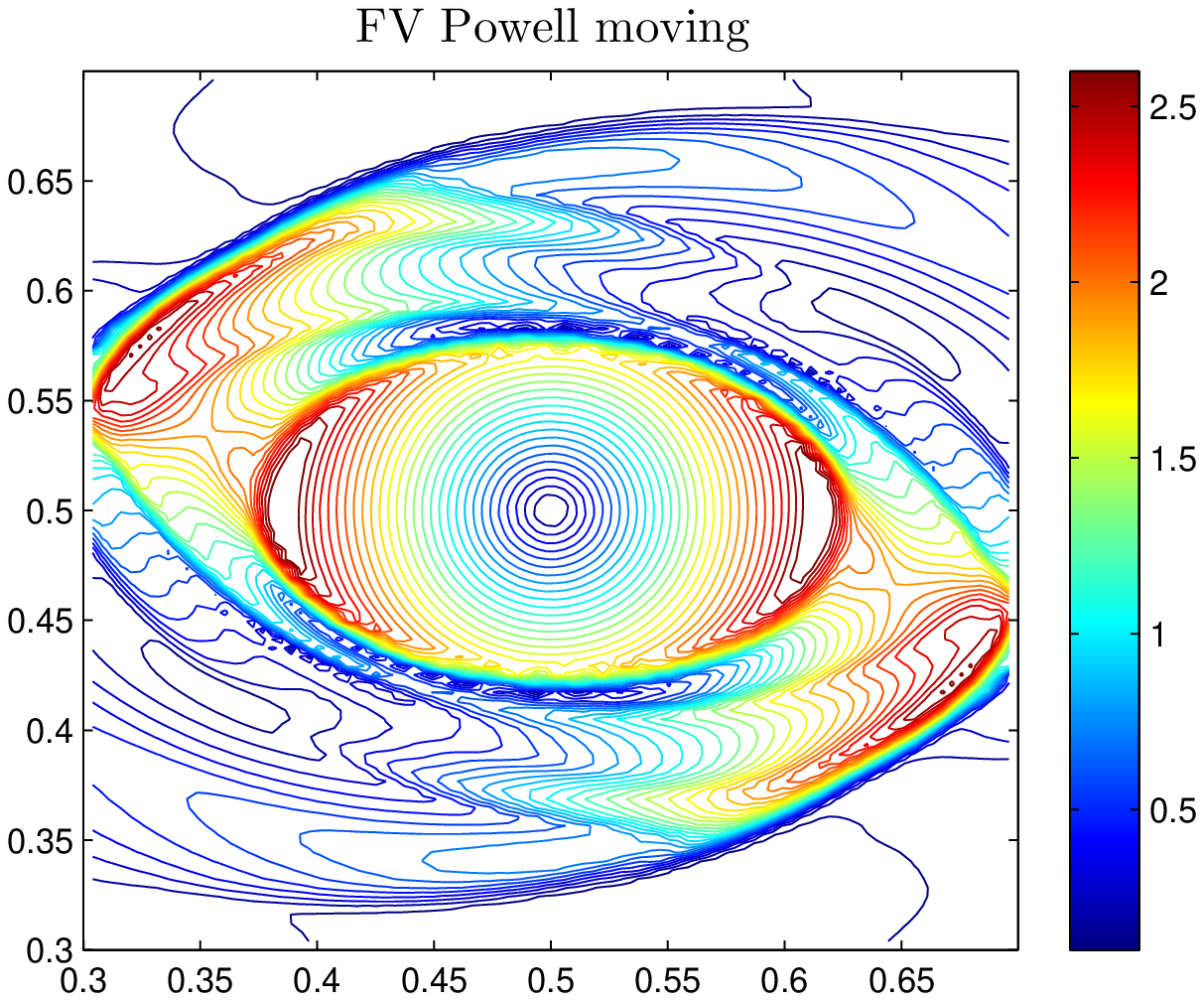}
\includegraphics[width=0.47\textwidth]{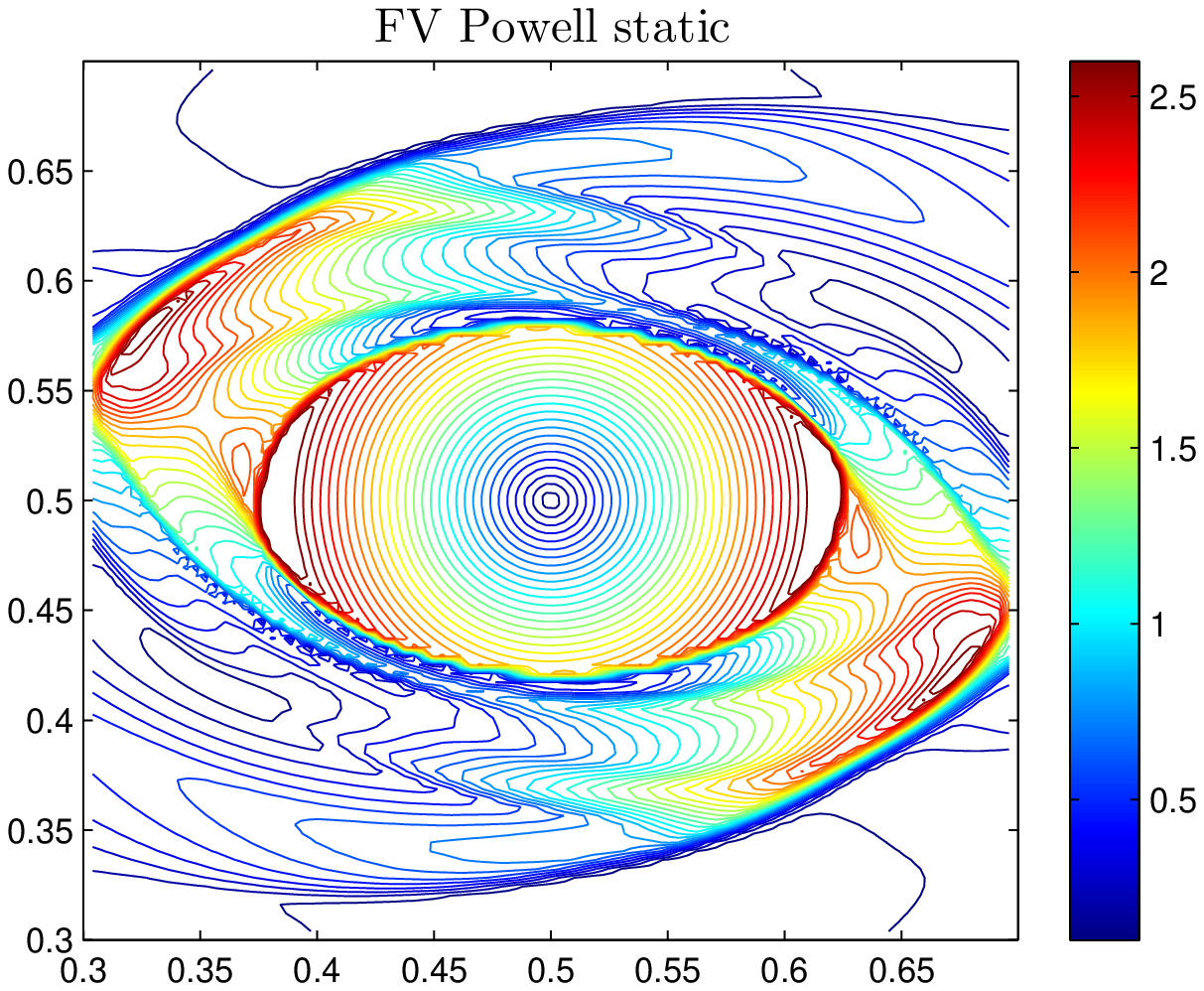}\\
\includegraphics[width=0.47\textwidth]{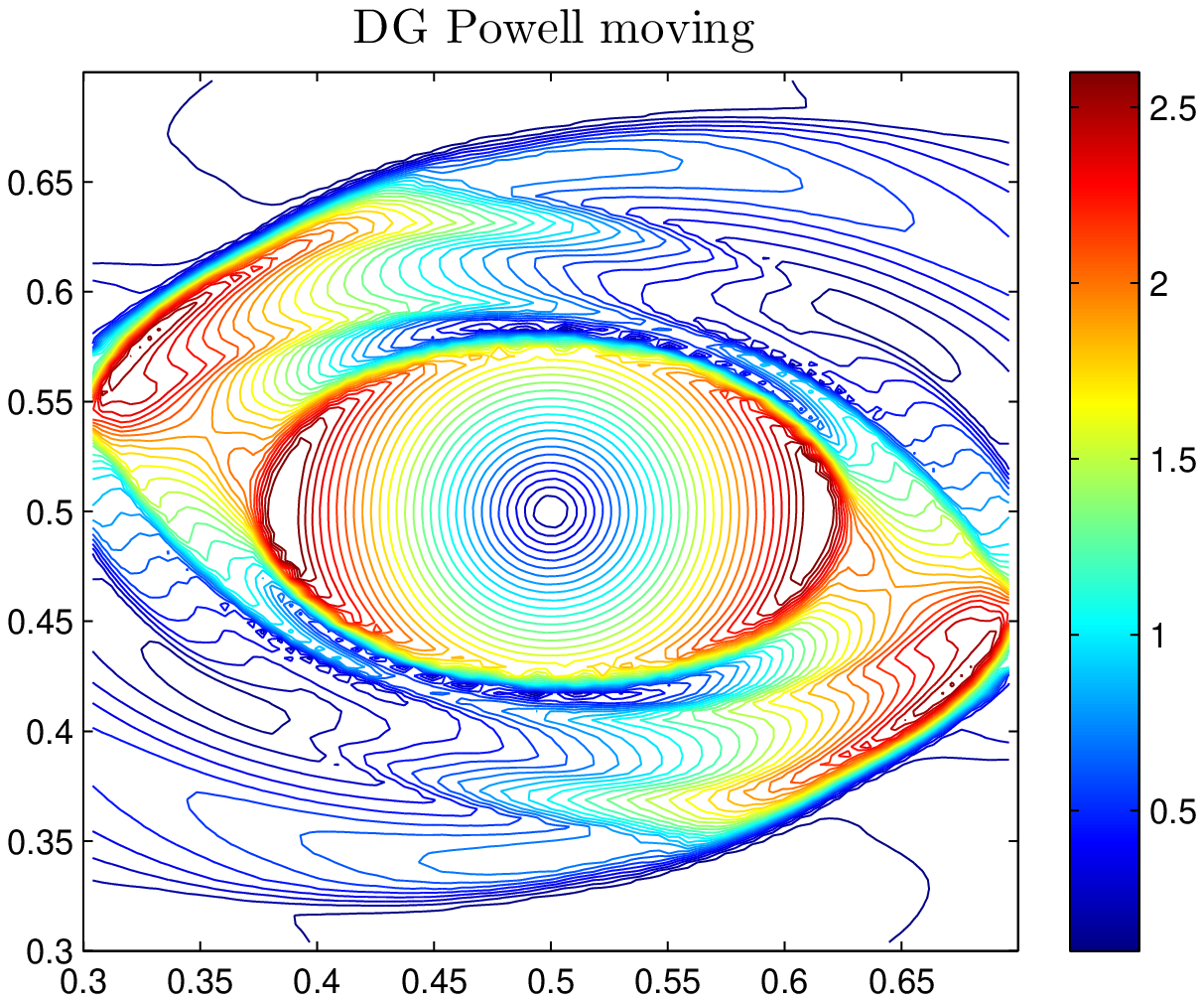}
\includegraphics[width=0.47\textwidth]{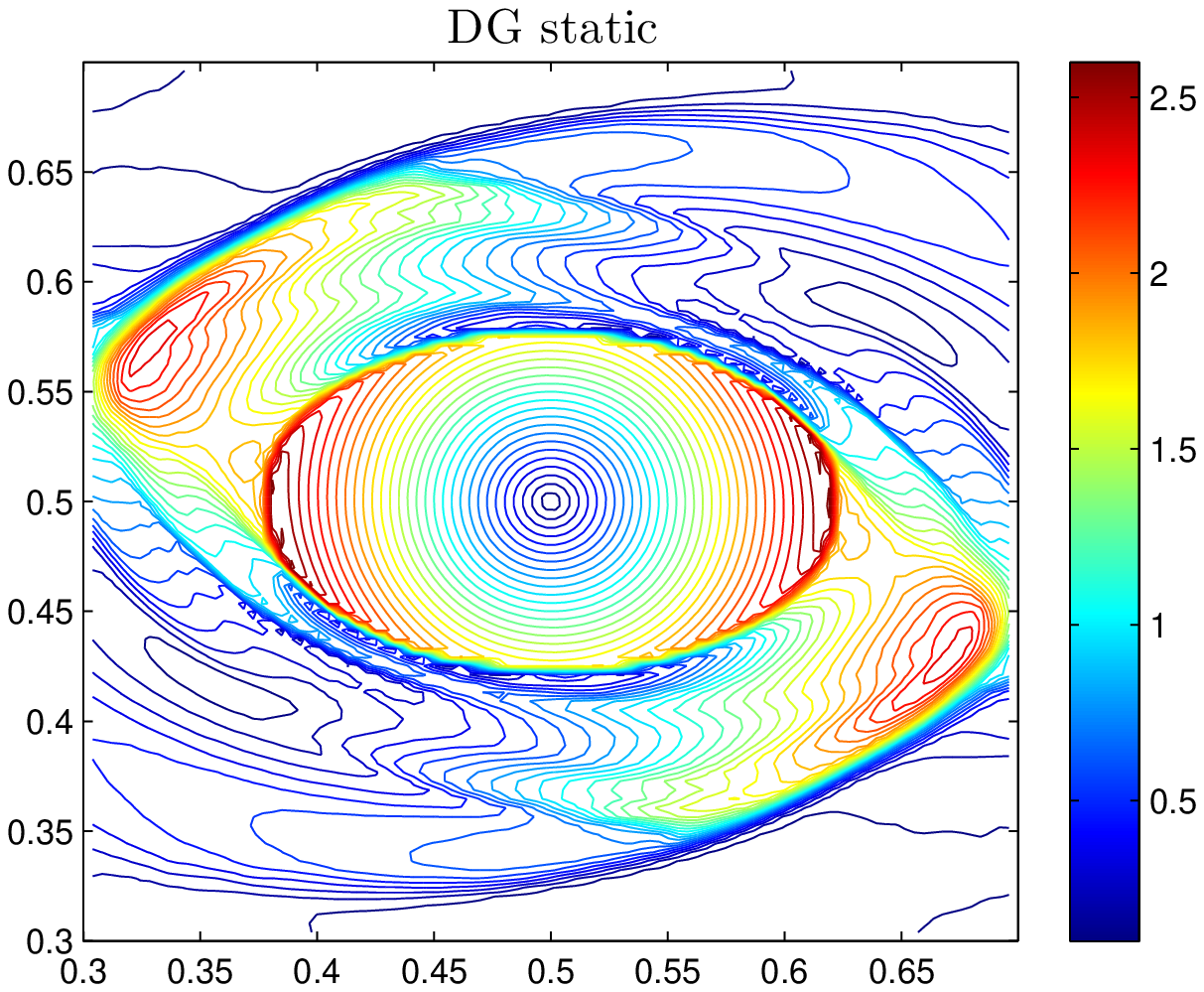}\\
\caption{Contour plots of Mach number in magnetic rotor at $t=0.295$ (resolution $512^{\,2}$) shown for static and moving FV methods with Powell cleaning, moving locally divergence-free DG method with Powell cleaning, and static locally divergence-free DG method without cleaning. The simulations do not
exhibit numerical artifacts typical of some MHD solvers that
employ cleaning schemes. The static locally divergence-free DG method does not require Powell cleaning.
Owing to our current choice of limiter for the moving DG approach, which takes a weighted average of local and stencil-determined gradients, we require Powell cleaning in this case.}
\label{fig:rotorA}
\end{figure*}

\begin{figure}
\centering
\includegraphics[width=0.47\textwidth]{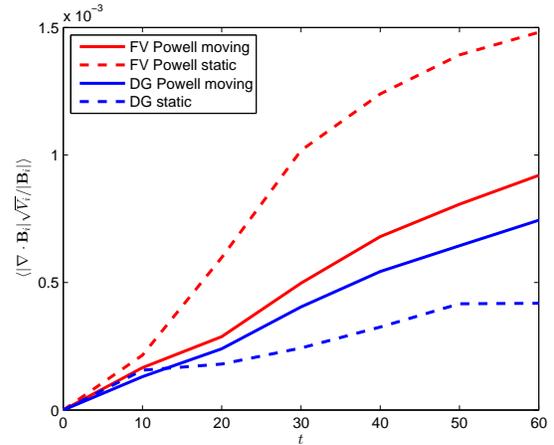}
\caption{Analysis of the global divergence of the magnetic field for
the four methods presented in Fig.~\ref{fig:rotorA} for the magnetic rotor problem. The DG methods demonstrate a better
handling of the global divergence errors, owing to their locally divergence-free formulation.}
\label{fig:rotorB}
\end{figure}

\begin{figure*}
\centering
\includegraphics[width=0.47\textwidth]{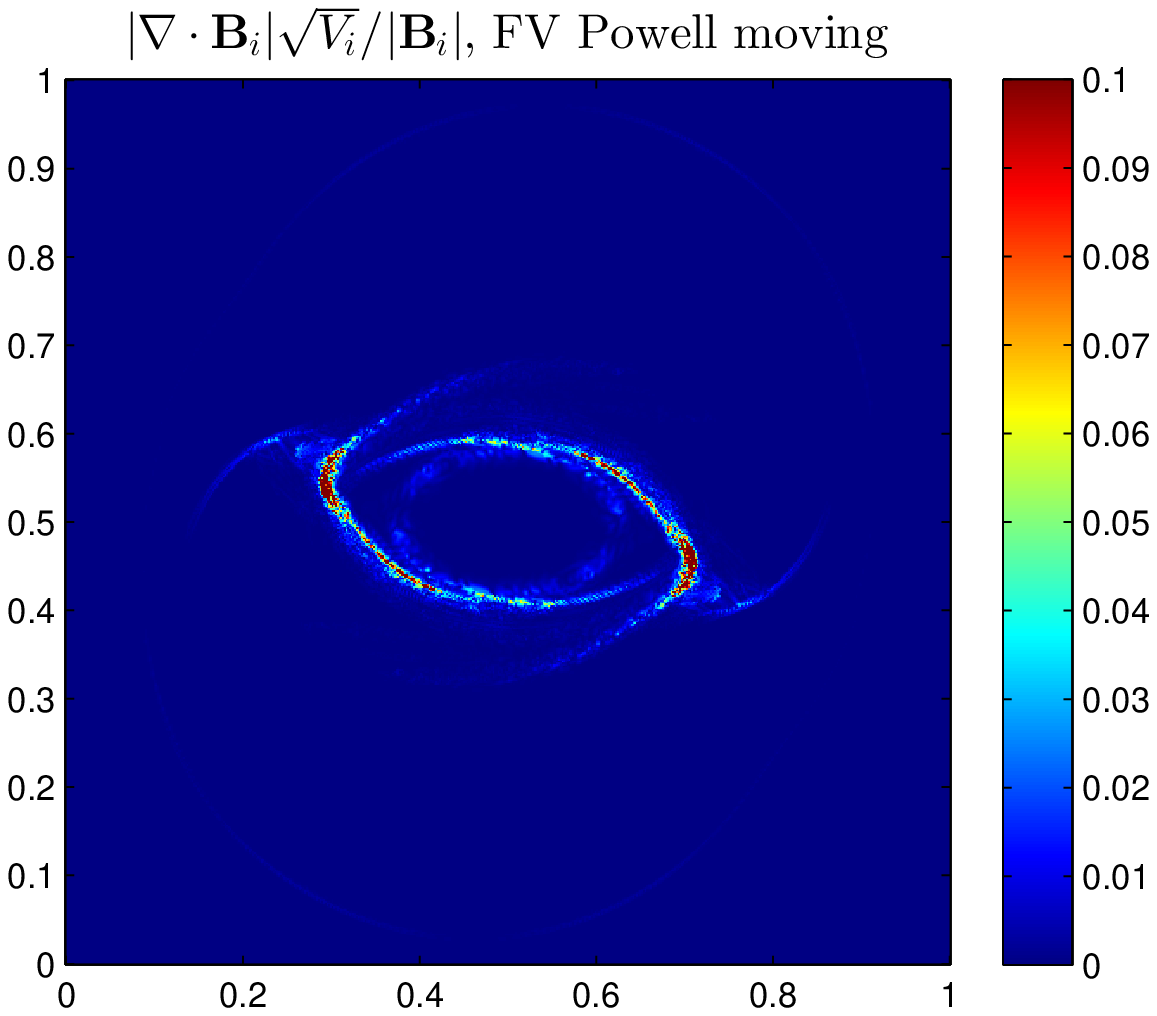}
\includegraphics[width=0.47\textwidth]{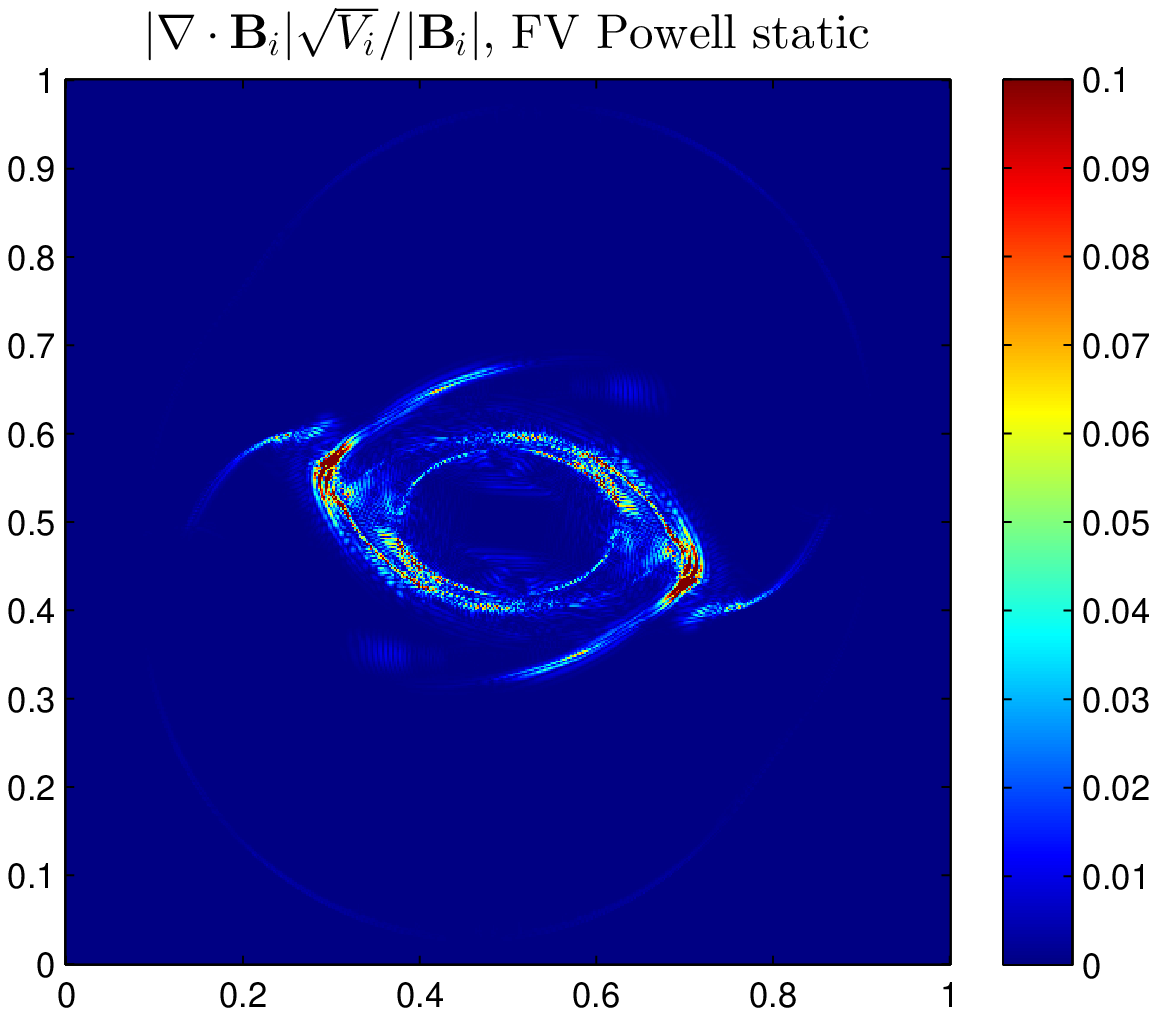}\\
\includegraphics[width=0.47\textwidth]{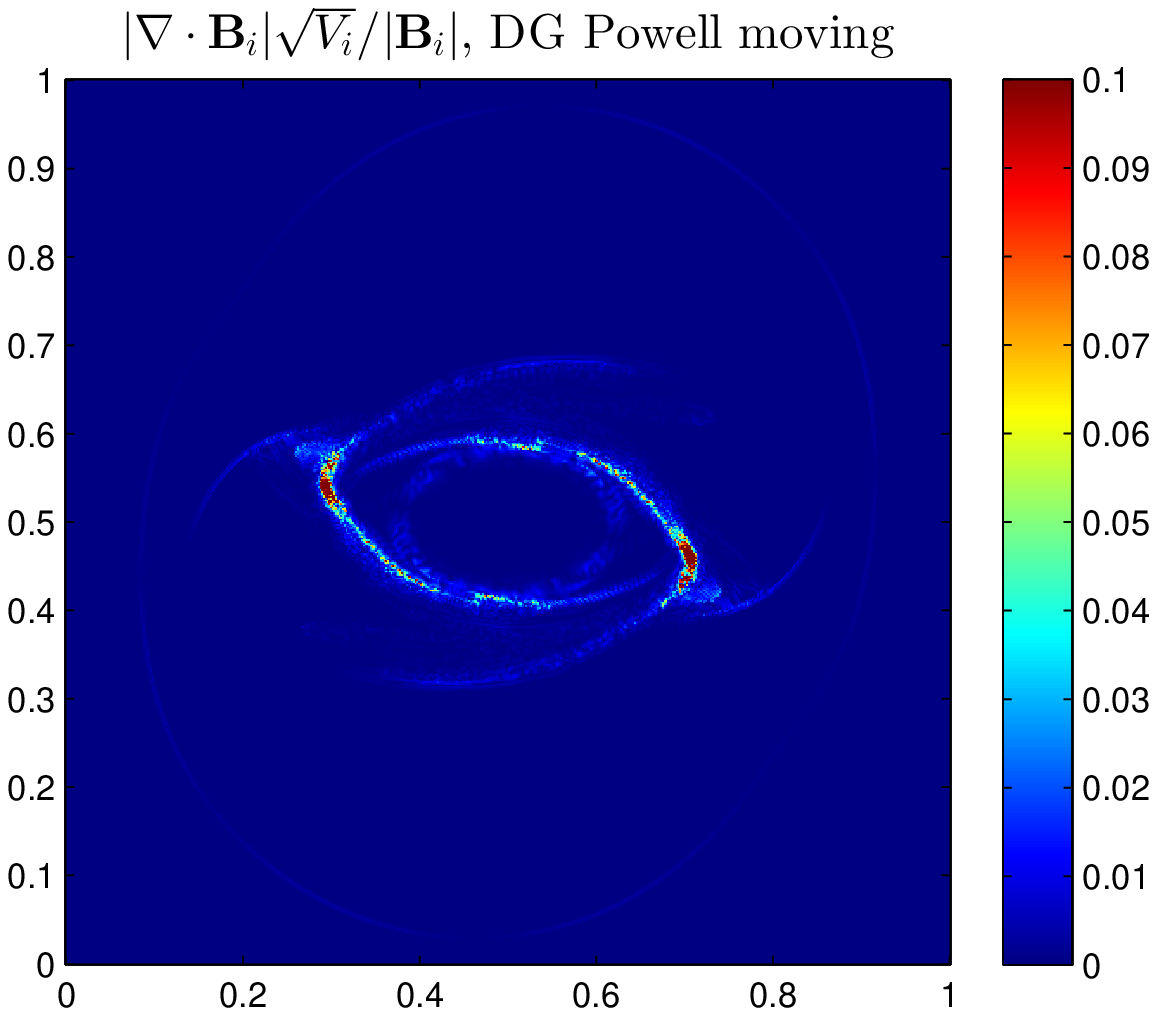}
\includegraphics[width=0.47\textwidth]{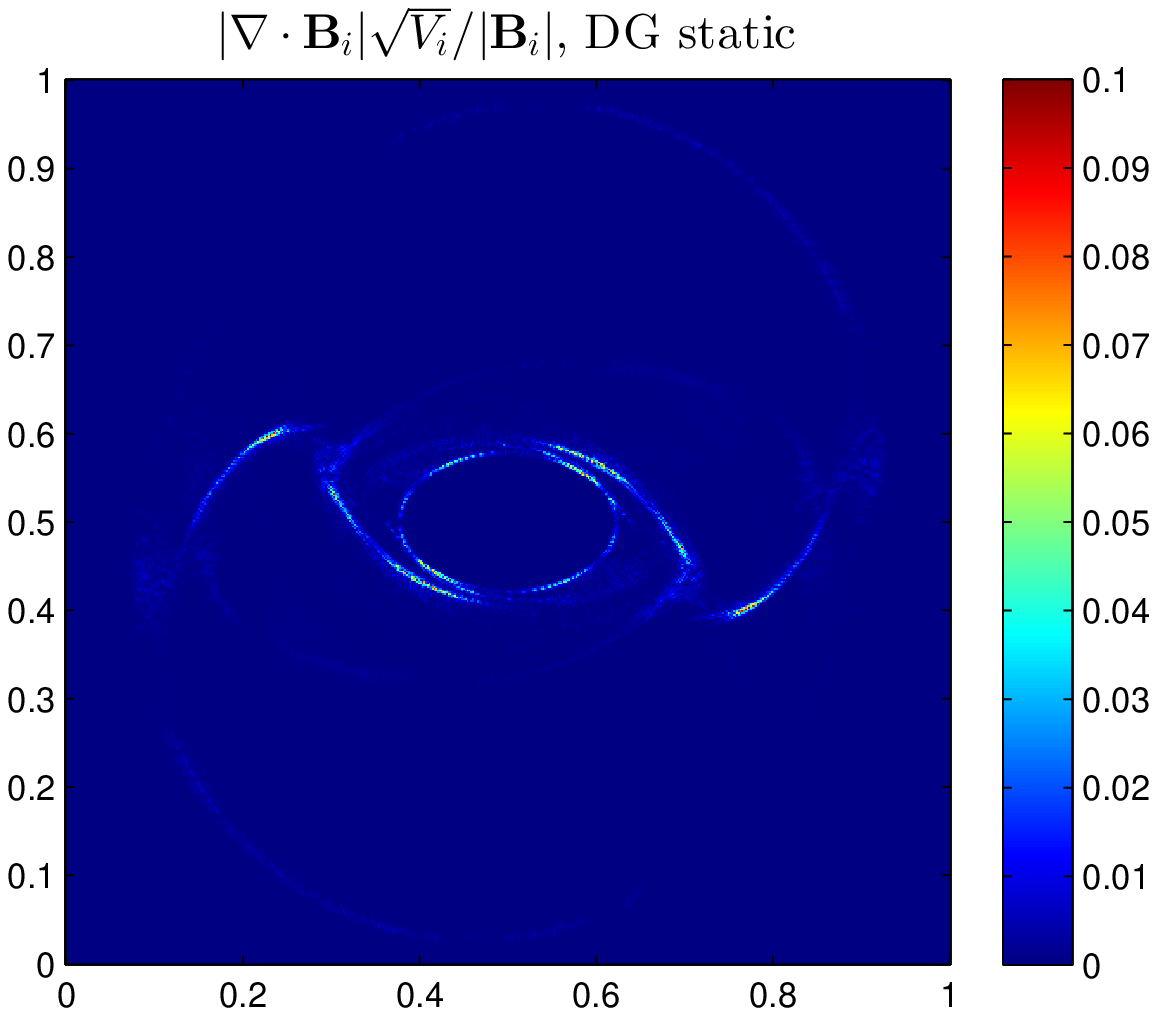}\\
\caption{Divergence errors in the magnetic field for magnetic rotor test corresponding to the plots in Fig.~\ref{fig:rotorA}.}
\label{fig:rotorC}
\end{figure*}

We show a zoom-in of the Mach number at the centre of the rotor in Fig.~\ref{fig:rotorA}, evolved with several of our schemes. No 
obvious unphysical artifacts are present in any of the simulations. 
The static mesh results show better-resolved features with some finer structures. This is due to the fact that in the moving mesh simulations, the mesh generating points move with the flow and there is actually a below-average density of mesh generating points in the center of the rotor, reducing the effective resolution in the zoom-in portion of the figure.  In principle, this could be overcome by allowing
local refinement of the mesh.
In Fig.~\ref{fig:rotorB} we show the global divergence errors of the magnetic field, and the divergence errors in each cell are presented in Fig.~\ref{fig:rotorC}. Even though the divergence errors did not have a
drastic impact on the solution, methods that minimize them
provide greater stability for solving arbitrary MHD problems. 
In the figure we see that the DG methods more tightly constrain 
the global divergence errors.
In particular, the static DG method is the most successful 
one at reducing the global divergence errors and,
in fact, does not require a cleaning scheme. It is followed by the moving DG method (which currently does require a Powell cleaning algorithm 
owing 
to the choice of limiter). The moving FV Powell method is third best, followed by the static Powell 
approach, which is considerably worse at constraining divergence errors compared to the static DG method which has no cleaning applied.

\subsection{Orszag-Tang vortex}\label{sec:orszagtang}
\begin{figure*}
\centering
\includegraphics[width=0.47\textwidth]{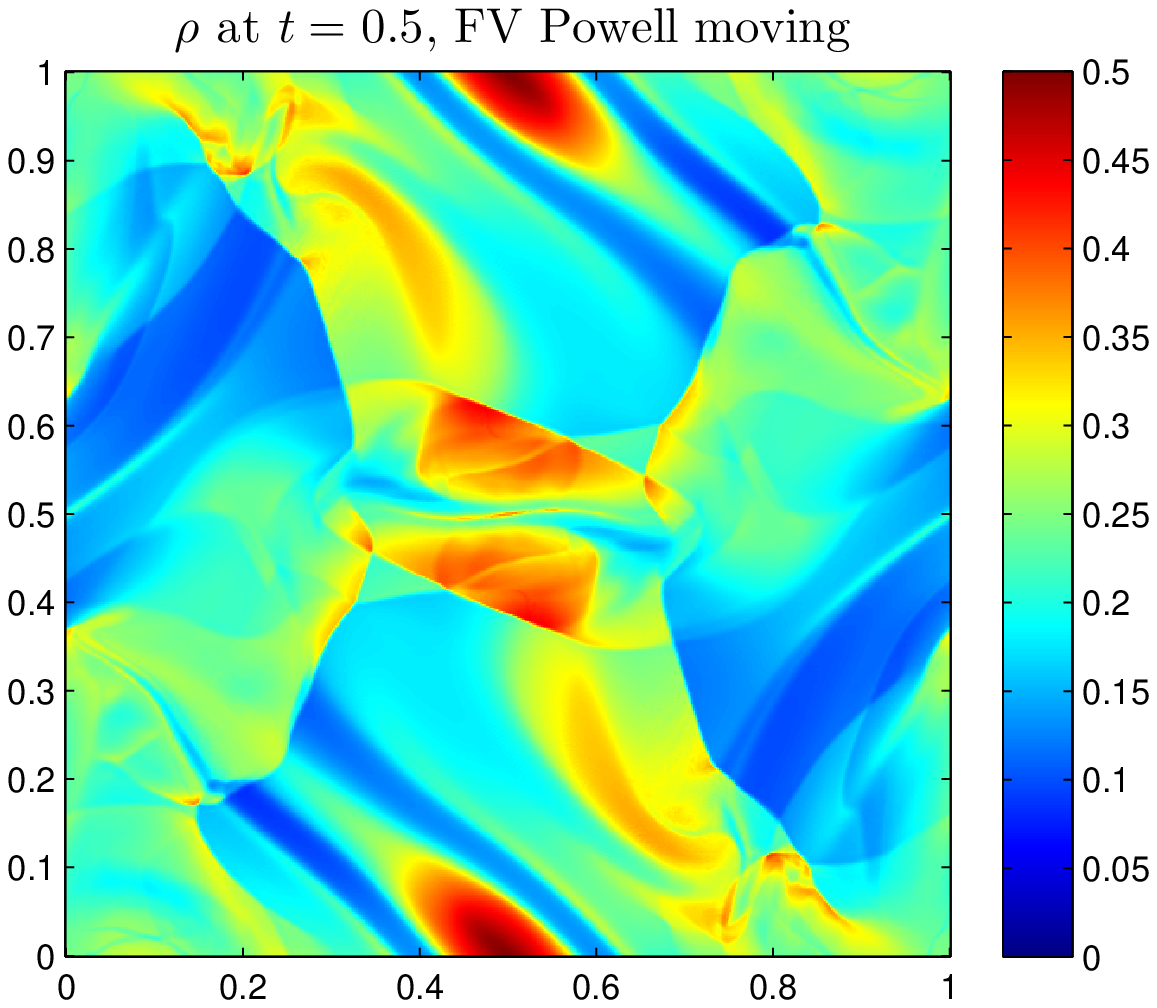}
\includegraphics[width=0.47\textwidth]{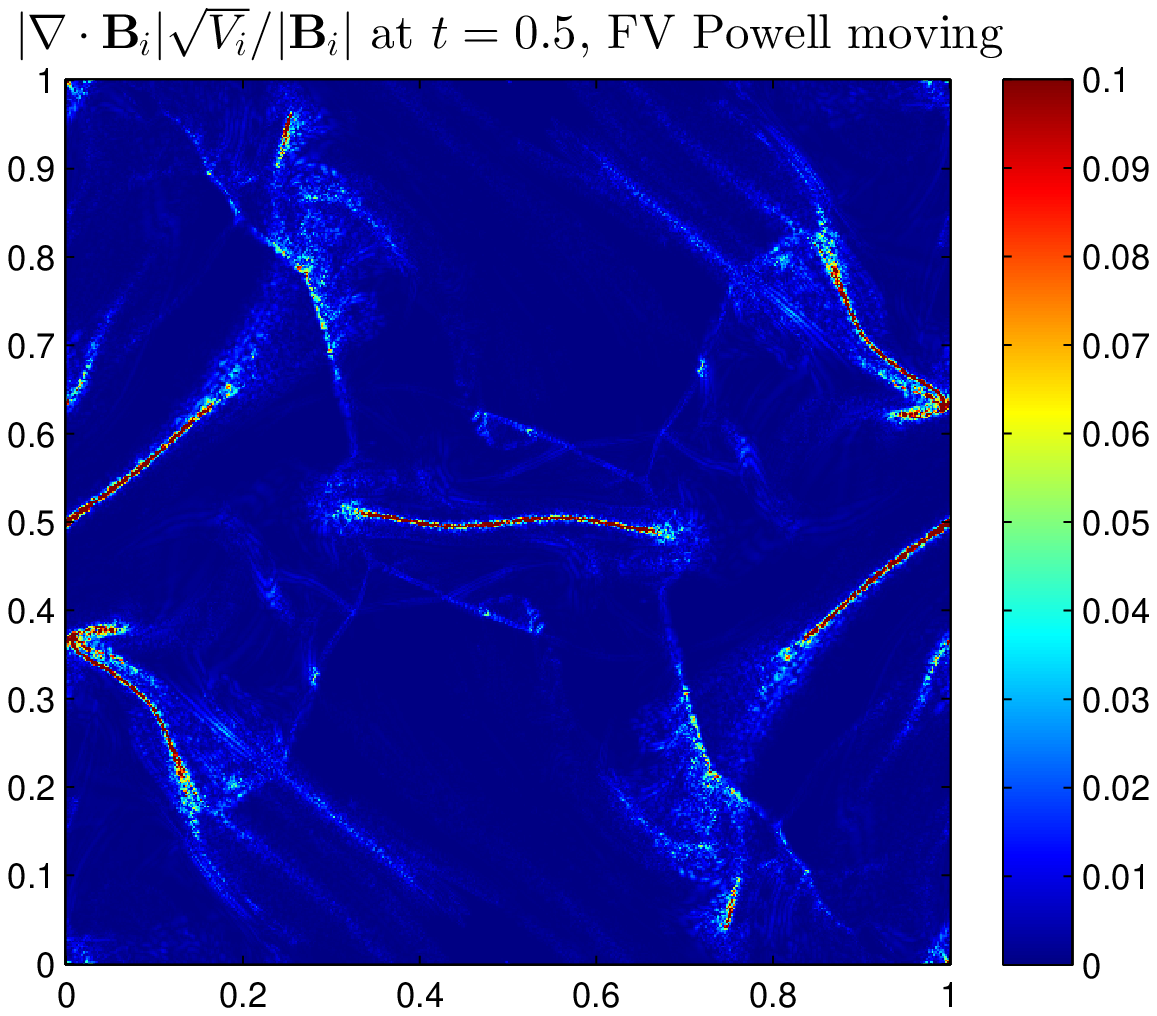}\\
\includegraphics[width=0.47\textwidth]{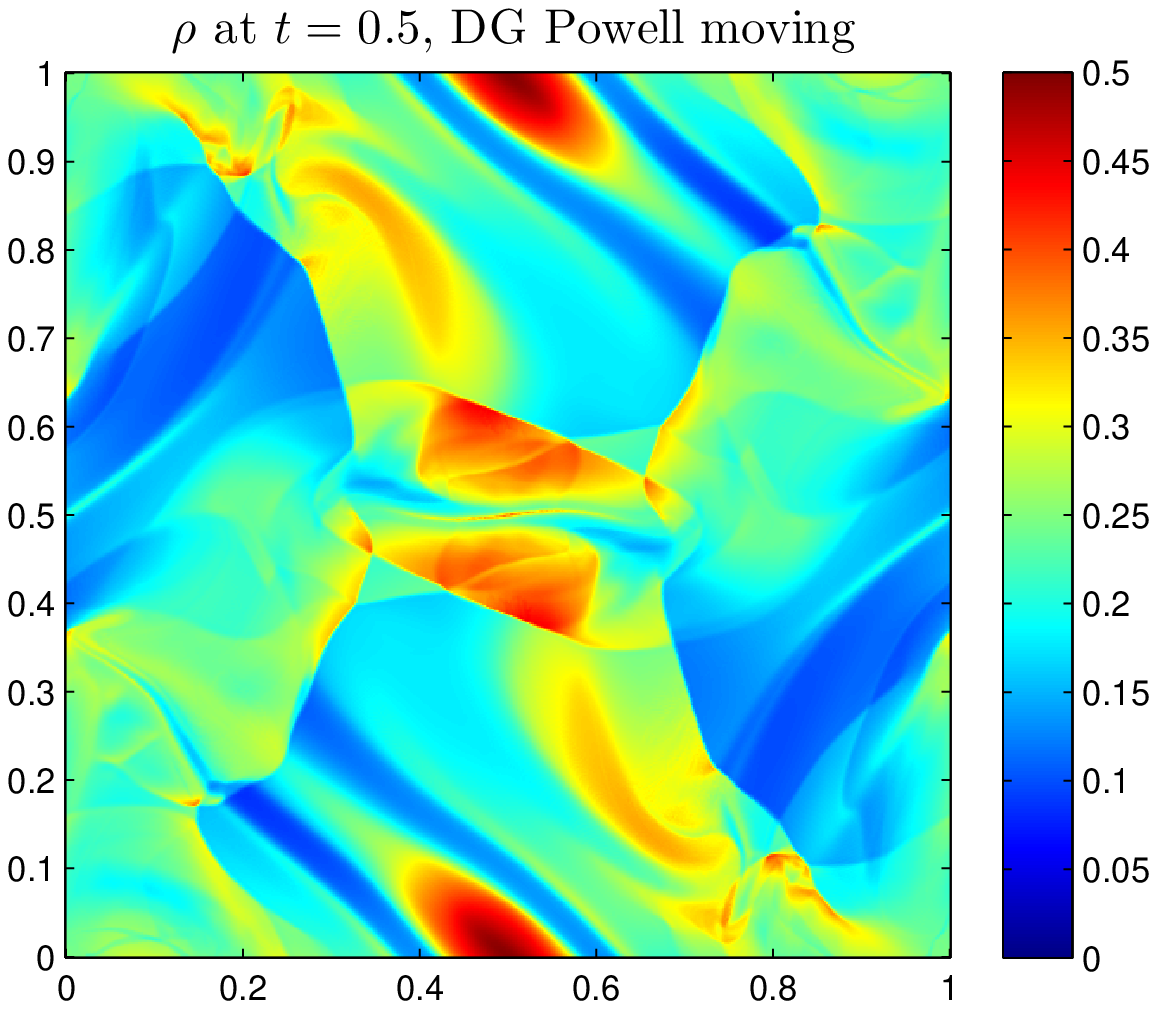}
\includegraphics[width=0.47\textwidth]{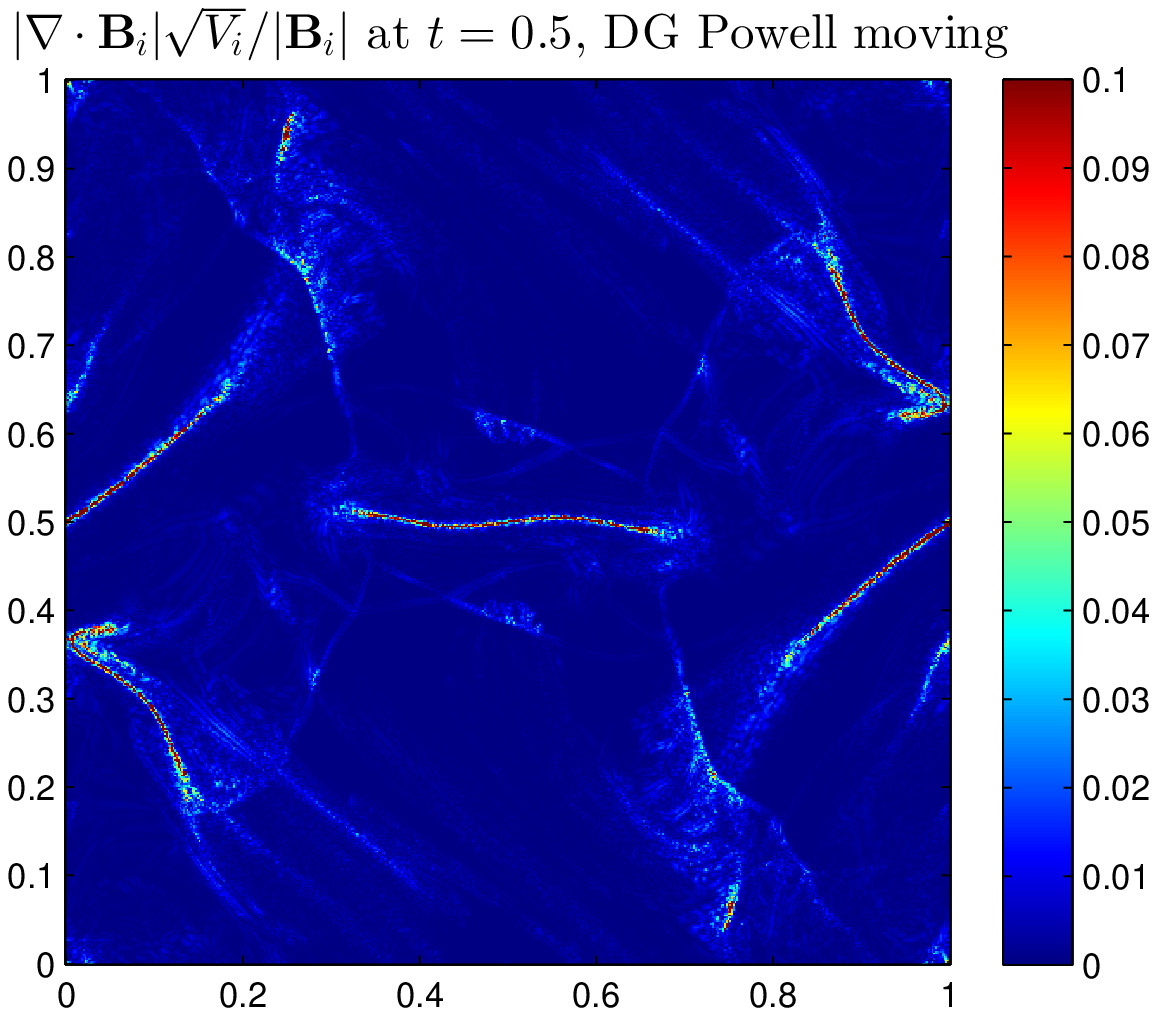}\\
\includegraphics[width=0.47\textwidth]{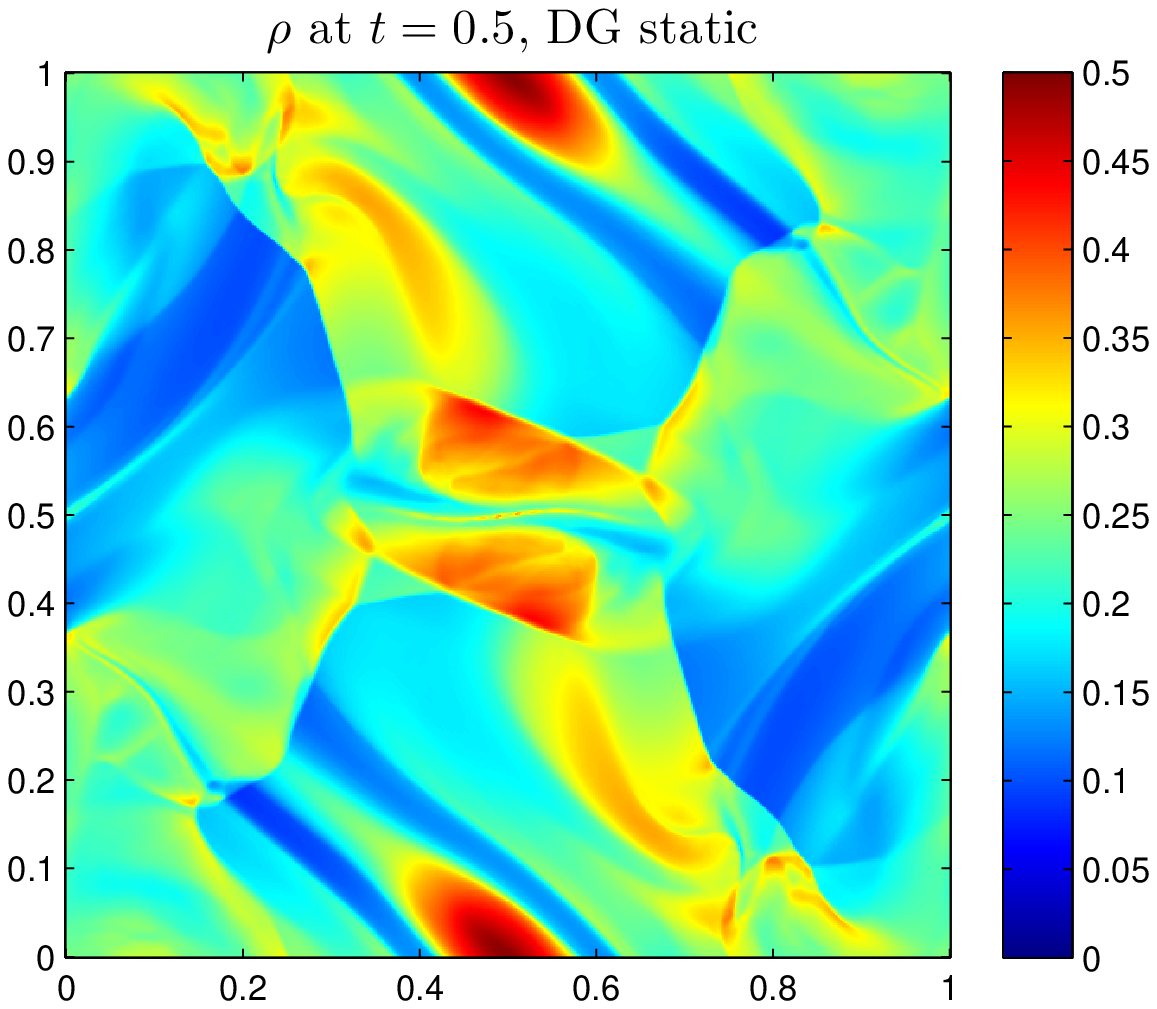}
\includegraphics[width=0.47\textwidth]{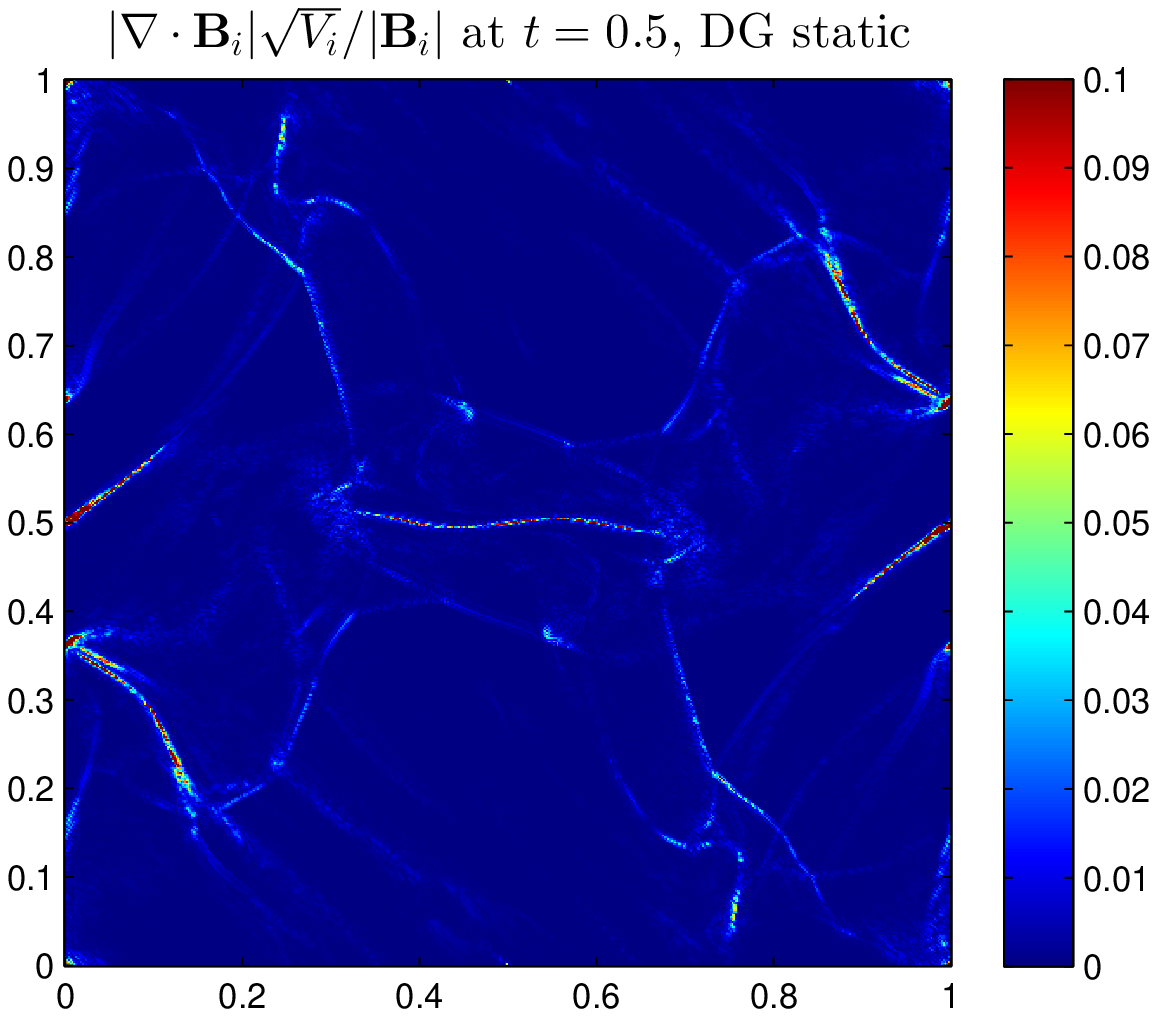}\\
\caption{Plots of density and local divergence errors in the Orszag-Tang test at $t=0.5$ (resolution $512^{\,\,2}$) for
moving FV method with Powell cleaning, moving locally divergence-free DG method with Powell cleaning, and static locally divergence-free DG method without cleaning, as labelled above each frame.
All simulations do a reasonable job
of arriving at an artifact-free solution. The static FV method with Powell cleaning (not shown) performs poorly and its solution becomes
corrupted by local divergence errors.}
\label{fig:orszagtangA}
\end{figure*}

\begin{figure}
\centering
\includegraphics[width=0.47\textwidth]{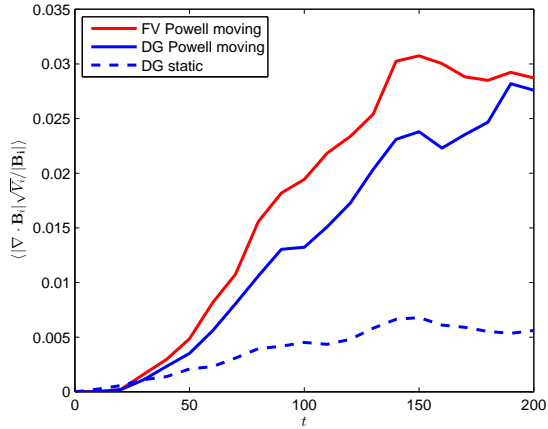}
\caption{Analysis of the global divergence of the magnetic field for
the three methods presented in Fig.~\ref{fig:orszagtangA} for the Orszag-Tang test. The DG methods demonstrate a better constraint
on the global divergence errors, owing to their locally divergence-free formulation.}
\label{fig:orszagtangB}
\end{figure}

As a final test, we consider the Orszag-Tang vortex \citep{1979JFM....90..129O},
which is an excellent test of supersonic MHD turbulence. We use the initial conditions as described by \cite{1991PhFlB...3...29P}:
\begin{equation}
\rho = \frac{\gamma^2}{4\pi} ,
\end{equation}
\begin{equation}
p = \frac{\gamma}{4\pi} ,
\end{equation}
\begin{equation}
\mathbf{v} = (-\sin(2\pi y), \sin(2\pi x)) ,
\end{equation}
\begin{equation}
\mathbf{B} = (-\sin(2\pi y), \sin(4\pi x)).
\end{equation}
The domain is a box of side length $1$
with periodic boundaries. The gas has adiabatic index $\gamma = 5/3$. We show the results of the simulations (density distribution and local cell $B$ field divergence errors) in Fig.~\ref{fig:orszagtangA} and Fig.~\ref{fig:orszagtangB}. The static DG method (which uses no Powell cleaning) maintains minimal divergence errors which plateau quickly. This is followed by the moving DG method with Powell cleaning and thirdly the FV moving method with Powell cleaning. We also ran a static FV simulation with the Powell cleaning scheme (not shown), which 
became unstable due to the growth of large magnetic field divergence errors which corrupted the solution. 

\section{Summary of comparison to the finite volume approach}\label{sec:disc1}

On a static mesh, the DG formulation has clear advantages over the FV formulation. 
Our tests show a significant reduction of 
errors and angular momentum diffusion, and increased effective resolution which better 
characterizes small-scale fluid instabilities and recovers a Kolmogorov-like power law for turbulent cascade 
to smaller scales. In addition, the locally divergence-free representation of the magnetic field allows the MHD equations to be solved in a robust and stable manner without large global divergence errors. No magnetic field cleaning scheme is required in
this case.

The DG method on a moving mesh also shows improvement over its FV counterpart in every test we performed. 
Numerical errors, angular momentum diffusion, and post-shock oscillations are reduced and 
the capability of resolving turbulence on small scales is enhanced. 
The advantages of DG over FV in our moving mesh
formulation are not always as great as for the 
static mesh case because we had to employ 
a modified slope limiter to prevent unphysical oscillations. We suspect that 
refined limiters would allow us to more fully exploit the advantages of DG over FV, which we leave
for future work.

\subsection{Memory consumption and CPU time}\label{sec:disc1a}

The DG method does require greater memory usage and more CPU time
than the FV approach. In
our own implementation, the memory allocated to store local cell
variables (fluid variables, fluid variable gradients, volumes,
moments) is increased by $40$~per~cent for 3D simulations (for both
the static and moving cases) due to the fact that in the DG method we
require the second-order moments of cells in addition to their volumes
and we also store moment-averaged derivative quantities $\mathbf{R}_e$ in
addition to primitive gradients. However, this results in only a small
net increase in total memory consumption in our implementation ($<10$~per
cent). A significant portion of the total memory is dedicated to
storing mesh connectivity information, which is the same for the DG
and FV formulations.

The CPU usage 
is increased by approximately $25$ per cent for static 3D DG runs and $35$ per cent for moving mesh 3D runs. This is due to the fact that in the DG method we are required to perform additional steps, specifically reading gradient information in the input, writing gradient information at every snapshot, calculating second-order cell moments with Gaussian quadrature in addition to cell volumes, calculating flux update terms for the moment-averaged gradient quantities at every timestep, and converting these gradients to primitive gradients.  For 
astrophysical applications that require self-gravity, the actual penalty in CPU consumption of the
DG approach relative to the FV method will be significantly less than this.

\section{Strengths of the DG method in astrophysical contexts}\label{sec:disc2}

The second-order DG method developed here shows improvements in
accuracy over the second-order FV MUSCL-Hancock approach without
significant increase in computation time or memory consumption. The
procedure is readily compatible with hierarchical time stepping and
mesh refinement.  The performance of our DG method on simple test problems suggest that it would improve simulations of cosmological
structure formation such as those performed with {\sc Arepo}
(e.g. \cite{2012MNRAS.425.2027K, 2012MNRAS.427.2224T, 2013MNRAS.429.3353N,
2013arXiv1305.2913V}).

The locally divergence-free representation of the magnetic field in
the DG method reduces global divergence errors in
$\nabla\cdot\mathbf{B}$. It is desirable to use this DG representation
of the solution in cases where CT is not applicable. Such is the case
for a moving mesh, where it is presently unclear whether the CT
approach can be adapted to an evolving unstructured mesh. The locally
divergence-free DG technique improves the current FV Powell scheme
used in {\sc Arepo} to solve the MHD equations
\citep{2012arXiv1212.1452P}. The method would be useful in studying,
for example, the potentially important role magnetic fields play in
accretion processes and explosions (e.g. \cite{2011ApJS..197...15D,
2012ApJ...755....7D, 2013arXiv1302.7306D}, where the moving mesh formulation is needed
to minimize large advection errors from bulk flows
\citep{2013arXiv1305.2195G}.

The DG method can be generalized to provide higher order accuracy on
arbitrary meshes and does not adversely affect the parallelizability
of current fluid solvers because cells continue to communicate with
only their nearest neighbours.  In this case, a higher order time
stepping scheme, such as Runge-Kutta, would be preferable, to maintain
the same order of accuracy in both the space and time domain.

Finally, we note that the DG method we have implemented shows
significant reduction of errors and angular momentum diffusion
compared to the FV method.  It could play an important role in
improving the current generation of AMR codes.

\section{Improving the slope limiter in future work}\label{sec:disc3}

We find that our moving DG formulation is sensitive to the choice of
slope limiter and we cannot use the same slope limiter as we do for
the static DG and moving and static FV method described in
\cite{2010MNRAS.401..791S}.  In addition to capturing shock
discontinuities, this limiter identifies and limits smooth extrema and
produces unphysical oscillations in the solution.  Our proposed
alternate limiter, designed as a very simple WENO-type limiter, works
well and provides stable results.  However some of the advantages
gained by DG over FV are not as great as in the static mesh case owing
to our choice of limiter, and so an investigation for more refined
limiters is clearly a priority for future efforts.  WENO and Hermite
WENO type approaches \citep{Luo:2007:HWL:1276530.1276745} seem to be a
promising avenue to explore.  These limiters replace solution
polynomials with reconstructed polynomials of the same order of
accuracy which also preserve cell averages (hence are fully
conservative).

\section{Concluding remarks}\label{sec:conclusion}

We have presented a new numerical procedure for solving the fluid and
MHD equations on moving and static meshes based on the DG method. The
technique is an attractive and competitive alternative to the
predominant FV approaches used in astrophysics.  The DG scheme we
developed, which is based on the centroidal Taylor basis expansion in
each cell, is in fact a generalization of the FV method, where local
gradients (and higher order derivatives in general) are evolved just
like fluid variables, instead of using a stencil to estimate their
values.  In this way, gradients are purely local. The second-order DG
procedure we developed does not significantly increase the runtime of
the simulations compared to the FV approach. The DG method can also be
readily extended to higher-order while keeping inter-element
communications minimal (elements only communicate with adjacent
elements with a common face), unlike FV schemes. This allows for
higher-order DG codes to be highly parallelizable. In addition, the DG
formulation is well-suited for unstructured meshes and mesh refinement
strategies, since the method is compact (the representation of the
solution on each element is independent).

On static meshes, second-order DG techniques demonstrate superior
accuracy over the same-order FV method. Particularly striking is the
reduction of angular momentum diffusion. As a result, DG schemes could
reduce the disadvantage Eulerian codes have compared to SPH (which
conserves angular momentum, but has difficulties in other areas such
as resolving fluid instabilities). Moreover, in the DG formulation the
magnetic field can be represented in a locally divergence-free form,
which leads to a stable scheme for solving the MHD equations without
the need of a cleaning scheme. The method shows superior control of
global B-field divergence errors over the Powell cleaning scheme. Of
course, CT schemes are preferred whenever possible because they
restrict the divergences of magnetic fields to zero to machine
precision. However, it is presently unknown how to extend the CT
procedure to arbitrary meshes or to arbitrary hierarchical time
stepping schemes. The locally divergence-free DG method is therefore
seen to have possible applications in AMR simulations with adaptive
highly flexible, hierarchical time stepping, required for large-scale
cosmological runs.

On moving meshes, our DG method also shows improvement over the FV
approach.  A challenge with the DG formulation on moving meshes is
that the solution can be sensitive to the choice of slope limiter, and
we cannot use the same limiter as we do for DG on a static mesh.
Performing slope limiting on a cell at a local minimum or local
maximum can produce unphysical oscillations in the DG gradient
solutions. To prevent this, we have currently implemented a slope
limiter that takes a weighted (by an oscillation factor) average of
the local DG unlimited slope and the slope obtained by a stencil
(projected onto a divergence-free basis in the case of magnetic
fields). This allows our method to be stable, and decreases
convergence errors, post-shock oscillations, and angular momentum
diffusion as well as enhancing resolution for describing turbulence
given a fixed number of cells. These improvements already make the new
procedure desirable over the FV approach. But we have not yet
maximally exploited all the advantages of the DG scheme as suggested
by the comparison of the DG and FV methods on a static mesh. We do
still require a Powell cleaning scheme for MHD simulations (on moving meshes only) since our
gradients are a combination of the DG gradients and stencil
gradients. Even so, we reduce global divergence errors compared to
moving mesh FV simulations, which helps promotes the stability of MHD
simulations on moving meshes.  Further refinements to the DG method
on a moving mesh and an exploration of different limiters form the
basis for future work in this area.

The significant advantages and desirable features the centroidal
Taylor-basis DG method offers over the FV approach can lead to
improvements in grid-based astrophysical simulations, even at the
second-order level of accuracy.  The numerical results we present
indicate the potential of the DG method to be a competitive procedure
for large-scale astrophysical problems.

\section*{Acknowledgments}
This material is based upon work supported by the National Science Foundation Graduate Research Fellowship under Grant No. DGE-1144152. Any opinion, findings, and conclusions or recommendations expressed in this material are those of the authors(s) and do not necessarily reflect the views of the National Science Foundation. The simulations in this paper were run on the Odyssey cluster supported by the FAS Science Division Research Computing Group at Harvard University. PM would like to thank V. Springel, D. Mu\~noz, D. Nelson, and P. Torrey for useful discussions about the {\sc Arepo} code architecture.  LH acknowledges support from NASA grant NNX12AC67G. MV acknowledges support from NASA through Hubble Fellowship grant HST-HF-51317.01.

\bibliography{mybib}{}

\bsp
\label{lastpage}
\end{document}